\documentclass[a4paper,11pt]{article}
\pdfoutput=1 

\usepackage{jheppub} 

\usepackage[T1]{fontenc} 

\usepackage{comment}
\usepackage{amsmath}    
\usepackage{amssymb}    %
\usepackage{graphicx}   
\usepackage{verbatim}   
\usepackage{color}      
\usepackage{subfigure}  
\usepackage{hyperref}
\usepackage{multirow}
\usepackage{comment}
\usepackage{pdflscape}
\usepackage{rotating}
\usepackage{booktabs,tabularx}
\usepackage{float}	
\usepackage{color}
\definecolor{rosso}{RGB}{210,0,0}
\renewcommand\arraystretch{1.1}

\def\LO{{\rm LO}}
\def\NLO{{\rm NLO}}
\def\LNLO{{\rm (N)LO}}
\def\NLOQCD{{\rm NLO}_{\rm QCD}}
\def\NLOEW{{\rm NLO}_{\rm EW}}
\def\LOQCD{{\rm LO}_{\rm QCD}}

\def\pt{p_{T}}
\def\pttt{p_{T}(t\bar t)}

\def\alphas{\alpha_s}

\def\wpm{ W^{\pm}}
\def\ttwm{t \bar t W^-}
\def\ttwp{t \bar t W^+}
\def\ttw{t \bar t W^{\pm}}
\def\ttt{t \bar t}
\def\ft{t \bar t t \bar t}

\def\beq{\begin{equation}}
\def\beqn{\begin{eqnarray}}
\def\eeq{\end{equation}}
\def\eeqn{\end{eqnarray}}
\def\beal{\begin{align}}
\def\endal{\end{align}}

\newcommand\mf{{\sc\small MadFKS}}
\newcommand\ml{{\sc\small MadLoop}}
\newcommand\ct{{\sc\small CutTools}}
\newcommand\nin{{\sc\small Ninja}}
\newcommand\coll{{\sc\small Collier}}

\newcommand{\MSb}{\overline{\rm MS}}

 \newcommand{\ord}[1]{\mathcal{O}(#1)}

\title{Large NLO corrections in $t\bar{t}W^{\pm}$ and $t\bar{t}t\bar{t}$ hadroproduction from supposedly subleading EW contributions}

\author[a]{Rikkert Frederix,}
\author[a]{Davide Pagani,}
\author[b,c,d]{and Marco Zaro}

\affiliation[a]{Technische Universit\"{a}t M\"{u}nchen, James-Franck-Str.~1, D-85748 Garching, Germany}
\affiliation[b]{Nikhef, Science Park 105, NL-1098 XG Amsterdam, The Netherlands}
\affiliation[c]{Sorbonne Universit\'es, UPMC Univ. Paris 06, UMR 7589, LPTHE, F-75005, Paris, France}
\affiliation[d]{CNRS, UMR 7589, LPTHE, F-75005, Paris, France}

\note{Preprint:  TUM-HEP-1106/17, NIKHEF/2017-57}

\abstract{We calculate the complete-NLO predictions for
  $t\bar{t}W^{\pm}$ and $t\bar{t}t\bar{t}$ production in
  proton--proton collisions at 13 and 100 TeV. All the non-vanishing
  contributions of $\mathcal{O}(\alpha_s^i \alpha^j)$ with $i+j=3,4$
  for $t\bar{t}W^{\pm}$ and $i+j=4,5$ for $t\bar{t}t\bar{t}$ are
  evaluated without any approximation.  For $t\bar{t}W^{\pm}$ we find
  that, due to the presence of $tW \to tW$ scattering, at 13(100) TeV
  the $\mathcal{O}(\alpha_s \alpha^3)$ contribution is about 12(70)\%
  of the LO, {\it i.e.}, it is larger than the so-called NLO EW
  corrections (the $\mathcal{O}(\alpha_s^2 \alpha^2)$ terms) and has
  opposite sign.  In the case of $t\bar{t}t\bar{t}$ production, large
  contributions from electroweak $t t \to t t$ scattering are already
  present at LO in the $\mathcal{O}(\alpha_s^3 \alpha)$ and
  $\mathcal{O}(\alpha_s^2 \alpha^2)$ terms. For the same reason we
  find that both NLO terms of $\mathcal{O}(\alpha_s^4 \alpha)$, {\it
    i.e.}, the NLO EW corrections, and $\mathcal{O}(\alpha_s^3
  \alpha^2)$ are large ($\pm 15 \%$ of the LO) and their relative
  contributions strongly depend on the values of the renormalisation
  and factorisation scales. However, large accidental cancellations
  are present (away from the threshold region) between these two
  contributions. Moreover, the NLO corrections strongly depend on the
  kinematics and are particularly large at the threshold, where even
  the relative contribution from $\mathcal{O}(\alpha_s^2 \alpha^3)$
  terms amounts to tens of percents. }

\begin{document} 
\maketitle
\flushbottom

\section{Introduction}
Precise predictions for  Standard-Model (SM)  processes at high-energy colliders are an essential ingredient for a correct and reliable comparison between experimental data and theories describing the  fundamental interactions of Nature. At the LHC and future colliders, the capability of performing further consistency checks for the SM as well as the possibility of identifying beyond-the-Standard-Model (BSM) effects  critically depend on the size of the theory uncertainties.

At high-energies,  SM calculations can be performed in a perturbative approach. Thus, the precision of the prediction for a generic observable can be successively  improved by taking into account higher-order effects. In particular, the so-called fixed-order calculations  consist in the perturbative expansion in powers of the two SM parameters $\alpha_s$ and $\alpha$.
The former parametrises strong interactions and its value is roughly 0.1 at the TeV scale or at the typical energy scales involved at the LHC. The latter parametrises electroweak (EW) interactions and its value is roughly 0.01. On the other hand, EW interactions also depend on the mass of the $W$ and $Z$ bosons (or alternatively on any other three independent parameters for the EW gauge sector)  and the masses of the fermions and the Higgs boson. 
 
Typically, the leading-order (LO) contribution for a specific process is given by the first non-vanishing terms of $\ord{\alpha_s^i \alpha^j}$, {\it i.e.}, those with the smallest value for $i+j$ and the largest value of $i$. For this reason, ``LO prediction'' in general refers to this level of accuracy, which is not sufficiently precise for almost all processes at the LHC. The calculation of next-to-LO (NLO) predictions in QCD, which consists in the inclusion of $\ord{\alpha_s^{i+1} \alpha^j}$ terms, can be performed automatically and with publicly available tools~\cite{Campbell:1999ah,Cullen:2011ac,Cullen:2014yla,Badger:2012pg,Cascioli:2011va,Actis:2012qn,Actis:2016mpe,Gleisberg:2008ta,Bevilacqua:2011xh,Hirschi:2011pa,Alwall:2014hca,Alioli:2010xd,Platzer:2011bc} for most of the processes. Recently, also NLO EW corrections, which consist of $\ord{\alpha_s^{i} \alpha^{j+1}}$ terms, have  been calculated via (semi-)automated tools~\cite{Cascioli:2011va,Kallweit:2014xda, Kallweit:2015dum,  Actis:2012qn, Actis:2016mpe, Biedermann:2017yoi, Alwall:2014hca, Frixione:2014qaa, Frixione:2015zaa, Frederix:2016ost,Chiesa:2015mya,Greiner:2017mft} for a large variety of processes.

 Being $\alpha<\alpha_s$,  NLO EW corrections are typically smaller than NLO QCD corrections at the  inclusive level, but they can be considerably enhanced at the differential level due to different kinds of effects such as weak Sudakov enhancements or  collinear photon final-state-radiation (FSR) in sufficiently exclusive observables. Thus, they have to be taken into account for a reliable comparison to data. For many production processes at the LHC, also next-to-NLO (NNLO) QCD corrections, the $\ord{\alpha_s^{i+2} \alpha^j}$ contributions, are essential and indeed many calculations have appeared in the recent years (see, {\it e.g.}, ref.~\cite{Heinrich:2017una} and references therein). Even the next-to-NNLO (${\rm N^3 LO}$) QCD calculation for the Higgs production cross section is now available \cite{Anastasiou:2015ema, Dreyer:2016oyx}. 

From a technical point of view, NLO QCD and EW corrections are simpler than NNLO corrections; they involve at most one loop or one additional radiated parton more than the LO calculation. However, they are not the only perturbative orders sharing this feature. Already starting from $2 \to 2$ processes with coloured and EW-charged initial- and final-state particles, such as dijet or top-quark pair hadroproduction, additional NLO terms appears. For these two processes, one-loop and real-emission corrections in the SM involve also $\ord{\alphas \alpha^2}$ and $\ord{\alpha^3}$ terms, which are neither part of the NLO QCD corrections nor of the NLO EW ones. Moreover,  Born diagrams originate also $\ord{\alphas \alpha}$ and $\ord{\alpha^2}$ contributions, which are typically not included in LO predictions. The sum of all these contributions yields the prediction at ``complete-NLO'' accuracy.

The complete-NLO results for dijet production at the LHC have been calculated in ref.~\cite{Frederix:2016ost} and for top-quark pair production in ref.~\cite{Czakon:2017wor}, the latter also combined with NNLO QCD corrections. Although one-loop contributions that are not part of NLO QCD and NLO EW corrections are present for many production processes at the LHC, calculations at this level of accuracy are rare, and those performed for dijet and top-quark pair production represent  an exception. The reason is twofold. First, being higher-order effects and $\alpha/\alpha_s\sim 0.1$ these corrections are expected to be smaller than standard NLO EW ones, and indeed they are for the case of dijet and top-quark pair production. Second, only with the recent automation of the calculation of EW corrections the necessary effort for calculating these additional orders has been reduced and therefore justified given their expected smallness.
Besides these reasons, in the subleading orders there can be new production mechanisms and care has to be taken to avoid process overlap. For example, the $\mathcal{O}(\alpha^2)$ contribution to dijet production contains hadronically decaying heavy vector bosons.

To our knowledge, the only other calculation where all the NLO effects beyond the NLO QCD and NLO EW accuracy have been considered is the case of vector-boson-scattering (VBS) for two positively charged $W$ bosons at the LHC including leptonic decays, namely the $pp\to \mu^+ \nu_\mu e^+ \nu_e jj$ process \cite{Biedermann:2017bss}. 
This complete-NLO prediction includes all the terms of $\ord{\alphas^i \alpha^j}$ with $i+j=6,7$ and $j\ge 4$, featuring both QCD-induced $W^+W^+jj$ production and electroweak $W^+ W^+$ scattering. Remarkably, at variance with dijet and top-quark pair production, the expected hierarchy of the different perturbative orders is not respected. Indeed, with proper VBS cuts the $\ord{\alpha^7}$ is by far the largest of the NLO contributions and moreover $\ord{\alpha^7}>\ord{\alpha^6 \alpha_s}>\ord{\alpha^5 \alpha_s^2}\sim\ord{\alpha^4 \alpha_s^3}$.

\medskip

In this article we want to give evidence that what has been found in ref.~\cite{Biedermann:2017bss}, {\it i.e.}, large contributions from supposedly subleading corrections, is not an exception due to the particularities of this process \cite{Biedermann:2016yds} and standard VBS selection cuts, which reduce the "QCD backgrounds".  It is rather a feature that may appear whenever the process considered involves the scattering of heavy particles in the SM, namely the $W$, $Z$ and Higgs bosons, but also top quarks. Indeed, although it is customary to expand in powers of $\alpha$, for these kind of processes $\ord{\alpha}$ corrections actually involve enhancements already at the coupling level,  {\it e.g.},  in the interactions among the top-quark, the Higgs boson and the longitudinal polarisations of the $W$ and $Z$ bosons. Thus, the $\ord{\alpha} \sim 0.01$ assumption is in general not valid and the expected hierarchy among perturbative orders may be not respected even at the  inclusive level. 

Here we focus on the case of the top quark and we explicitly show two different cases in which the expected hierarchy is  not respected: the $\ttw$ and $\ft$ production processes, which are already part of the current physics program at the LHC \cite{CMS-PAS-TOP-17-005,Aaboud:2016xve,CMS-PAS-TOP-17-009}. To this purpose we perform the calculation of the complete-NLO predictions of these two processes at 13 and 100 TeV in proton--proton collisions. All the seven $\ord{\alphas^i \alpha^j}$ contributions with $i+j=3,4$ and $j \ge 1$ for $\ttw$ production and all the eleven $\ord{\alphas^i \alpha^j}$ contributions with $i+j=4,5$ are calculated exactly without any approximation.  For both processes the calculation has been performed in a completely automated way via an extension of the code {\sc\small MadGraph5\_aMC@NLO}  \cite{Alwall:2014hca}. This extension has  already been validated for the NLO EW case in refs.~\cite{Frixione:2015zaa, Badger:2016bpw} and in ref.~\cite{Frederix:2016ost,Czakon:2017wor} for the calculation of the complete-NLO corrections. The code will soon be released and further documented in a detailed dedicated paper~\cite{mg5amcEW}.

Complete-NLO corrections involve large contributions for both the $\ttw$ and $\ft$ production processes, but very different structures underlie the two calculations.
Indeed, while large EW effects  in $\ttw$ production originate from the $t W \to t W$ scattering, which appears only via NLO corrections,  in $\ft$ production large EW effects  are already present at LO, due to the electroweak $t t  \to t t$ scattering. 

It has been noted in ref.~\cite{Dror:2015nkp} that EW $pp \to \ttw j$ production involves $t W\to t W$ scattering via the $gq \to \ttw q'$ channel. Even though ref.~\cite{Dror:2015nkp} focusses on BSM physics in $tW \to tW$ scattering, this contribution is sizeable already in the SM and is part of the NLO contributions of $\ord{\alpha_s \alpha^3}$ to the inclusive $\ttw$ production. It is {\it not}  part of the NLO EW corrections, which are of $\ord{\alpha_s^2 \alpha^2}$ and have already been calculated in ref.~\cite{Frixione:2015zaa}. However, while in the case of  $pp \to \ttw j$ production the final-state jet must be reconstructed, this is not necessary for the inclusive $pp\to\ttw$ process. In fact, we will argue that the $tW \to tW$ scattering component can be enhanced over the irreducible background from inclusive $\ttw$ production by applying a central jet veto.

Recently it was suggested that $\ft$ production can be used as a probe of the top-quark Yukawa coupling ($y_t$), as discussed in the tree-level analysis presented in ref.~\cite{Cao:2016wib}. Performing an expansion in power of $y_t$ one finds that  $\ord{y_t^2}$ and $\ord{y_t^4}$ contributions to $\ft$ production are not much smaller than purely-QCD induced terms (and in general non-Yukawa induced contributions) and therefore $\ft$ production is quite sensitive to the value of the top Yukawa coupling. Expanding the LO prediction in powers of $\alpha$, the $\ord{y_t^2}$ and $\ord{y_t^4}$ terms are fully included in the $\ord{\alpha_s^3 \alpha}$ and $\ord{\alpha_s^2 \alpha^2}$ terms. These perturbative orders are even larger than their Yukawa-induced components, and they also feature large cancellations at the inclusive level.
It is therefore interesting to compute NLO corrections to all these terms, since we expect them to be large as well. Indeed, we find that they are much larger than the values expected from a naive $\alpha_s$ and $\alpha$ power counting. On the other hand, even larger cancellations are present among NLO terms, although not over the whole phase space.

The structure of the paper is the following. In sec.~\ref{calc:frame} we describe the calculations   and we introduce a more suitable notation for referring to the various $\ord{\alphas^i \alpha^j}$ contributions.  In sec.~\ref{sec:results} we provide numerical results at the inclusive and differential levels for complete-NLO predictions for proton--proton collisions at 13 and 100 TeV. We discuss in detail the impact of the individual $\ord{\alphas^i \alpha^j}$ contributions. The common input parameters are described in sec.~\ref{sec:inputs}, while $pp\to\ttw$ and $pp\to\ft$ results are described in secs.~\ref{sec:ttw} and \ref{sec:4t}, respectively. Conclusions are given in sec.~\ref{sec:conclusions}.

\section{Calculation framework for $\ttw$ and $\ft$ production at complete-NLO}
\label{calc:frame}

Performing an expansion in powers of $\alpha_s$ and $\alpha$,  a generic observable for the processes $pp \to \ttw(+X)$ and  $pp \to \ft(+X)$ can be expressed as
\noindent
\begin{align}
\Sigma^{\ttw}(\alpha_s,\alpha) = \sum_{m+n\geq 2} \alpha_s^{m} \alpha^{n+1} \Sigma_{m+n+1,n}^{\ttw} \label{expansion:wp}\, , \\
 \Sigma^{\ft}(\alpha_s,\alpha) = \sum_{m+n\geq 4} \alpha_s^m \alpha^n \Sigma_{m+n,n}^{\ft} \label{expansion:4t}\, ,
\end{align}
\noindent
respectively, where  $m$ and $n$ are  positive integer numbers and we have used the notation introduced in refs.~\cite{Alwall:2014hca, Frixione:2014qaa}. 
For $\ttw$ production, LO contributions consist of $\Sigma_{m+n+1,n}^{\ttw}$  terms with $m+n=2$ and are induced by tree-level diagrams only. NLO corrections are given by the terms with $m+n=3$ and are induced by the interference of diagrams from the all the possible Born-level and one-loop amplitudes  as well all the possible interferences among tree-level diagrams involving one additional quark, gluon or photon emission. Analogously, for $\ft$ production, LO contributions consist of $\Sigma_{m+n,n}^{\ft}$  terms with $m+n=4$ and NLO corrections are given by the terms with $m+n=5.$ 
In this work we calculate all the perturbative orders entering at the complete-NLO accuracy, {\it i.e.}, $m+n=2,3$ for  $\Sigma^{\ttw}(\alpha_s,\alpha) $ and $m+n=4,5$ for  $\Sigma^{\ft}(\alpha_s,\alpha) $.

Similarly to ref.~\cite{Frederix:2016ost},  we introduce a more user-friendly notation for referring to the different $\Sigma_{m+n+1,n}^{\ttw}$ and $\Sigma_{m+n,n}^{\ft}$ quantities.
At LO accuracy, we can denote the $\ttw$ and $\ft$ observables  as $\Sigma^{\ttw}_{\rm LO}$ and $\Sigma^{\ft}_{\rm LO}$ and further redefine the perturbative orders entering these two quantities as
\begin{align}
\Sigma^{\ttw}_{\rm LO}(\alpha_s,\alpha) &= \alpha_s^2 \alpha \Sigma_{3,0}^{\ttw} + \alpha_s \alpha \Sigma_{3,1}^{\ttw} + \alpha^2 \Sigma_{3,2}^{\ttw} \nonumber\\
 &\equiv \Sigma_{\rm LO_1} + \Sigma_{\rm LO_2} + \Sigma_{\rm LO_3}\, , \label{eq:blobsttwLO} \\
 \Sigma^{\ft}_{\rm LO}(\alpha_s,\alpha) &= \alpha_s^4  \Sigma_{4,0}^{\ft} + \alpha_s^3 \alpha \Sigma_{4,1}^{\ft} + \alpha_s^2 \alpha^2 \Sigma_{4,2}^{\ft} + \alpha_s^3 \alpha \Sigma_{4,3}^{\ft} + \alpha^4 \Sigma_{4,4}^{\ft} \nonumber\\
 &\equiv \Sigma_{\rm LO_1} + \Sigma_{\rm LO_2} + \Sigma_{\rm LO_3}+ \Sigma_{\rm LO_4}+ \Sigma_{\rm LO_5}\, .
 \label{eq:blobs4tLO}
\end{align}
\noindent
In a similar fashion the NLO corrections and their single perturbative orders can be defined as
\begin{align}
\Sigma^{\ttw}_{\rm NLO}(\alpha_s,\alpha) &= \alpha_s^3 \alpha \Sigma_{4,0}^{\ttw} + \alpha_s^2 \alpha^2 \Sigma_{4,1}^{\ttw} + \alpha_s \alpha^3 \Sigma_{4,2}^{\ttw} + \alpha^4 \Sigma_{4,3}^{\ttw} \nonumber\\
 &\equiv \Sigma_{\rm NLO_1} + \Sigma_{\rm NLO_2} + \Sigma_{\rm NLO_3} + \Sigma_{\rm NLO_4}\, ,
\label{eq:blobsttwNLO}
\\
\Sigma^{\ft}_{\rm NLO}(\alpha_s,\alpha) &= \alpha_s^5  \Sigma_{5,0}^{\ft} + \alpha_s^4 \alpha^1 \Sigma_{5,1}^{\ft} + \alpha_s^3 \alpha^2 \Sigma_{5,2}^{\ft} + \alpha_s^2 \alpha^3 \Sigma_{5,3}^{\ft} + \alpha_s^1 \alpha^4 \Sigma_{5,4}^{\ft} +  \alpha^5 \Sigma_{5,5}^{\ft} \nonumber\\
 &\equiv \Sigma_{\rm NLO_1} + \Sigma_{\rm NLO_2} + \Sigma_{\rm NLO_3} + \Sigma_{\rm NLO_4} + \Sigma_{\rm NLO_5} + \Sigma_{\rm NLO_6}\, .
\label{eq:blobs4tNLO}
\end{align}
\noindent

In the following we will use the symbols $\Sigma_{\LNLO_i}$ or interchangeably their shortened aliases $\LNLO_i$ for referring to the different perturbative orders. Clearly the $\Sigma_{\LNLO_i}$ terms in $\ttw$ production, eqs.~\eqref{eq:blobsttwLO} and \eqref{eq:blobsttwNLO}, and in $\ft$ production, eqs.~\eqref{eq:blobs4tLO} and \eqref{eq:blobs4tNLO}, are different quantities. 
One should bear in mind that, usually, with the term ``LO'' one refers only to $\LO_1$, which here we will also denote as $\LOQCD$, while an observable at NLO QCD accuracy is  $\Sigma_{\rm LO_1} + \Sigma_{\rm NLO_1}$, which we will also denote as $\LOQCD + \NLOQCD$. The so-called NLO EW corrections which are of $\ord{\alpha}$ w.r.t.~the  $\LO_1$, are the $\Sigma_{\rm NLO_2}$ terms, so we  will also  denote it as $\NLOEW$.  Since in this article we will  use the $\LNLO_i$ notation,  the term ``LO''  will refer to the sum of all the $\LO_ i$ contributions rather than $\LO_1$ alone. The prediction at complete-NLO accuracy, which is the sum of all the $\LO_i$ and $\NLO_ i$ terms, will be  also denoted as ``$\LO + \NLO$''.

\medskip

\begin{figure}[t]
\centering
\includegraphics[width=0.32\textwidth]{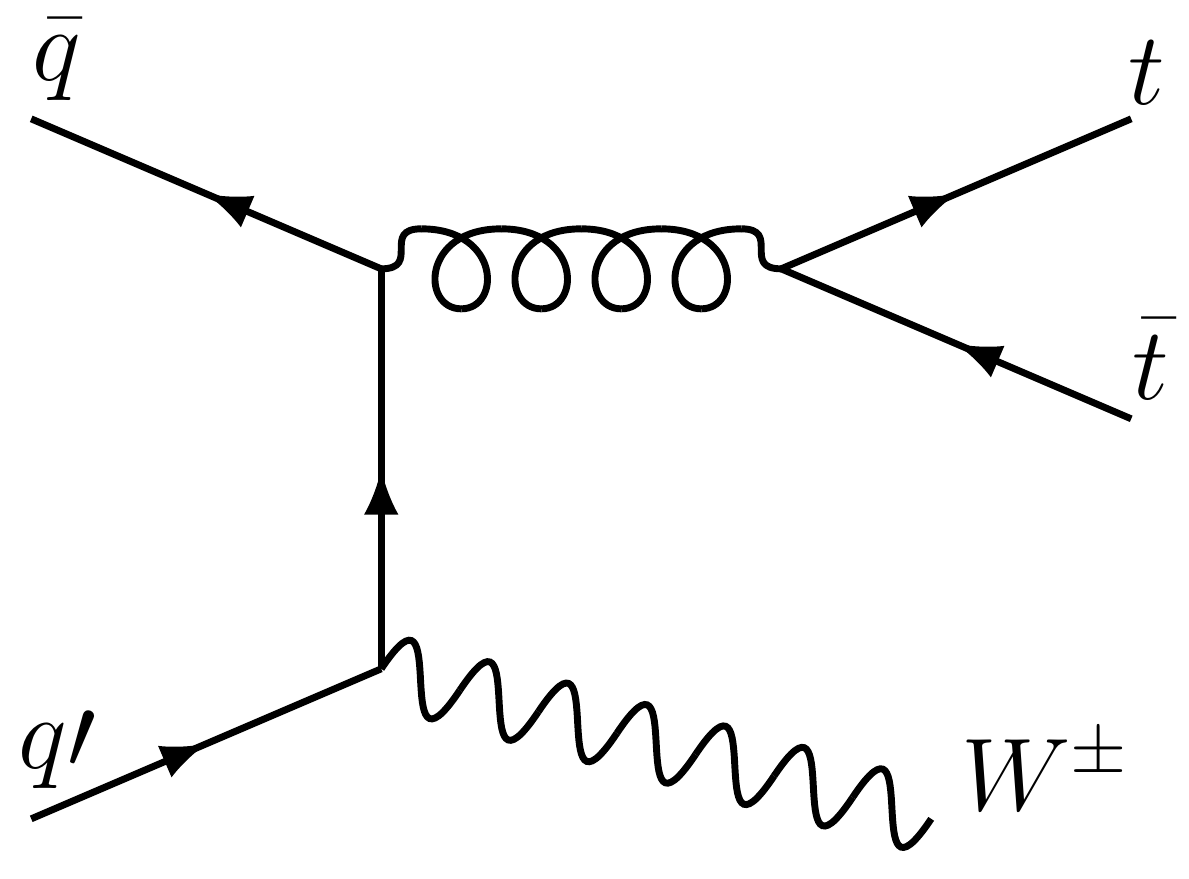}
\hspace*{2.cm}
\includegraphics[width=0.32\textwidth]{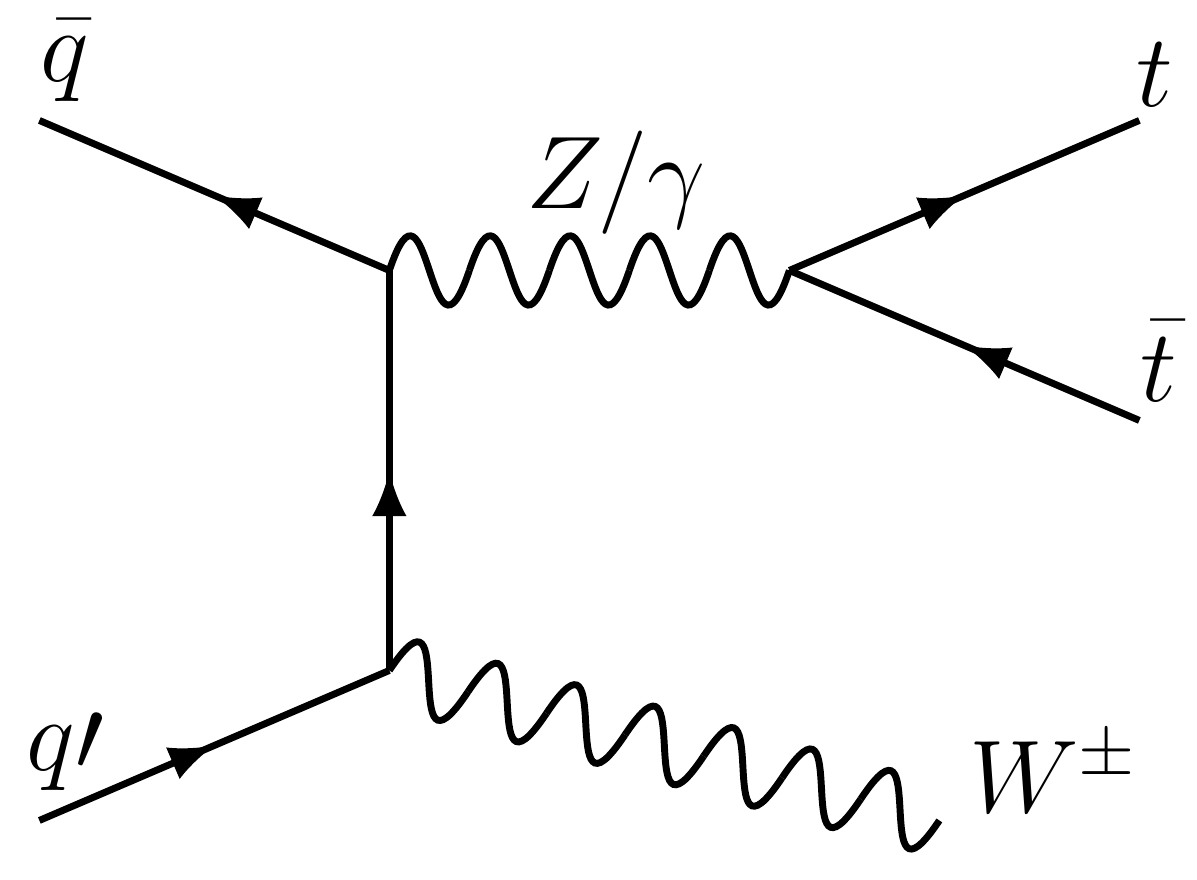}
\caption{Representative diagrams for the  Born $\bar{q}q'\to\ttw$ amplitude. The left diagram is of $\ord{\alphas \alpha^{1/2}}$, the right one is of $\ord{ \alpha^{3/2}}$. }
\label{fig:bornttw}
\end{figure}

\begin{figure}[t]
\centering
\includegraphics[width=0.32\textwidth]{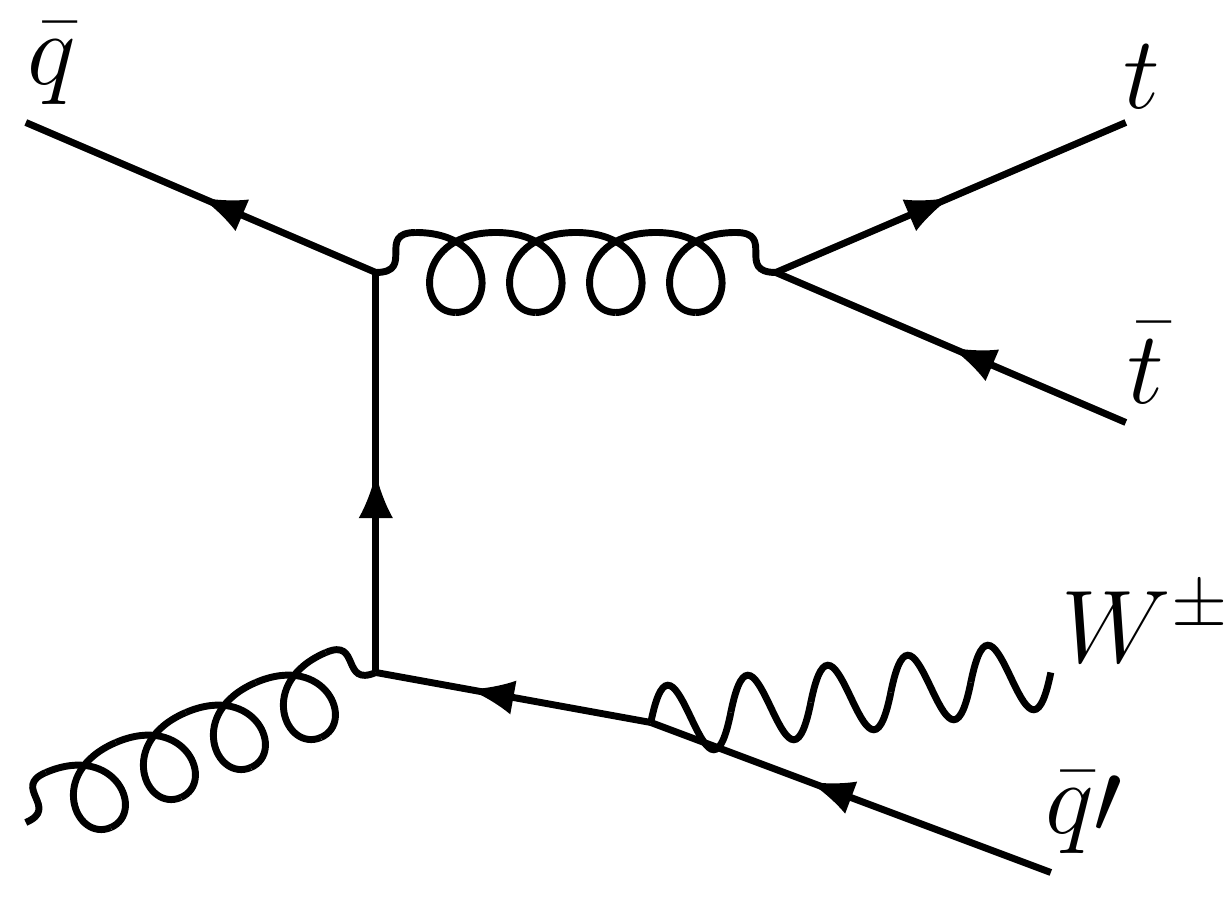}
\hspace*{2.cm}
\includegraphics[width=0.32\textwidth]{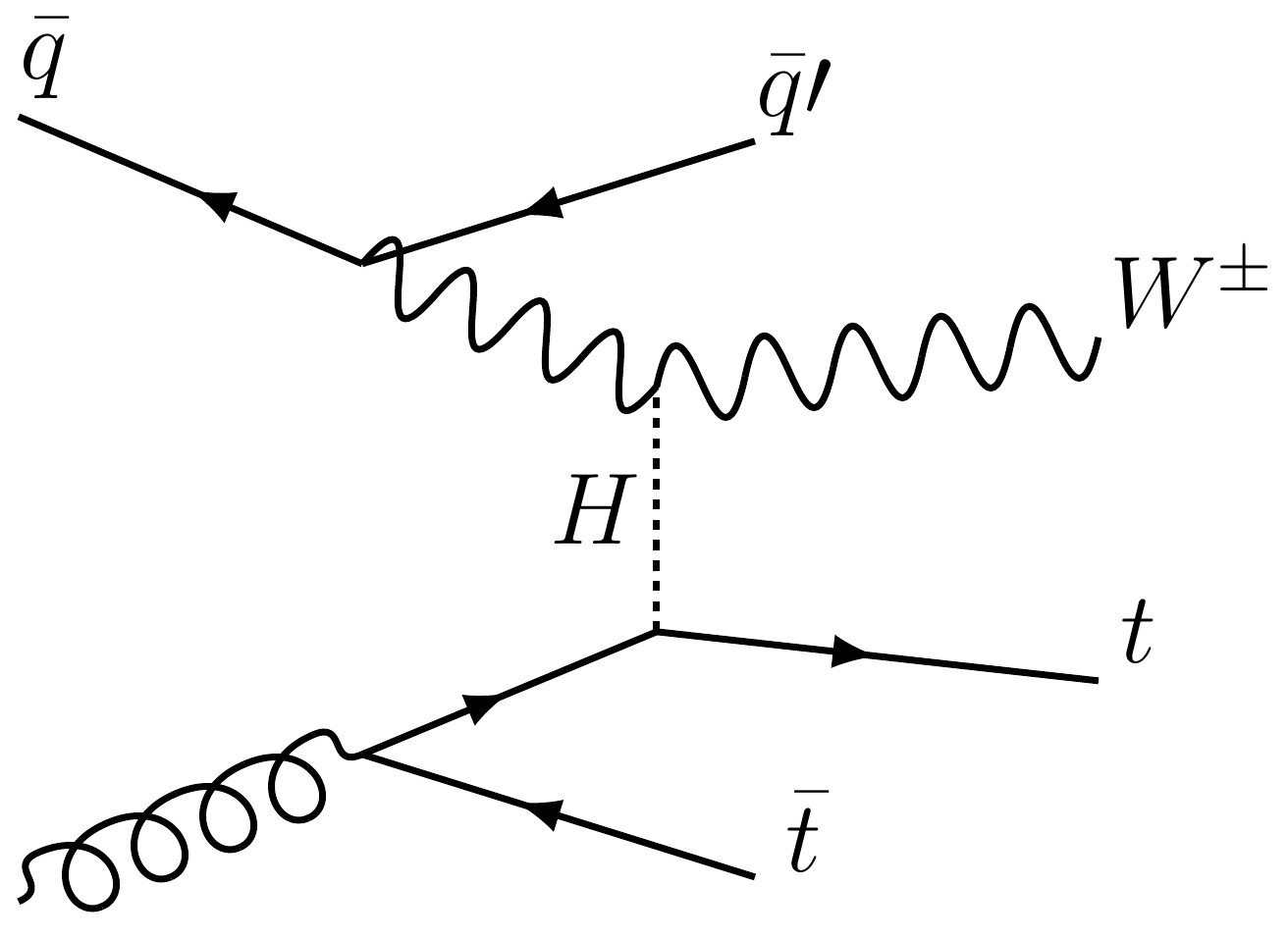}
\caption{Representative diagrams for the  $\bar qg\to\ttw \bar q'$ real-emission amplitudes. The left diagram is of $\ord{\alphas^{3/2} \alpha^{1/2}}$ and leads to $\log^2(\pt^2(\ttt)/m_W^2)$ terms in the $\NLO_1$ contribution. The right one  is of $\ord{\alphas^{1/2} \alpha^{3/2}}$, involves the $t W \to t W$ scattering  and contributes to the $\NLO_3$.}
\label{fig:realttw}
\end{figure}

We now turn to the description of the structures underlying the calculation of $\ttw$ and $\ft$ predictions at complete-NLO accuracy. We start with  $\ttw$ production, which is in turn composed by $\ttwp$ and $\ttwm$ production, and then we move to $\ft$ production.

 In $\ttwp$($\ttwm$)production, tree-level diagrams originate only from $u \bar d$($\bar{u} d$) initial states ($u$ and $d$ denote generic up- and down-type quarks), where a $W^+(W^-)$ is radiated from the $u(d)$ quark and the $\ttt $ pair is produced either via a gluon or a photon/$Z$ boson (see Fig.~\ref{fig:bornttw}). The former class of diagrams leads to the $\LO_1$ via squared amplitude, the latter to   $\LO_3$. The interference between these two classes of diagrams is absent due to colour, thus $\LO_2$ is analytically zero. Conversely, all the $\NLO_i$ contributions are non-vanishing. 

 The $\NLO_1$ is in general large, it  has been calculated in refs.~\cite{Hirschi:2011pa,Garzelli:2012bn,Campbell:2012dh, Maltoni:2014zpa} and studied in detail in ref.~\cite{Maltoni:2015ena}, where giant $K$-factors for the $\pttt$ distribution have been found. Large QCD corrections are induced also by the opening of the $gq \to \ttw q'$ channels, which depend on the gluon luminosity and  are therefore enhanced for high-energy proton--proton collisions. Moreover, the $\pttt$ distribution receives an additional $\log^2(\pt^2(\ttt)/m_W^2)$ enhancement in the $qg$ initial-state subprocess (see left diagram in Fig.~\ref{fig:realttw} and ref.~\cite{Maltoni:2015ena} for a detailed discussion). Also, the impact of soft-gluon emissions is non-negligible and their resummed contribution has been calculated in refs.~\cite{Li:2014ula,Broggio:2016zgg,Kulesza:2017hoc} up to next-to-next-to-leading-logarithmic accuracy. The $\NLO_2$ has been calculated for the first time in ref.~\cite{Frixione:2015zaa} and further phenomenological studies have been provided in ref.~\cite{deFlorian:2016spz}. In a boosted regime, due to Sudakov logarithms,  the $\NLO_2$ contribution can be as large as the NLO QCD scale uncertainty.
 
The $\NLO_3$ and $\NLO_4$ contributions are calculated for the first time here. In particular, the $\NLO_3$ contribution is expected to be sizeable since it contains $gq \to \ttw q'$ real-emission channels that involve EW $t W \to t W$ scattering (see right diagram in Fig.~\ref{fig:realttw}), which as pointed out in ref.~\cite{Dror:2015nkp} can be quite large. Moreover, as in the case of $\NLO_1$,  due to the initial-state gluon this channel becomes even larger by increasing the energy of proton--proton collisions.\footnote{In $\ttt Z$($\ttt H$) production the $\NLO_3$ contributions feature $t H \to t H$($t Z \to t Z$) scattering in $gq \to \ttt Z q$($gq \to \ttt H q$) real-emission channels. However, at variance with $\ttw$ production, the $gg$ initial state is available at $\LOQCD$. Thus, the $qg$ luminosity is not giving an enhancement and the relative impact from $\NLO_3$ is smaller than in $\ttw$ production.} The $t W \to t W$ scattering is present also in the $\NLO_4$ via the $\gamma q \to \ttw q'$, however in this case its contribution is suppressed by a factor $\alpha/\alpha_s$  and especially by the smaller luminosity of the photon. In addition to the real radiation of quarks, also the $q\bar{q'} \to \ttw g$ and $q\bar{q'} \to \ttw \gamma$ processes contribute to  the $\NLO_3$ and $\NLO_4$, respectively. Concerning virtual corrections, the $\NLO_4$ receives contributions only from one-loop  amplitudes of $\ord{\alpha^{5/2}}$, interfering with $\ord{\alpha^{3/2}}$ Born diagrams. Instead, the $\NLO_3$ receives contributions both from    $\ord{\alpha^{5/2}}$  and $\ord{\alpha_s \alpha^{3/2}}$   one-loop  amplitudes  interfering with   $\ord{\alphas \alpha^{1/2}}$ and $\ord{ \alpha^{3/2}}$   Born diagrams, respectively. Clearly, due to the different charges, $\NLO_{i}$ terms are different for the $\ttwp$ and $\ttwm$ case, however, since we did not find large qualitative differences at the numerical level, we  provide only inclusive results for $\ttw$ production. 

\begin{figure}[t]
\centering
\includegraphics[width=0.31\textwidth]{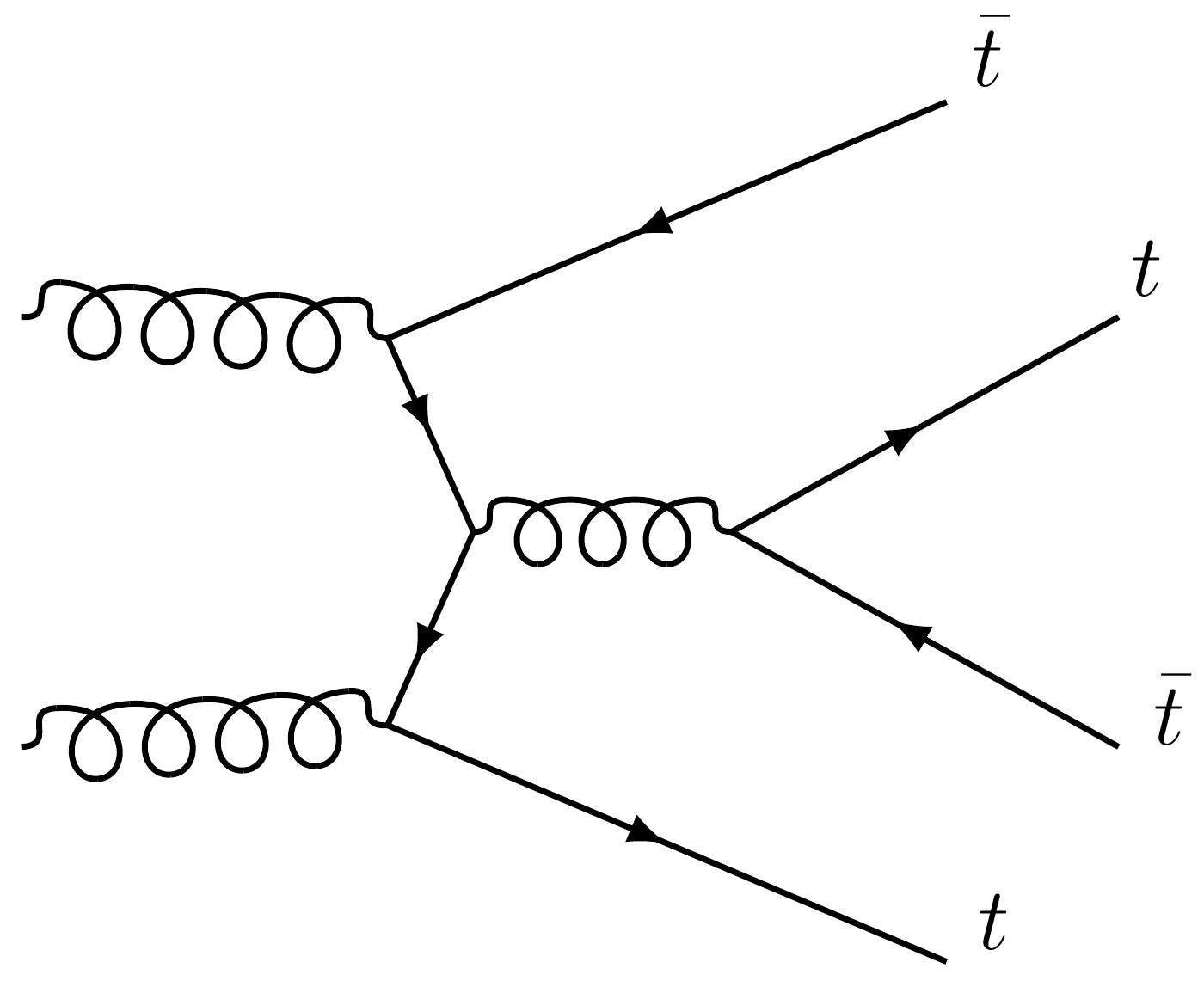}
\hspace*{2.cm}
\includegraphics[width=0.31\textwidth]{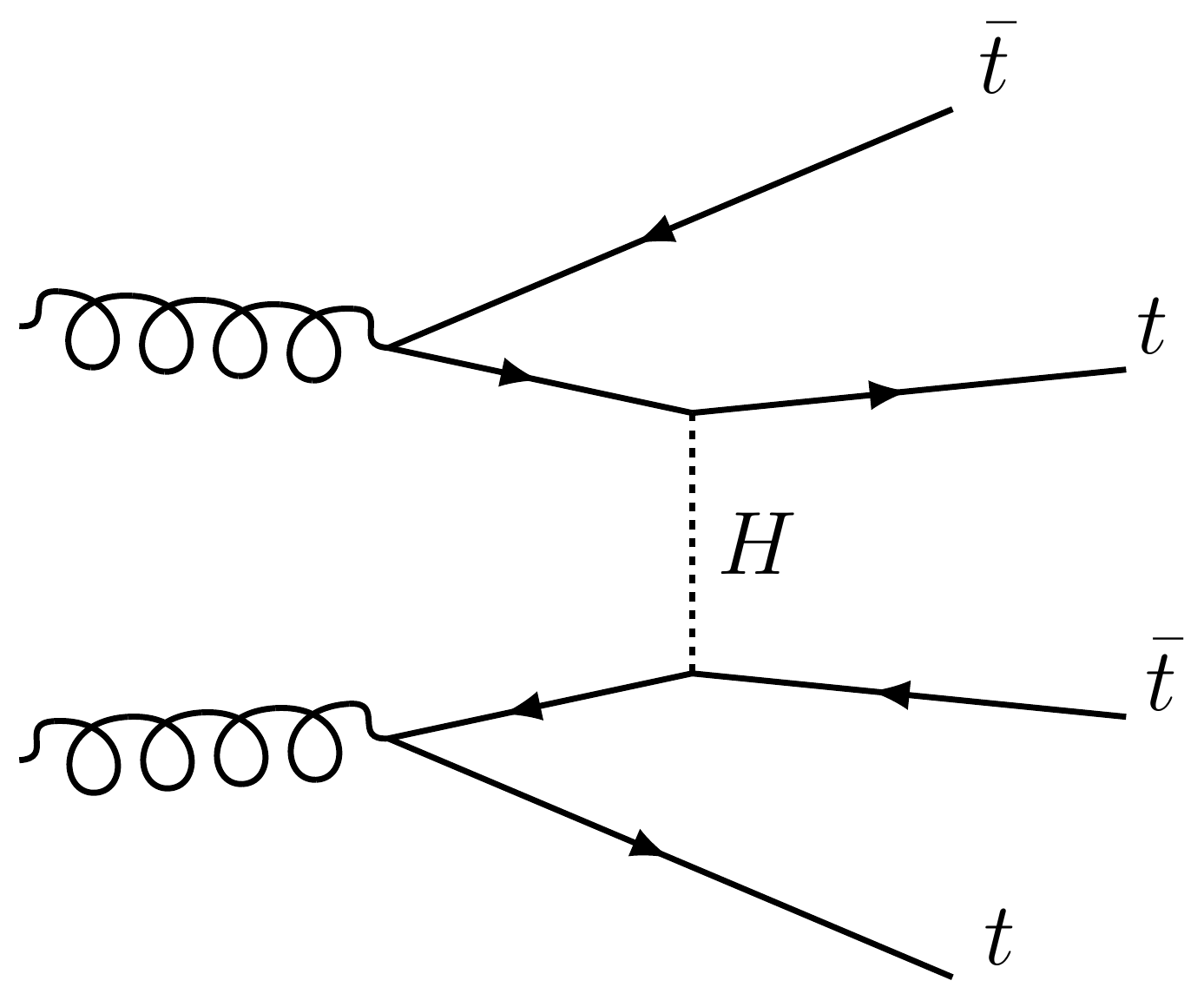}
\caption{Representative diagrams for the  Born $gg\to\ft$ amplitude. The left diagram is of $\ord{\alphas ^{2}}$, the right one is of $\ord{ \alpha_s \alpha}$. Both diagrams involve $t t \to t t$ scattering contributions.}
\label{fig:born4t}
\end{figure}

We now turn to the case of $\ft$ production, whose calculation involves a much higher level of complexity. While the $\NLO_1$ contribution have already been calculated in refs.~\cite{Bevilacqua:2012em,Alwall:2014hca} and studied in detail in ref.~\cite{Maltoni:2015ena}, all the other $\LNLO_i$ contributions are calculated for the first time here.

\begin{figure}[t]
\centering
\includegraphics[width=0.31\textwidth]{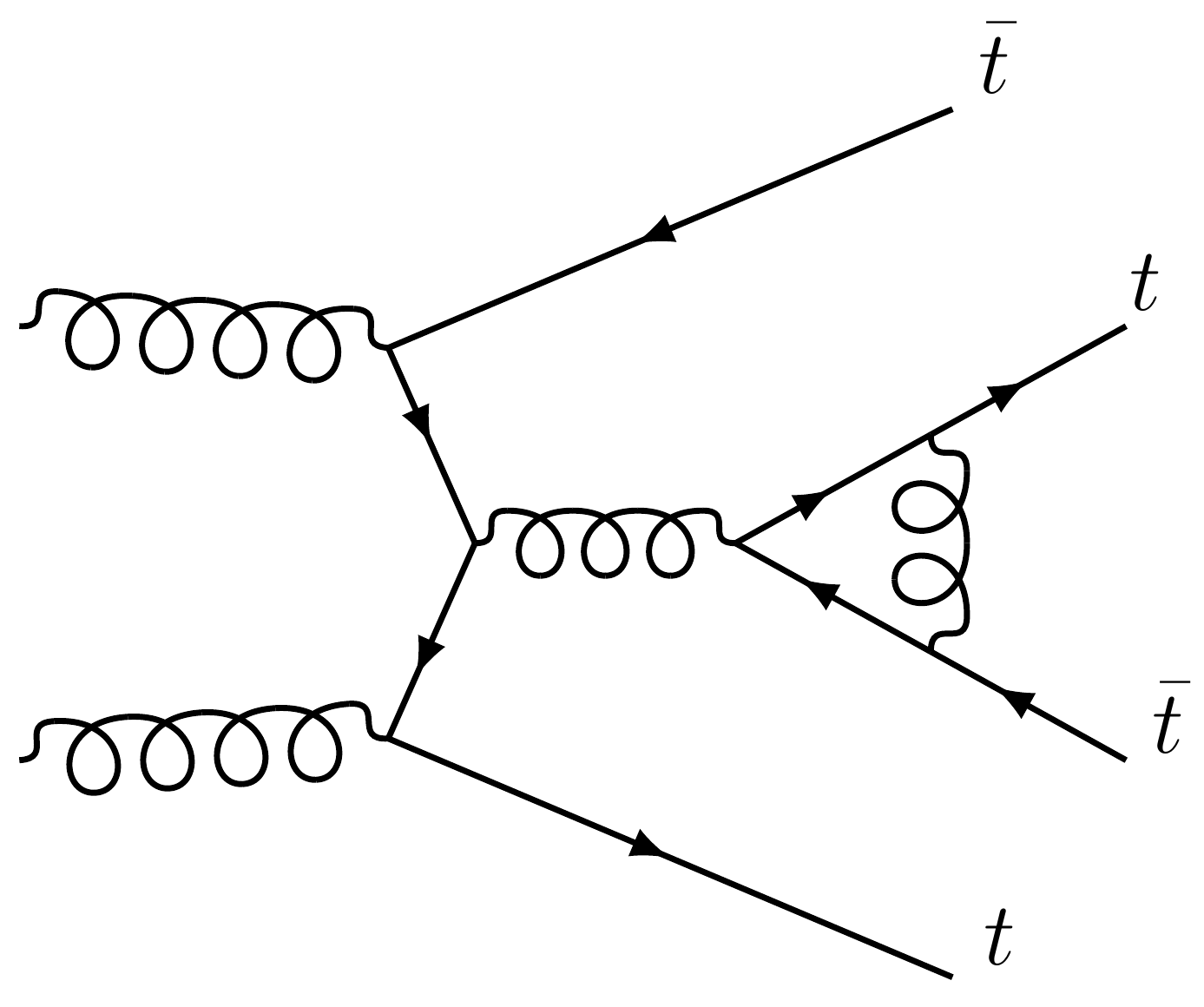}
\hspace*{0.2cm}
\includegraphics[width=0.31\textwidth]{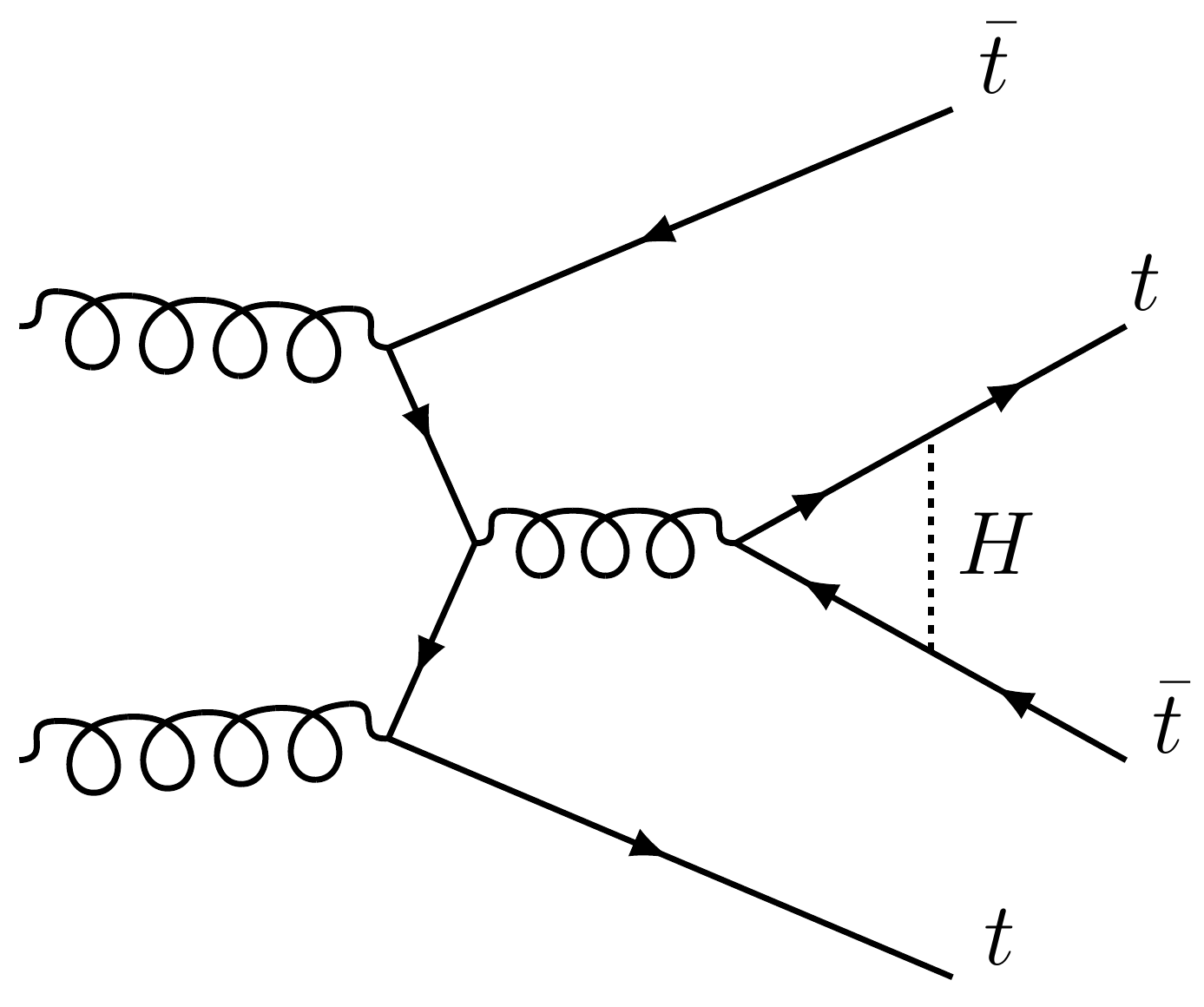}
\hspace*{0.2cm}
\includegraphics[width=0.31\textwidth]{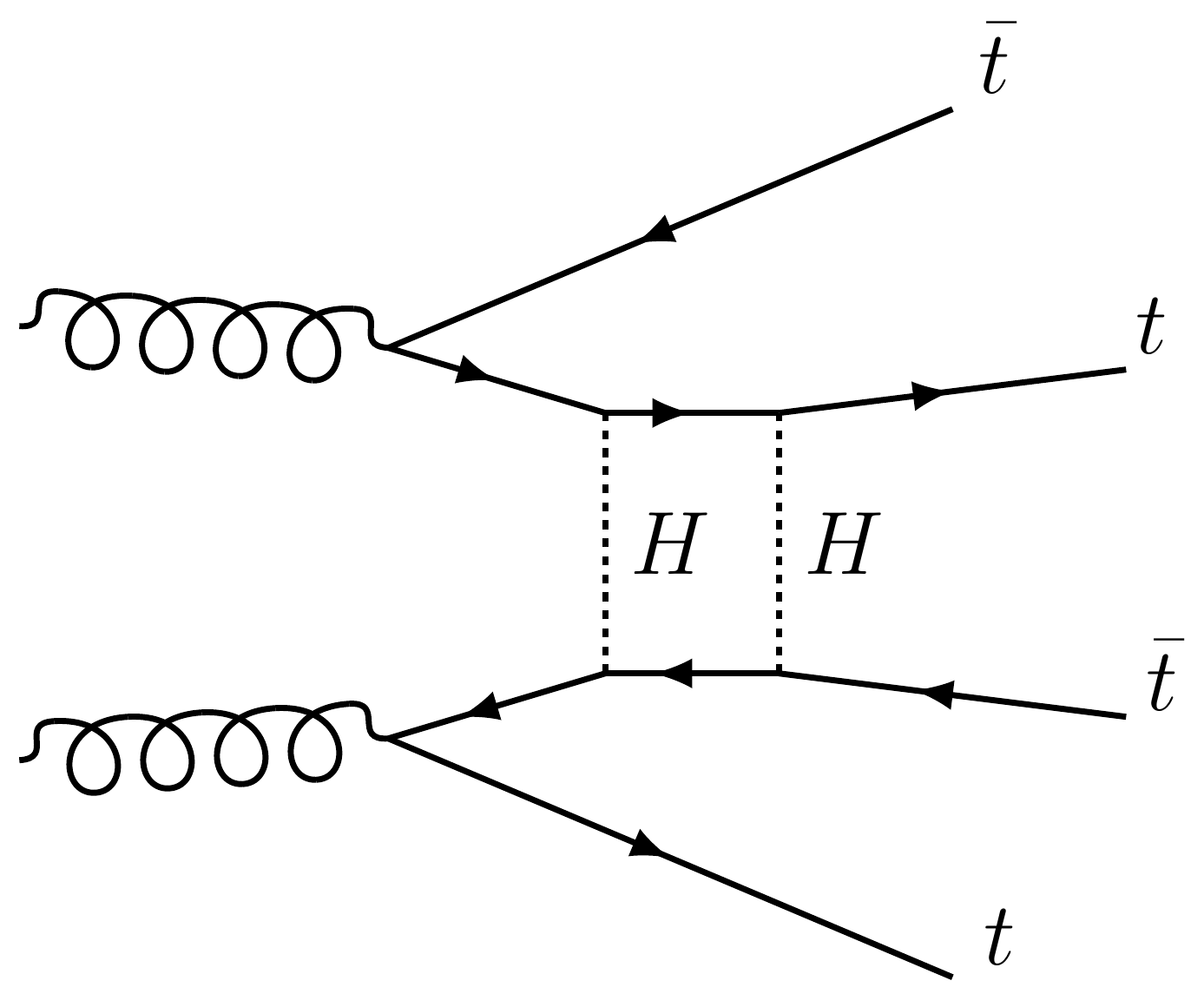}
\caption{Representative diagrams for the  one-loop $gg\to\ft$ amplitude. The left diagram is of $\ord{\alphas ^{3}}$, the central one is of $\ord{ \alpha_s^2 \alpha}$ and the right one is of $\ord{ \alpha_s \alpha^2}$. The interferences of these diagrams with those shown in Fig.~\ref{fig:born4t} lead to contributions to $\NLO_1$, $\NLO_2$, $\NLO_3$ and $\NLO_4$.}
\label{fig:loop4t}
\end{figure}

The  $gg\to \ft$ Born amplitude contains only $\ord{\alpha_s^2}$ and $\ord{\alpha_s \alpha}$ diagrams, while the $q\bar{q} \to \ft$ Born amplitude contains also $\ord{\alpha^2}$ diagrams. Thus the $gg$ initial state contributes to  $\LO_i$ with $i\leq 3$ and the  $q \bar q$ initial states contribute
to all the $\LO_i$. Also the $\gamma g$ and $\gamma \gamma$ initial states are available at the Born level; they  contributes to  $\LO_i$ with respectively $i\geq 2$ and $i\geq 3$. However,  their contributions are suppressed by the size of the photon parton distribution function (PDF). Representative $gg\to \ft$ Born diagrams are shown in Fig.~\ref{fig:born4t}.
As already mentioned in the introduction, $\LO_2$ and $\LO_3$  are larger than the values naively expected from $\alpha_s$ and $\alpha$ power counting, {\it i.e.},  $\LO_2\gg (\alpha/\alpha_s)\times \LOQCD$ and $\LO_3\gg(\alpha/\alpha_s)^2\times \LOQCD$. 
Thus, $\NLO_2$, $\NLO_3$ and also $\NLO_4$ are expected to be non-negligible, especially $\NLO_2$, $\NLO_3$ because they  involve ``QCD corrections''\footnote{As discussed in ref.~\cite{Frixione:2014qaa}, this classification of terms entering at a given order is not well defined;  some diagrams can be viewed both as a ``QCD  correction''   and an ``EW correction'' to different tree-level diagrams. Nevertheless, this intuitive classification is useful for understanding the underlying structure of such calculations. For this reason we use these expressions within quotation marks.\label{footnote2}   }  to  $\LO_2$ and $\LO_3$ contributions, respectively.  As discussed in ref.~\cite{Maltoni:2015ena}, the $\ft$ production cross-section is mainly given by the $gg$ initial state, for this reason we expect $\LO_4$, $\LNLO_5$ and $\NLO_6$  to be negligible. Representative $gg\to \ft$ one-loop diagrams are shown in Fig.~\ref{fig:loop4t}.
Although suppressed by the photon luminosity, also the $\gamma g$ and $\gamma \gamma$ initial states contribute to  $\NLO_i$ with  $i\geq 2$ and $i\geq 3$ respectively, 

Note that, for both the $pp\to\ttw$ and $pp\to\ft$ processes, we do
not include the (finite) contributions from the real-emission of heavy
particles ($W^\pm$, $Z$ and $H$ bosons and top quarks), sometimes
called the ``heavy-boson-radiation (HBR) contributions''.  Although
they can be formally considered as part of the inclusive predictions
at complete-NLO accuracy, these finite contributions are typically
small and generally lead to very different collider
signatures.\footnote{HBR contributions to $\NLO_2$ in $\ttw$
  production have been provided in ref.~\cite{Frixione:2015zaa}. }

Eqs.~\eqref{eq:blobsttwNLO}~and~\eqref{eq:blobs4tNLO} define the NLO
corrections in an additive approach. Another possibility would be
applying the corrections multiplicatively, which is not uncommon when
combining NLO QCD and NLO EW corrections. The difference between the
two approaches only enters at the NNLO-level and is formally beyond
the accuracy of our calculations. The typical example where the
multiplicative approach is well-motivated is when the $\NLO_1$
corrections are dominated by soft-QCD physics, and the $\NLO_2$
corrections by large EW Sudakov logarithms. Since these two
corrections almost completely factorise, it can be expected that the
mixed NNLO $\ord{\alpha_s\alpha}$ corrections to $\LO_1$ are dominated
by the product of the $\ord{\alpha_s}$ and $\ord{\alpha}$
corrections, {\it i.e.}, the $\NLO_1$ and $\NLO_2$ contributions. Hence, in this case, the dominant contribution to the
mixed NNLO corrections can be taken into account by simply combining NLO corrections in the
 multiplicative approach.  However, for $\ttw$ production, the $\NLO_1$
terms are dominated by hard radiation, as we argued above. Therefore,
even though the $\NLO_2$ is dominated by large Sudakov logarithms, the
multiplicative approach leads to uncontrolled NNLO terms. Moreover,
due to the opening of the $tW \to tW$ scattering, the same would apply also for a multiplicative combination with the $\NLO_3$. A similar argument is present for $\ft$ production: for $i
\le 3$, the $\NLO_i$ terms are dominated by ``QCD corrections'' on top
of the $\LO_i$ terms. Since the various  $\LO_i$
have clearly different underlying structures due to the possibility of
EW $tt \to tt$ scattering, also in this case there is no reason for believing that their NLO
corrections  factorise at NNLO and therefore that mixed NNLO corrections are
dominated by products of $\NLO_i$ corrections. Hence, for both the $pp
\to \ttw$ and $pp \to \ft$ processes, not only the multiplicative approach  is
not leading to improved predictions, but there are clear indications to the fact that this approximation introduces uncontrolled terms. Thus,  we use only the additive one.

Before discussing the numerical results of the complete-NLO
predictions in the next section, we would like to mention that the
calculation for $\ft$ production shows a remarkably rich structure for
the $\NLO_3$ and $\NLO_4$ contributions. As already said, the
$q\bar{q} \to \ft$ Born amplitude contains $\ord{\alpha_s^2}$,
$\ord{\alpha_s \alpha}$ and $\ord{\alpha^2}$ diagrams, and for this
reason, the $q \bar q \to \ft$ process contributes to $\LO_3$ via both
the square of its $\ord{\alpha_s \alpha}$ Born amplitude and the
interference of its $\ord{\alpha_s^2}$ and $\ord{\alpha^2}$ Born
amplitudes. In order to have such a double structure at the leading
order, it is necessary to have at least six external particles that
are all coloured and EW interacting at the same time. Since each
$\NLO_i$ is given by ``QCD corrections'' on top of the $\LO_i$ and by
``EW corrections'' on top of the $\LO_{i-1}$, the $\NLO_3$ and
$\NLO_4$ virtual corrections to $q\bar{q} \to \ft$ extend this double
structure to \emph{three} different interference (or squared) terms:
two originating from $\LO_3$ and one from either $\LO_2$ (in the case
of $\NLO_3$) or $\LO_4$ (in the case of $\NLO_4$). This is the first
time that a calculation with such a triple structure for the virtual
corrections has been performed.

\section{Numerical results}
\label{sec:results}
In this section, we present numerical results for the complete-NLO predictions for the $\ttw$ and $\ft$ production processes.
As mentioned in the introduction, we used an extension of the {\sc\small MadGraph5\_aMC@NLO} framework for all our numerical studies. This extension has already been used for the calculation of complete-NLO corrections as already mentioned in the introduction. In {\sc\small MadGraph5\_aMC@NLO}, infra-red singularities are dealt  with via the FKS method~\cite{Frixione:1995ms,
Frixione:1997np} (automated in the module \mf~\cite{Frederix:2009yq,
Frederix:2016rdc}). One-loop amplitudes are computed by dynamically switching  between
different  kinds of techniques for integral reduction: the OPP~\cite{Ossola:2006us},
Laurent-series expansion~\cite{Mastrolia:2012bu},
and tensor integral reduction~\cite{Passarino:1978jh,Davydychev:1991va,Denner:2005nn}.
These techniques have been automated in the module \ml~\cite{Hirschi:2011pa}, which is used for the generation of the amplitudes
and in turn exploits \ct~\cite{Ossola:2007ax}, \nin~\cite{Peraro:2014cba,
Hirschi:2016mdz} and \coll~\cite{Denner:2016kdg}, together with an in-house
implementation of the {\sc OpenLoops} optimisation~\cite{Cascioli:2011va}.

\subsection{Input parameters}
\label{sec:inputs}
In the following we specify the common set of input parameters that are used in the $pp \to \ttw$ and $pp \to \ft$ calculations. 
The masses of the heavy SM particles are set to
\begin{equation}
m_t = 173.34 \text{ GeV}\, , \quad m_H = 125 \text{ GeV} \, , \quad m_W = 80.385 \text{ GeV} \, , \quad m_Z = 91.1876 \text{ GeV} \,,
\end{equation}
while all the other masses are set equal to zero. We employ the on-shell renormalisation for all the masses and set all the decay widths equal to zero. The renormalisation of $\alphas$ is performed in the $\MSb$-scheme with five active flavours,\footnote{With the unit CKM matrix no $b$ quarks are present in the initial state for $\ttw$ production, while for $\ft$ their relative effect w.r.t.~LO$_{1}$ is at or below the per-mil level.} while the EW input parameters and the associated condition for the renormalisation of $\alpha$ are in the $G_\mu$-scheme, with
\begin{equation}
G_\mu = 1.16639 \cdot 10^{-5} \text{ GeV}^{-2} \,.
\end{equation}
The CKM matrix is set equal to the $3 \times 3$ unity matrix.

We employ dynamical definitions for the renormalisation ($\mu_r$) and factorisation ($\mu_f$) scales. In particular, their common central value $\mu_c$ is defined as 
\beqn
\mu_c&=&\frac{H_{T}}{2}\, ~~{\rm for~~}\ttw\,, \\
\mu_c&=&\frac{H_{T}}{4}\, ~~{\rm for~~}\ft\,,
\eeqn
where
\beqn
H_{T}&\equiv& \sum_{i=1,N(+1)  }m_{T,i}\, , \label{eq:HT}
 \eeqn
and $m_{T,i}\equiv\sqrt{m^2_i+\pt^2(i)}$ are the transverse masses of the $N(+1)$ final-state particles. Our scale choice for $\ft$ production is motivated by the study in ref.~\cite{Maltoni:2015ena}. Theoretical uncertainties due to the scale definition are estimated via the independent variation of $\mu_r$ and $\mu_f$ in the interval $\{\mu_c/2,2\mu_c\}$. In order to show the scale dependence of $\LNLO_i/\LOQCD$ relative corrections we will also consider the diagonal variation $\mu_r=\mu_f$, simultaneously in  the numerator and the denominator. This scale dependence does not directly indicate scale uncertainties, but it will be very useful in our discussion.

Concerning the PDFs, we use the  {\tt LUXqed\_plus\_PDF4LHC15\_nnlo\_100}  set \cite{Manohar:2016nzj,Manohar:2017eqh}, which is in turn based on the {\sc \small PDF4LHC} set \cite{Butterworth:2015oua, Ball:2014uwa, Harland-Lang:2014zoa, Dulat:2015mca}. This PDF set includes NLO QED effects in the DGLAP evolution and especially the most precise determination of the photon density.

\subsection{Results for $pp\to\ttw$ production}
\label{sec:ttw}

We start by presenting predictions for $pp\to\ttw$ total cross sections at 13 and 100 TeV proton--proton collisions with and without applying a jet veto and then we discuss results at the differential level.
The total cross sections at 13~TeV for $\ttw$ production are shown in Tab.~\ref{table:wpm} at different accuracies, namely, $\LOQCD$,  $\LOQCD + \NLOQCD$,  $\LO$ and  $\LO + \NLO$.  We also show for each value its relative scale uncertainty and we provide the ratio of the predictions at $\LO + \NLO$ and $\LOQCD + \NLOQCD$ accuracy. Analogous results at 100 TeV are  displayed in Tab.~\ref{table:wpm100}. Numbers in parentheses refer to the case in which we apply a jet veto,  rejecting all the events with
\beqn
\pt(j)>100~\text{GeV}~~{\rm and}~~|y(j)|<2.5\, ,
\label{jetveto}
\eeqn
where also  hard photons are considered as a jet.\footnote{We explicitly verified  that vetoing only quark and gluons, but not photons, leads to differences below the percent level. Moreover, from an experimental point of view, vetoing jets that are not isolated photons would be simply an additional complication.} The purpose of this jet veto will become clear in the discussion below. 
Further details about the size of the individual $\LNLO_i$ terms are provide in Tab.~\ref{table:wpmorders} (13 TeV) and Tab.~\ref{table:wpmorders100} (100 TeV), where we show predictions for the quantities
\beqn
\delta_{\LNLO_i}(\mu)=\frac{\Sigma_{\LNLO_i}(\mu)}{\Sigma_{\LOQCD}(\mu)}\, ,
\label{delta}
\eeqn where $\Sigma(\mu)$ is simply the total cross section evaluated
at the scale $\mu_f=\mu_r=\mu$.  In Tabs.~\ref{table:wpmorders} and
\ref{table:wpmorders100} we do not show the result for $\LO_1\equiv
\LOQCD$, since it is by definition always equal to one, regardless of
the value of $\mu$. We want to stress that results in
Tabs.~\ref{table:wpmorders} and \ref{table:wpmorders100} do not show
directly scale uncertainties; the value of $\mu$ is varied
simultaneously in the numerator and the denominator of $\delta$. The
purpose of studying $\delta$ as a function of $\mu$ will become clear
below when we discuss the different dependence in $\delta_{\NLO_1}$
versus $\delta_{\NLO_2}$ and $\delta_{\NLO_3}$.

\begin{table}[t]
\small
\renewcommand{\arraystretch}{1.5}
\begin{center}
\begin{tabular}{c c c c c c}
\toprule
$\sigma[\textrm{fb}]$ & LO${}_{\textrm{QCD}}$ & $\LOQCD+\NLOQCD$ & LO & $\LO+\NLO$ & $\frac{\LO+\NLO}{\LOQCD+\NLOQCD}$\\
\midrule
$\mu=H_T/2$ & $ 363 {}^{+24 \%}_{-18 \%}$ & $ 544 {}^{+11 \%}_{-11 \%}\, ( 456 {}^{+ 5 \%}_{- 7 \%})$ & $ 366 {}^{+23 \%}_{-18 \%}$ & $ 577 {}^{+11 \%}_{-11 \%}\, ( 476 {}^{+ 5 \%}_{- 7 \%})$ & $ 1.06\, ( 1.04 )$ \\
\bottomrule
\end{tabular}
\caption{Cross sections for $\ttw$ production at 13 TeV in various
  approximations. The numbers in parentheses are obtained
  with the jet veto of eq.~\eqref{jetveto} applied.}
\label{table:wpm}
\end{center}
\end{table}

\begin{table}[t]
\small
\renewcommand{\arraystretch}{1.5}
\begin{center}
\resizebox{\columnwidth}{!}{%
\begin{tabular}{c c c c c c}
\toprule
$\sigma[\textrm{pb}]$ & LO${}_{\textrm{QCD}}$ & $\LOQCD+\NLOQCD$ & LO & $\LO+\NLO$ & $\frac{\LO+\NLO}{\LOQCD+\NLOQCD}$\\
\midrule
$\mu=H_T/2$ & $ 6.64 {}^{+28 \%}_{-21 \%}$ & $ 16.58 {}^{+17 \%}_{-15 \%}\, ( 11.37 {}^{+11 \%}_{-12 \%})$ & $ 6.72 {}^{+27 \%}_{-21 \%}$ & $ 20.86 {}^{+15 \%}_{-14 \%}\, ( 14.80 {}^{+11 \%}_{-11 \%})$ & $ 1.26\, ( 1.30 )$ \\
\bottomrule
\end{tabular}
}
\caption{Same as in Tab.~\ref{table:wpm} but for 100~TeV.}
\label{table:wpm100}
\end{center}
\end{table}


From Tabs.~\ref{table:wpm} and \ref{table:wpm100} it can be seen that
the $\LOQCD$ predictions, both at 13 and 100 TeV, have a scale
dependence that is larger than  20\%. Including the $\LO_i$
contributions with $i>1$ changes the cross section by about 1\% and
leaves also the scale dependence almost unchanged. As discussed in
sec.~\ref{calc:frame}, the $\LO_2$ is exactly zero due to colour, thus
this small correction is entirely coming from the $\LO_3$
contribution. In Tabs.~\ref{table:wpmorders} and
\ref{table:wpmorders100} it can be seen that the scale dependence of
this $\LO_3$ contribution is slightly different from the $\LO_1$. The
factorisation scale dependence is almost identical for the $\LO_1$ and
$\LO_3$ terms (both are $q\bar{q}'$ initiated and have similar
kinematic dependence), thus this difference is entirely due to the
variation of the renormalisation scale, which, at leading order, only
enters the running of $\alpha_s$. The $\LO_1$ has two powers of
$\alpha_s$ while the $\LO_3$ has none. The value of $\alpha_s$
decreases with increasing scales, and therefore, it is no surprise
that $\delta_{\LO_3}$ increases with
larger values for the scales.

As already known, in $\ttw$ production NLO QCD corrections are large
and lead to a reduction of the scale uncertainty. Indeed, for the central scale choice, 
the total cross section at 13~TeV increases by 50\% when including the $\NLOQCD$
contribution, and a massive 150\% correction is present at
100~TeV. The reduction in the scale dependence is about a factor two
for 13~TeV, resulting in an 11\% uncertainty. On the other hand, given
the large $\NLOQCD$ corrections,  at 100~TeV the resulting scale
dependence at $\LOQCD+\NLOQCD$ is larger than at 13~TeV, remaining at
about 16\%. Comparing these pure-QCD predictions to the complete-NLO
cross sections ($\LO+\NLO$) we see that the latter are about 6\%
larger at 13~TeV, while the relative scale dependencies are identical.  At 100 TeV, even though the relative scale dependence at complete-NLO is 1-2
percentage points smaller than at $\LOQCD+\NLOQCD$, in absolute terms it
is actually larger. This effect is due to the large increase of about 26\%  induced by $\LNLO_i$ terms with $i>1$. Indeed, this increase is
mostly coming from the contribution of the $tW \to tW$ scattering, which appears at $\NLO_3$ via the quark real-emission and  has a
Born-like scale dependence. However, this dependence  is relatively small since the $\NLO_3$ involves only a single power of $\alpha_s$.

\begin{table}[t]
\small
\begin{center}
\begin{tabular}{c r@{\,}l r@{\,}l r@{\,}l}
\toprule
$\delta[\%]$ & \multicolumn{2}{c}{$\mu= H_T/4$} & \multicolumn{2}{c}{$\mu=H_T/2$} & \multicolumn{2}{c}{$\mu = H_T$} \\
\midrule 
LO${}_2$ & \multicolumn{2}{c}{-} & \multicolumn{2}{c}{-} & \multicolumn{2}{c}{-} \\
LO${}_3$ & \multicolumn{2}{c}{$0.8$} & \multicolumn{2}{c}{$0.9$} & \multicolumn{2}{c}{$1.1$} \\
\midrule 
NLO${}_1$ & $34.8$ & ($7.0$) & $50.0$ & ($25.7$) & $63.4$ & ($42.0$) \\
NLO${}_2$ & $-4.4$ & ($-4.8$) & $-4.2$ & ($-4.6$) & $-4.0$ & ($-4.4$) \\
NLO${}_3$ & $11.9$ & ($8.9$) & $12.2$ & ($9.1$) & $12.5$ & ($9.3$) \\
NLO${}_4$ & $0.02$ & ($-0.02$) & $0.04$ & ($-0.02$) & $0.05$ & ($-0.01$) \\

\bottomrule
\end{tabular}
\caption{$\sigma_{\LNLO_i}/\sigma_{\LOQCD}$ ratios for $\ttw$
  production at 13 TeV for various values of $\mu=\mu_r=\mu_f$.}
\label{table:wpmorders}
\end{center}
\end{table}

\begin{table}[t]
\small
\begin{center}
\begin{tabular}{c r@{\,}l r@{\,}l r@{\,}l}
\toprule
$\delta[\%]$ & \multicolumn{2}{c}{$\mu= H_T/4$} & \multicolumn{2}{c}{$\mu=H_T/2$} & \multicolumn{2}{c}{$\mu = H_T$} \\
\midrule 
LO${}_2$ & \multicolumn{2}{c}{-} & \multicolumn{2}{c}{-} & \multicolumn{2}{c}{-} \\
LO${}_3$ & \multicolumn{2}{c}{$0.9$} & \multicolumn{2}{c}{$1.1$} & \multicolumn{2}{c}{$1.3$} \\
\midrule 
NLO${}_1$ & $159.5$ & ($69.8$) & $149.5$ & ($71.1$) & $142.7$ & ($73.4$) \\
NLO${}_2$ & $-5.8$ & ($-6.4$) & $-5.6$ & ($-6.2$) & $-5.4$ & ($-6.1$) \\
NLO${}_3$ & $67.5$ & ($55.6$) & $68.8$ & ($56.6$) & $70.0$ & ($57.6$) \\
NLO${}_4$ & $0.2$ & ($0.1$) & $0.2$ & ($0.2$) & $0.3$ & ($0.2$) \\

\bottomrule
\end{tabular}
\caption{$\sigma_{\LNLO_i}/\sigma_{\LOQCD}$ ratios for $\ttw$
  production at 100 TeV for various values of $\mu=\mu_r=\mu_f$.}
\label{table:wpmorders100}
\end{center}
\end{table}

In Tabs.~\ref{table:wpmorders} and \ref{table:wpmorders100} we can see
that $\delta_{\NLO_1}\equiv \delta_{\NLOQCD}$ is strongly $\mu$
dependent, while this is not  the case for $\delta_{\NLO_i}$
with $i>1$. In fact, this behaviour is quite generic and not
restricted to $\ttw$ production; it can be observed for a wide class
of processes. The $\mu$ dependence in $\delta_{\NLO_1}$ leads to the
reduction of the scale dependence of $\LOQCD + \NLOQCD$ results
w.r.t.~the $\LOQCD$ ones. On the contrary, the $\delta_{\NLO_i}$
quantities with $i>1$ are typically quite independent of the value of
$\mu$. The reason is the following. The ${\NLO_i}$ contributions are
given by ``QCD corrections'' to ${\LO_{i}}$ contributions as well ``EW
corrections'' to the ${\LO_{i-1}}$ ones. The former involve explicit
logarithms of $\mu$ due the renormalisation of both $\alpha_s$ and
PDFs, while the latter contain only explicit logarithms of $\mu$ due
the $\ord{\alpha}$ PDFs counterterms. Indeed, in the $G_\mu$-scheme,
or other schemes such as $\alpha(0)$ or $\alpha(m_Z)$, the numerical
input for $\alpha$ does not depend on an external renormalisation
scale. Moreover, the $\ord{\alpha}$ PDF counterterms induce a much
smaller effect than those of QCD, since they are
$\ord{\alpha/\alpha_s}$ suppressed and do not directly involve the
gluon PDF. Thus, for a generic process, since a ${\LO_{i}}$
contribution is typically quite suppressed w.r.t.~the ${\LO_{i-1}}$
one ---or even absent, as {\it e.g.} for (multi) EW vector boson
production--- the scale dependence of $\delta_{\NLO_i}$ with $i>1$ is
small. For this reason it is customary, and typically also reasonable,
to quote NLO EW corrections independently from the scale
definition. As can be seen in Tabs.~\ref{table:wpmorders} and
\ref{table:wpmorders100} this is also correct for $\ttw$, but as we
will see in the next section the situation is quite different for
$\ft$ production, where also the $\delta_{\LNLO_i}(\mu)$
quantities with $i>1$ strongly depend on the value of $\mu$.

By considering the $\mu$ dependence of the $\delta_{\NLO_1}(\mu)$
contributions in Tabs.~\ref{table:wpmorders} and
\ref{table:wpmorders100}, we see a different behaviour in the two
tables. At 13~TeV the scale dependence of $\delta_{\NLOQCD}(\mu)$
increases with increasing scales. This is to be expected: the $\LO_1$
contribution has a large renormalisation-scale dependence, resulting
in a rapidly decreasing cross section with increasing scales. In order
to counterbalance this, the scale dependence of the $\NLO_1$
contribution must be opposite so that the scale dependence at NLO QCD accuracy is
reduced. On the other hand, at 100~TeV, the scale dependence of the
$\delta_{\NLO_1}(\mu)$ decreases with increasing scales, suggesting
that the scale dependence at $\LOQCD+\NLOQCD$ is actually larger than
at $\LOQCD$. As can be seen in Tab.~\ref{table:wpm100} this does not
appear to be the case. The reason is that contrary to 13~TeV, at
100~TeV collision energy the $\LOQCD$ has not only a large
renormalisation-scale dependence, but also the factorisation-scale one
is sizeable. In fact, the scale dependence in Tab.~\ref{table:wpm100}
is dominated by terms in which $\mu_r$ and $\mu_f$ are varied in
opposite directions, {\it i.e.}, $\{\mu_r,\mu_f\}=\{2 \mu_c,\mu_c /2\}$ and $\{2 \mu_c,\mu_c /2\}$. However, in Tab.~\ref{table:wpmorders100} we only consider
the simultaneous variation of $\mu_r$ and $\mu_f$. If we had estimated
the scale uncertainty in Tabs.~\ref{table:wpm} and \ref{table:wpm100} by only varying $\mu=\mu_r=\mu_f$, we would actually
have seen an increment of the uncertainties in moving from $\LOQCD$ to
$\LOQCD+\NLOQCD$.

The NLO EW corrections, the $\NLO_2$ contribution, are negative and
have a $-$4-6\% impact w.r.t.~the $\LO_1$ cross section. This is well
within the $\LOQCD+\NLOQCD$ scale uncertainties. The opening of the
$tW \to tW$ scattering enhances the $\NLO_3$ contribution
enormously. In fact, it is much larger than the $\NLO_2$ terms,
yielding a $+$12\% effect at 13~TeV and almost a $+$70\% increase of
the cross section at 100~TeV, both w.r.t.~$\LOQCD$. While at 13~TeV
this is still within the $\LOQCD+\NLOQCD$ scale uncertainty band, this
is not at all the case at 100~TeV. Indeed, it is these $\NLO_3$
contributions that are responsible for the enhancement in the cross
sections at the complete-NLO level as compared to the $\LOQCD+\NLOQCD$
ones, as presented in the last column of Tabs.~\ref{table:wpm} and
\ref{table:wpm100}. Hence, they must be included for accurate
predictions for $pp\to \ttw$ cross sections. Conversely, the $\NLO_4$
contributions are at the sub-percent level and can be neglected in all
phenomenologically relevant studies.

Applying a jet veto, such as the one of eq.~\eqref{jetveto}, impacts
only the real-emission corrections for $\ttw$ production.  All the
$\LO_i$ terms remain unaffected and, since the dominant NLO real-emission
contributions for this process are positive, the $\NLO_i$ cross sections decrease. This
is also what one expects from a physical point of view: the jet veto
cuts away part of the available phase space, resulting in a decrease
in the number of expected events. Indeed, in
Tabs.~\ref{table:wpmorders} and \ref{table:wpmorders100} we can see
that this is the case (for all values of $\mu$). On the other hand,
not all the $\NLO_i$ are affected in the same way by the jet veto. The
$\NLOQCD$ contribution is reduced by a large amount, about a factor
two for the central value of the scales, while the reduction in the
other $\NLO_i$ cross sections is much smaller. The reason for this
difference is the following: a large fraction of the $\NLO_1$
contribution originates from hard radiation, mainly due to the opening
of the quark-gluon luminosity and the double logarithmic enhancement
due to the radiation of a relatively soft/collinear $W$ boson from a
hard quark jet, {\it c.f.}, the left diagram of
Fig.~\ref{fig:realttw}. Instead, the $\NLO_2\equiv \NLOEW$ is dominated by ``EW
corrections'' to $\LO_1$ and, therefore, does not involve a large increase due to
the opening of the $qg$ initiated real-emission contributions. Hence, the effect from the
jet veto is strongly reduced. On the other hand, the $\NLO_3$ does contain the enhancement
from the gluon luminosity and is completely dominated by the $tW \to
tW$ scattering, which is part of the real-emission contributions, see
the right diagram of Fig.~\ref{fig:realttw} and the discussion in
sec.~\ref{calc:frame}. Even so, these contributions are not very
strongly affected by the jet veto, since the jet in $tW \to tW$
scattering is going mostly in the forward directions, which are
unaffected by the central jet veto of eq.~\eqref{jetveto}. The jet
veto may be customised in order to enhance or suppress the $\NLO_i$
contributions, {\it e.g.}, to study the impact of $tW \to tW$
scattering in more detail. However, it should be noted that a stronger
jet veto would further suppress the $\NLO_i$ contributions, but it may also
lead to unreliable results at fixed-order, due to the presence of
unresummed large and negative contributions from QCD Sudakov
logarithms. We leave a detailed study of the effects of various jet
vetoes for future work.

On the total cross sections, see Tabs.~\ref{table:wpm} and
\ref{table:wpm100}, the effect of the jet veto is not only manifest in
the reduction of the $\LOQCD+\NLOQCD$ and $\LO+\NLO$ cross sections,
but also in their greatly-reduced scale uncertainties. The latter are
almost halved for the 13~TeV cross sections and reduced to about 11\%
at 100~TeV. This is another confirmation that the $\NLOQCD$ is
dominated by hard radiation due to the opening of additional
production channels, which have a large tree-level induced scale
dependence. This reduction of the uncertainties coming from scale
variations means that the difference between the purely QCD
calculation and complete-NLO predictions becomes of the same order as
the scale uncertainties (at 13~TeV) or even considerably larger (at
100~TeV). Hence, with the jet veto applied, it becomes even more
important to include the $\NLO_3$ contribution for a reliable
prediction of the cross section for $\ttw$ hadroproduction. We stress
that the inclusion of only NLO EW corrections leads to a smaller shift
and in the opposite direction.

\subsubsection*{Differential distributions}

\begin{figure}[!h]
\vspace*{-2.2cm}
\centering
\includegraphics[width=0.43\textwidth]{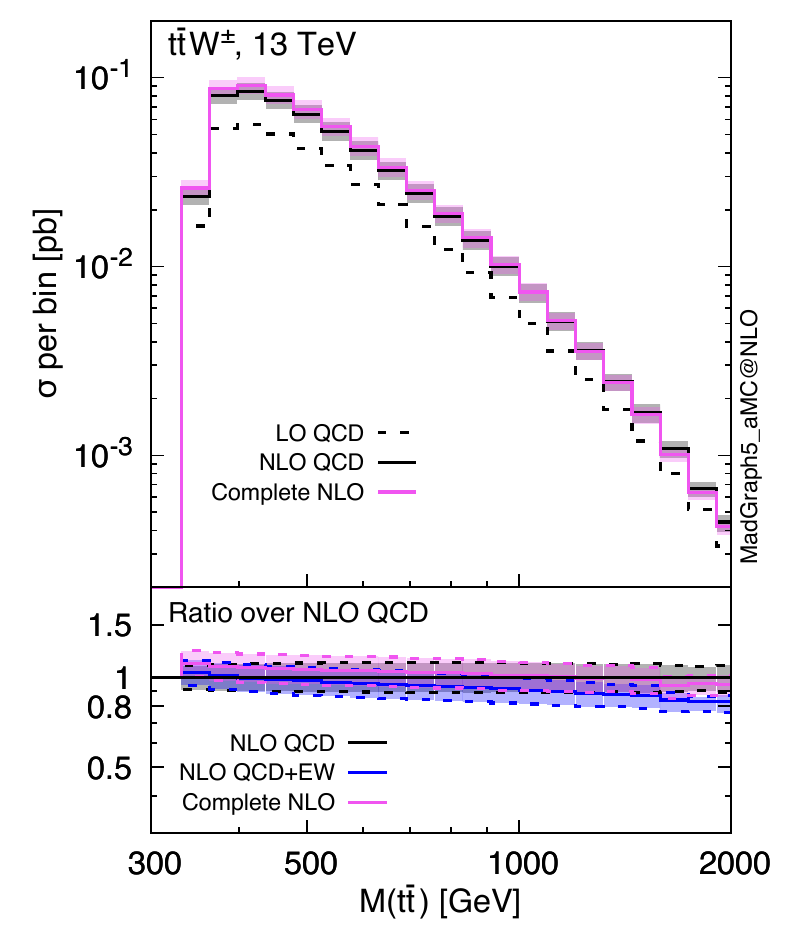}
\vspace{0.1cm}
\includegraphics[width=0.43\textwidth]{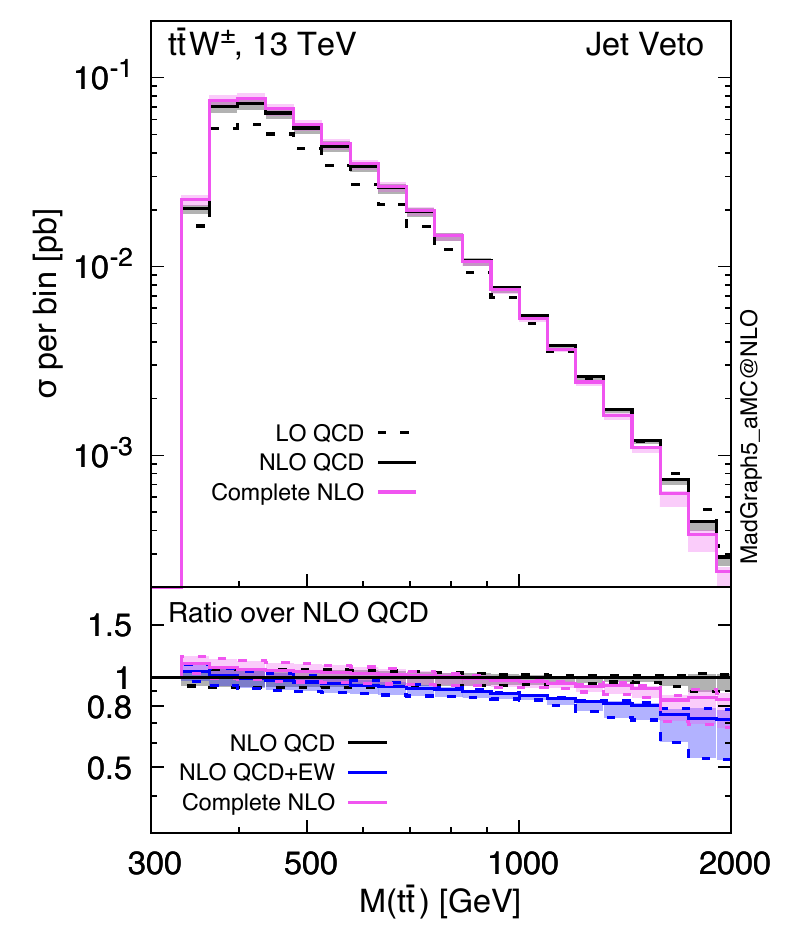}
\vspace{0.1cm}
\includegraphics[width=0.43\textwidth]{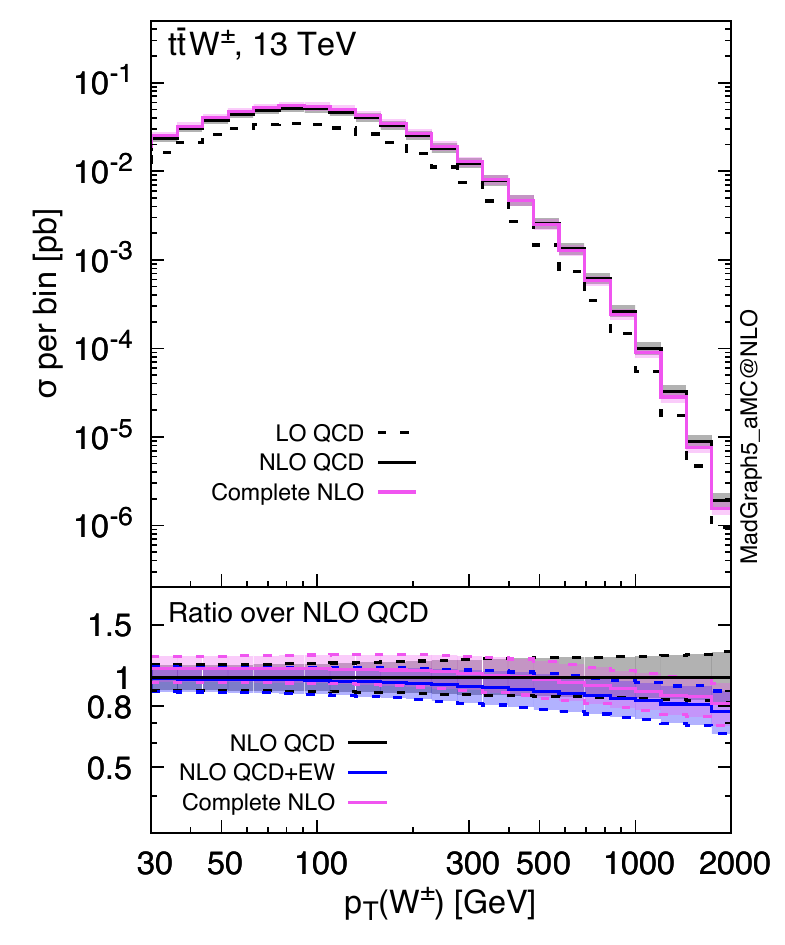}
\includegraphics[width=0.43\textwidth]{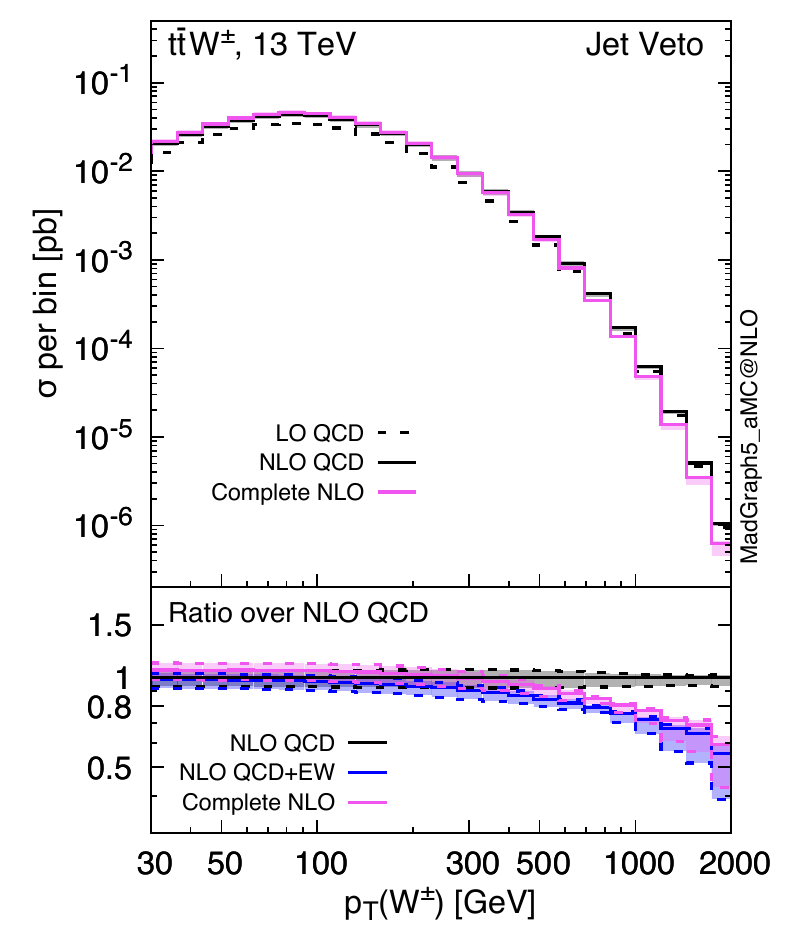}
\includegraphics[width=0.43\textwidth]{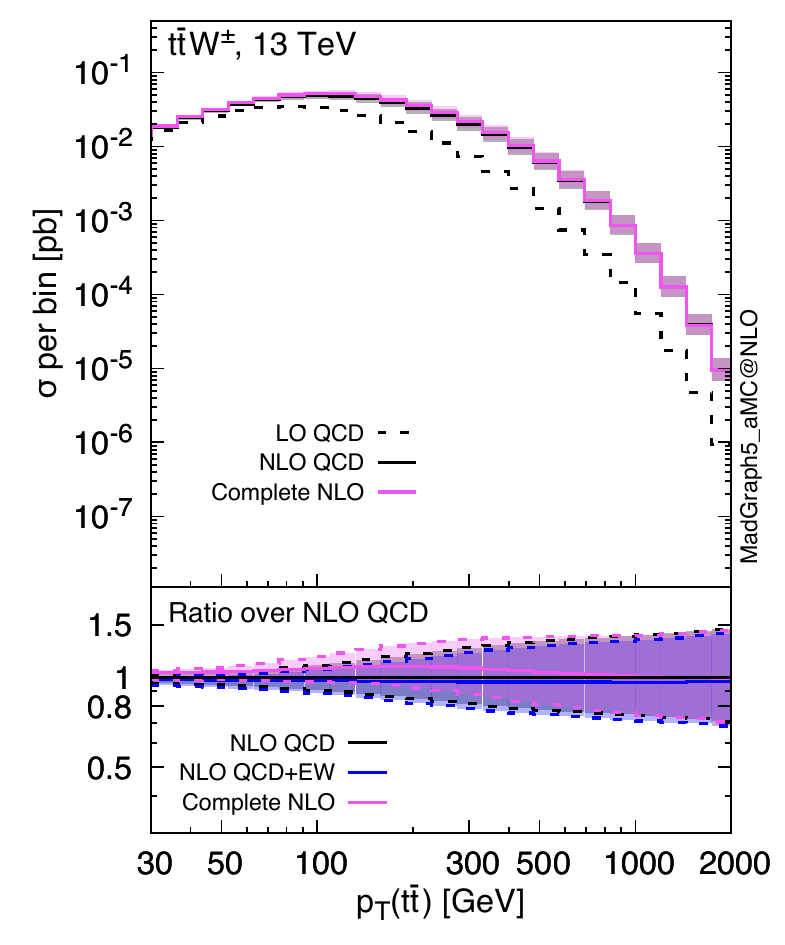}
\includegraphics[width=0.43\textwidth]{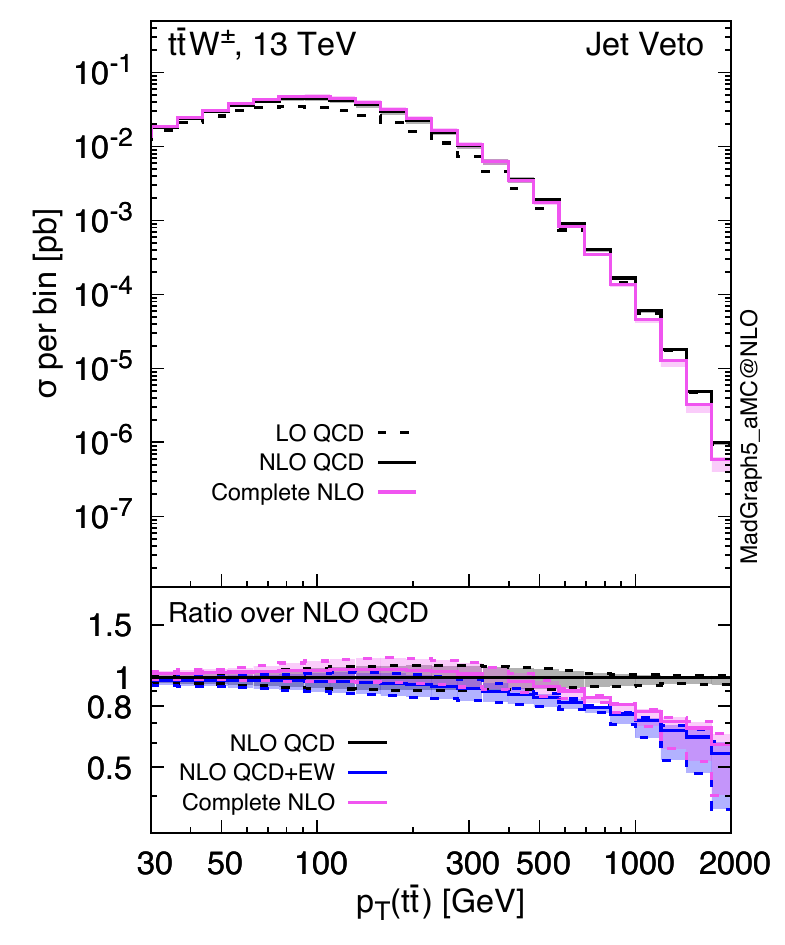}
\vspace*{-0.5cm}
\caption{Differential distributions for $\ttw$ production at 13 TeV. 
 For the plots on the right,  the jet veto of eq.~\eqref{jetveto} has been applied. The main panels show the  scale-uncertainty bands for $\LOQCD+\NLOQCD$ (black) and $\LO+\NLO$ (pink), and central value of $\LOQCD$; In the lower inset the scale-uncertainty bands are normalised to the  $\LOQCD+\NLOQCD$ central value and also the $\LOQCD+\NLOQCD+\NLOEW$ prediction (blue) is displayed.}
\label{fig:ttw13}
\end{figure}

\begin{figure}[!h]
\vspace*{-2.cm}
\centering
\includegraphics[width=0.43\textwidth]{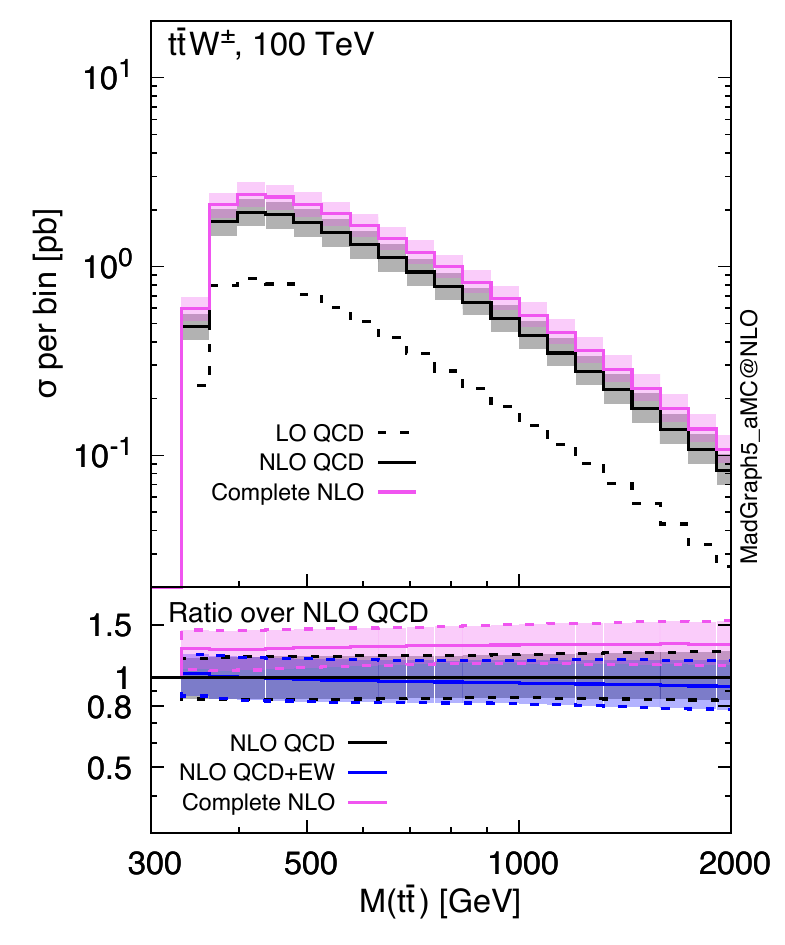}
\vspace{0.1cm}
\includegraphics[width=0.43\textwidth]{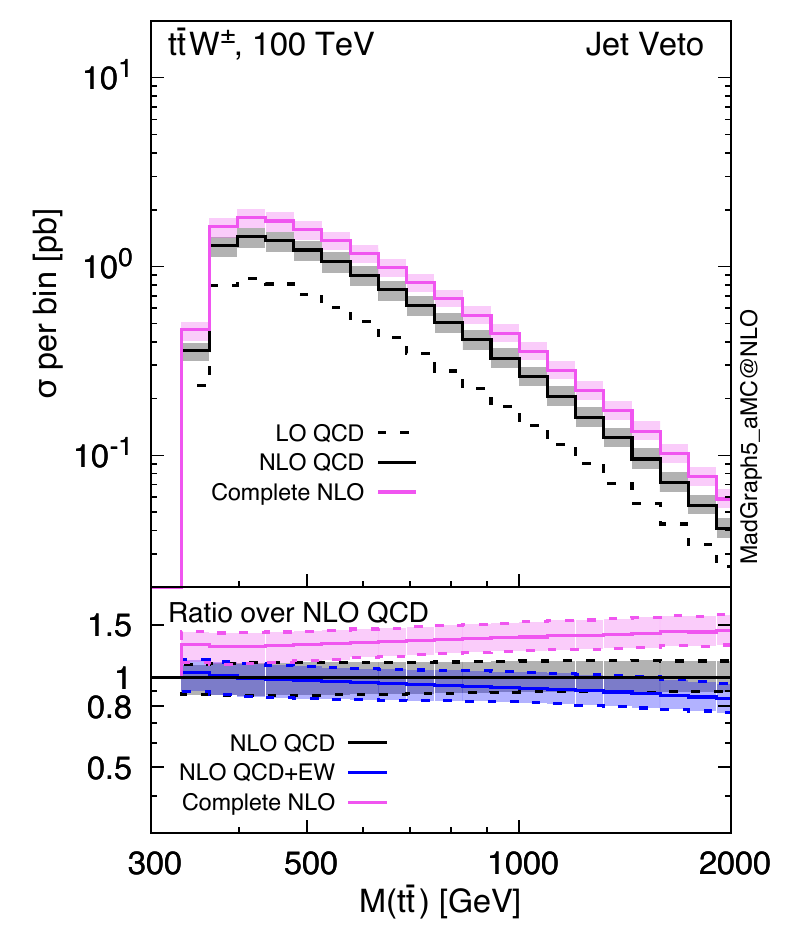}
\vspace{0.1cm}
\includegraphics[width=0.43\textwidth]{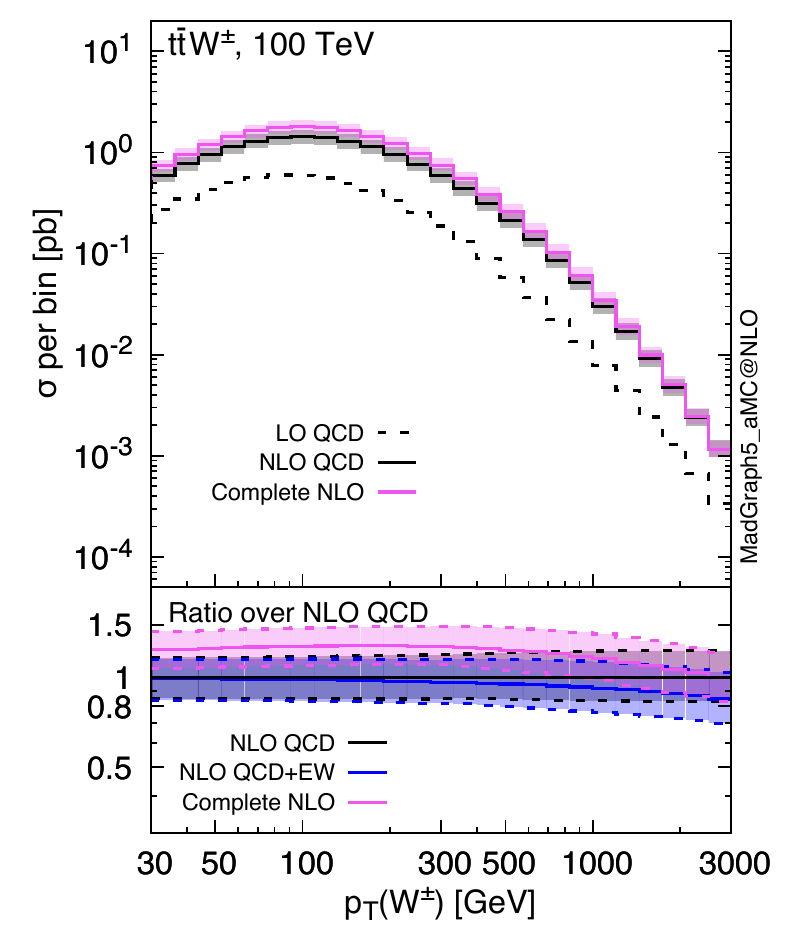}
\includegraphics[width=0.43\textwidth]{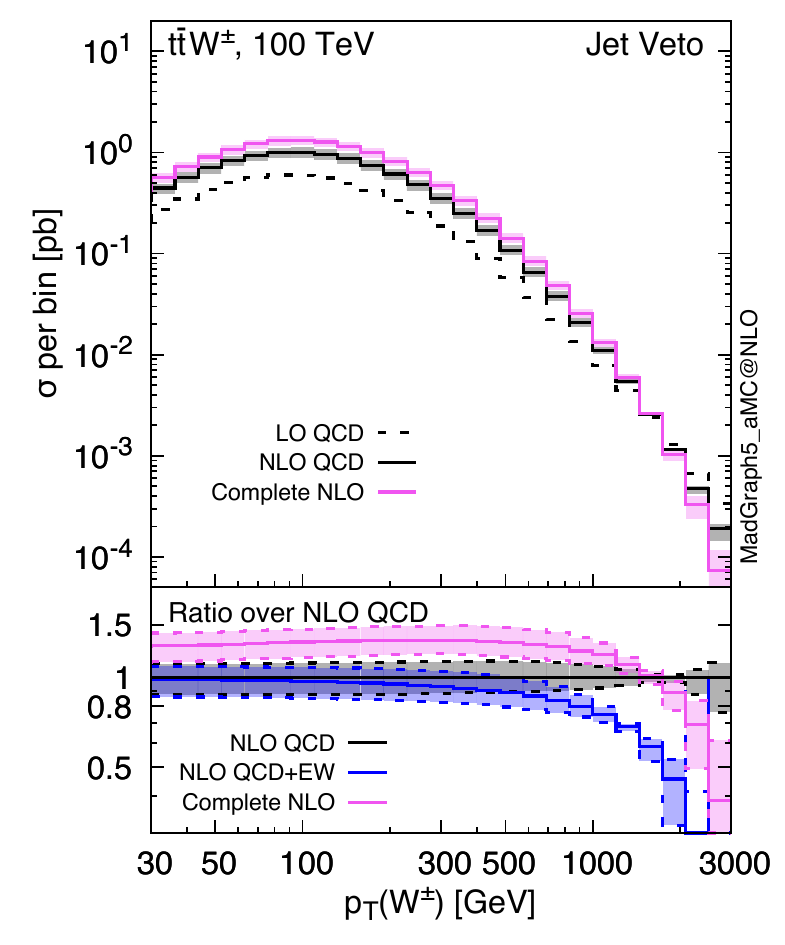}
\includegraphics[width=0.43\textwidth]{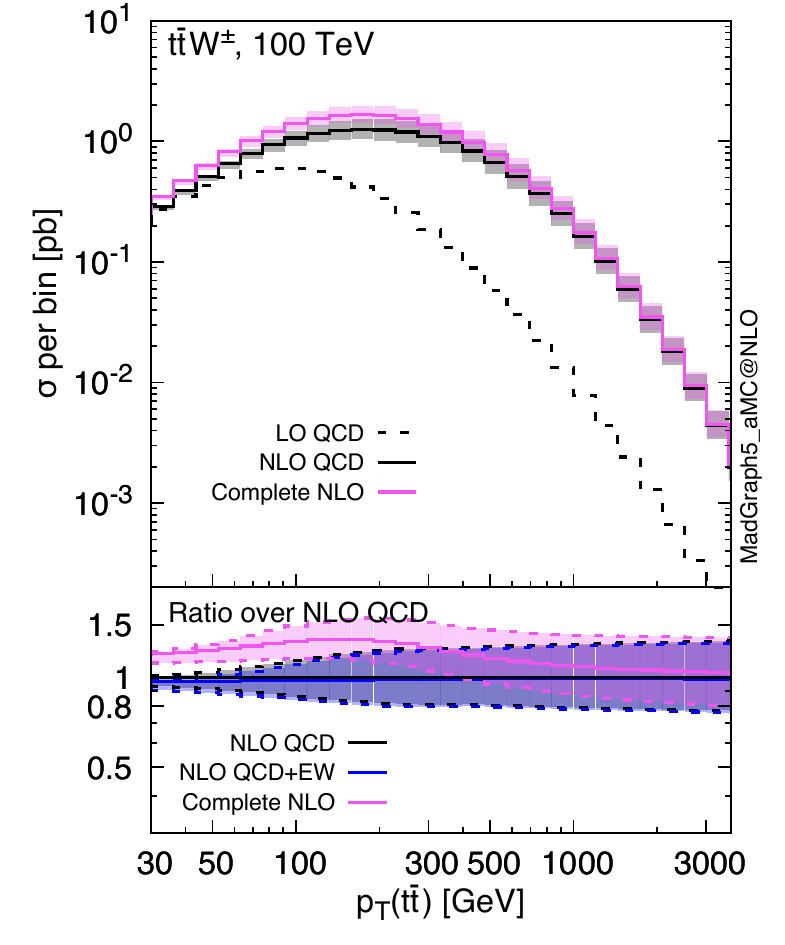}
\includegraphics[width=0.43\textwidth]{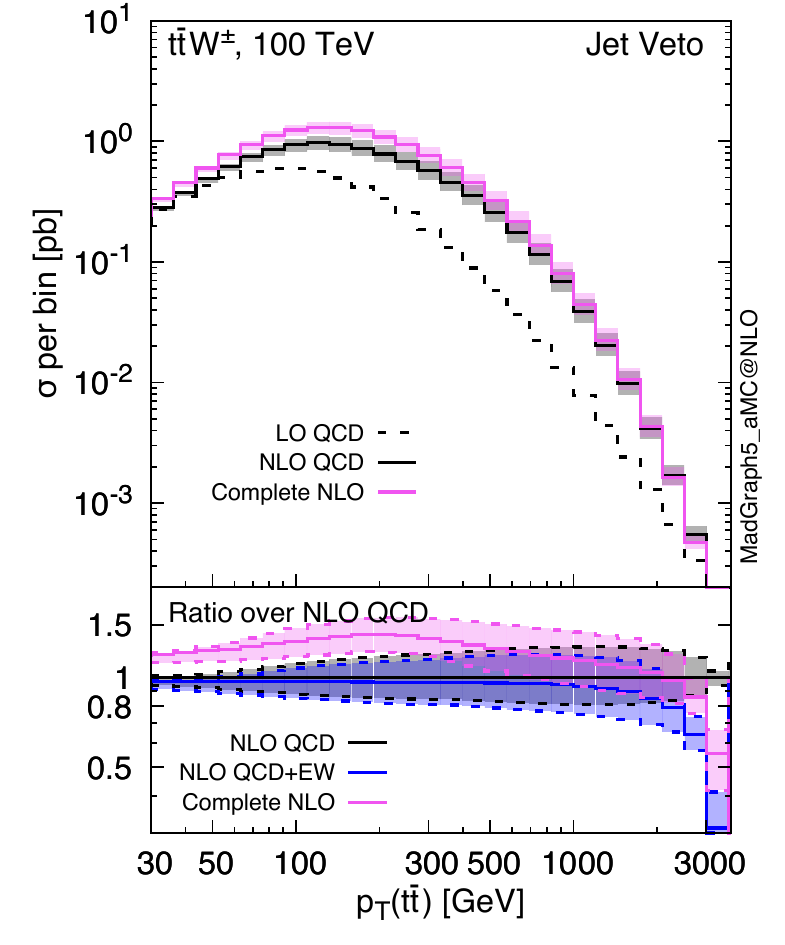}
\caption{Same as Fig.~\ref{fig:ttw13} but for 100 TeV collisions.}
\label{fig:ttw100}
\end{figure}

Results for three representative distributions, $m(\ttt)$,
$\pt(W^\pm)$ and $\pt(\ttt)$, are shown for 13~TeV in
Fig.~\ref{fig:ttw13} and for 100~TeV in Fig.~\ref{fig:ttw100}. We
consider the observables without (the plots on the left) and with (the
plots on the right) the jet veto of eq.~\eqref{jetveto}.  Each plot
has the following layout. The main panel shows distributions at NLO
QCD (black) and complete-NLO (pink) accuracy, including scale
variation uncertainties. For reference, we include also the $\LOQCD$
central value ($\mu=\mu_c\equiv H_T/2$) as a black-dashed
line.\footnote{Comparisons among the scale uncertainties of the
  $\LOQCD$ and $\LOQCD + \NLOQCD$ result have been documented in
  detail for 13 and 100~TeV in refs.~\cite{Maltoni:2015ena}
  and~\cite{Mangano:2016jyj}, respectively.} The lower insets show
three different quantities, all normalised to the central value of the
$\LOQCD + \NLOQCD$ prediction. The grey band is the $\LOQCD + \NLOQCD$
prediction including scale-uncertainties and the pink band is the one
at complete-NLO accuracy, {\it i.e.}, they are the same quantities in
the main panel but normalised. The blue band is instead what is
typically denoted as the result at ``NLO~QCD~+~EW'' accuracy, namely,
the $\LOQCD + \NLOQCD + \NLOEW$ prediction.  Via the comparison of
these three quantities one can see at the same time the difference
between results at NLO QCD and complete-NLO accuracy but also their
differences with NLO~QCD~+~EW results, which have already been
presented in refs.~\cite{Frixione:2015zaa}. 

At 13~TeV and without the jet veto (left plots of
Fig.~\ref{fig:ttw13}), the predictions for the three observables at
the various levels of accuracy presented, coincide within their
respective scale uncertainties. For the $m(\ttt)$ and, in particular
the $\pt(W^\pm)$, we see that the NLO EW corrections are negative and
increase (in absolute value) towards the tails of the two
distributions as expected from EW Sudakov logarithms coming from the
virtual corrections. Only in the very tail of the distributions, close
to $m(\ttt)\sim 2000$~GeV and $\pt(W^{\pm})\sim 2000$~GeV the uncertainty
bands of the NLO~QCD and NLO~QCD~+~EW predictions no longer overlap. As expected
from the inclusive results, the complete-NLO results increase the
NLO~QCD~+~EW predictions such that they move again closer to the
NLO~QCD central value. Indeed, the NLO~QCD and the complete-NLO bands
do overlap for the complete phase-space range plotted. Moreover, the
difference between the NLO~QCD~+~EW predictions and the complete-NLO
is close to a constant for these two observables. Conversely, applying
the jet veto changes the picture. First, it is quite apparent that the
relative impact of the NLO EW corrections is increased significantly,
reaching up to $-$40\% in the tail of the $\pt(W^\pm)$ distribution,
as compared to only $-$20\% without the jet veto. The reason is
obvious: the jet veto reduces the large contribution from the
$\NLOQCD$, hence, relatively speaking the $\NLOEW$ becomes more
important. In other words, while the $\NLOQCD$ has a large contribution
from the real-emission corrections, and are therefore greatly affected
by the jet veto, in this region of phase space the $\NLOEW$ is dominated by the EW Sudakov logarithms,  which are not influenced by
the jet veto.  The other important effect coming from the jet veto is the
reduction of the scale uncertainties: as we have already seen at the inclusive level,
this reduction is about a factor two for 13~TeV. For the $m(\ttt)$ and
$\pt(W^\pm)$ this also appears to be the case over the complete
kinematic ranges plotted for the NLO~QCD predictions. At small and
intermediate ranges, this is also the case for the NLO~QCD~+~EW and
the complete-NLO results. On the other hand, in the far tails, the
uncertainty bands from the NLO~QCD~+~EW and, to a slightly lesser
extend, the complete-NLO are increased. Again, this is no surprise,
since, as we have just concluded, these predictions contain a large
contribution from EW Sudakov corrections in the $\NLOEW$, which have
the same large scale uncertainty as the $\LO_1$. Given that,
relatively speaking, these $\NLOEW$ contributions become significantly
more important with the jet veto, also the scale uncertainties become
significantly larger.

For the third observable, $\pt(\ttt)$, the situation is extreme. This
is mainly due to the fact that the $\NLOQCD$ corrections are not
constant over the phase space as was the case for $m(\ttt)$ and
$\pt(W^\pm)$. Rather, due to terms of order
$\alpha_s\log^2(\pt^2(\ttt)/m_W^2)$ the $\NLOQCD$ greatly enhances the
$\LOQCD$ predictions for moderate, and, in particular, large
$\pt(\ttt)$. This enhancement originates from the real-emission $\ttt
W^{\pm} q$ final-states, where a soft and collinear $W^{\pm}$ can be
emitted from the final-state quark (see left diagram in
Fig.~\ref{fig:realttw}). Thus, while at the Born level the $\ttt$ pair
is always recoiling against the $W^{\pm}$ boson, at NLO QCD accuracy,
for large $\pt(\ttt)$ values, it mainly recoils against a jet that is
emitting the $W^{\pm}$ boson. More details about this mechanism can be
found in ref.~\cite{Maltoni:2015ena}. For this reason, without a jet
veto, at NLO QCD accuracy very large corrections and scale
uncertainties are present for large $\pt(\ttt)$ values. Indeed, the
dominant $\NLOQCD$ contribution, the soft and collinear emission of a
$W^{\pm}$ boson from a final-state quark, is very large and does not
lead to a reduction of the scale dependence.\footnote{The size of the
  $\NLOQCD$ contribution is the difference between the dashed and the
  solid black line.} Moreover, since the $\NLOQCD$ are by far the
dominant contributions, the effects from $\LNLO_i$, with $i>1$ are
completely negligible at large transverse momenta.
Only for intermediate transverse momenta, 80~GeV~$<\pt(\ttt)<$~400~GeV, we
see a small effect in the comparison of NLO~QCD and
complete-NLO.

On the other hand, with a jet veto, the $\NLOQCD$ contribution (and
therefore also the scale uncertainties) is strongly reduced. Indeed,
when the jet veto is applied, hard-jets and the corresponding
logarithmic enhancements are not present, and the $\ttt$ pair is mostly
recoiling directly against the $W^\pm$ boson, making the predictions
for $\pt(\ttt)$ and $\pt(W^\pm)$ very similar. The only difference is
in the comparison of the NLO QCD and the complete-NLO predictions. For
the $\pt(W^\pm)$ observable, this difference is basically a constant
in the region 30~GeV~$<\pt(W^\pm)<$~400~GeV. On the other hand, for
$\pt(\ttt)$ we see that the $\NLO_3$ contribution is not a constant:
there is a reduction at small transverse momenta. Indeed, one would
expect from $tW \to tW$ scattering that the transverse momenta of the
top pair is typically  larger than in the (N)LO${}_1$, due to the
$t$-channel enhancement (between the $\ttt$ and the $W^\pm j$ pairs) at large transverse momenta. This is somewhat
washed-out for the $\pt(W^\pm)$ since it is the $W$ boson together
with the jet that receive this enhancement.

At 100~TeV, see Fig.~\ref{fig:ttw100}, the differences between the
various predictions are qualitatively different from 13~TeV. The
reason is that the opening of the $qg$-induced contributions in
$\NLO_1$ and the $tW \to tW$ scattering contribution in $\NLO_3$ are
much more dramatic. The central value of the complete-NLO predictions
is typically outside of the NLO QCD band even though the scale
uncertainties are larger at 100~TeV than at 13~TeV. Moreover, with the
jet veto, the bands generally do not even touch, apart from where they
cross at large $\pt(W^\pm)$ and $\pt(\ttt)$. 

Without a jet veto, on the basis of all the previous considerations, also NLO corrections on top of the $\ttw j$ final state may be relevant for $\ttw$ inclusive production. Indeed sizeable effects are expected from QCD and EW corrections on top of the dominant $\alpha_s\log^2(\pt^2(\ttt)/m_W^2)$ contribution and the large $\NLO_3$ one, both arising from the $qg$ initial state. The former would lead also to a reduction of the scale dependence in the tail of the $\pt(\ttt)$ distribution, which is dominated by the $\ttw j$ final state. However, these contributions are part of the NNLO corrections to the inclusive $\ttw$ production and therefore are not available and not included in our calculation. A possible way for estimating these effects is merging $\ttw $ and $\ttw j$ (and $\ttw \gamma$) final states at NLO accuracy. In the case of NLO QCD corrections a study in this direction has been suggested for $\ttw$ production in ref.~\cite{Maltoni:2015ena}. For $\NLOEW$ and subleading $\NLO_i$ corrections a fully-consistent technology is not yet available to perform this kind of study.


\begin{figure}[t]
\centering
\includegraphics[width=0.45\textwidth]{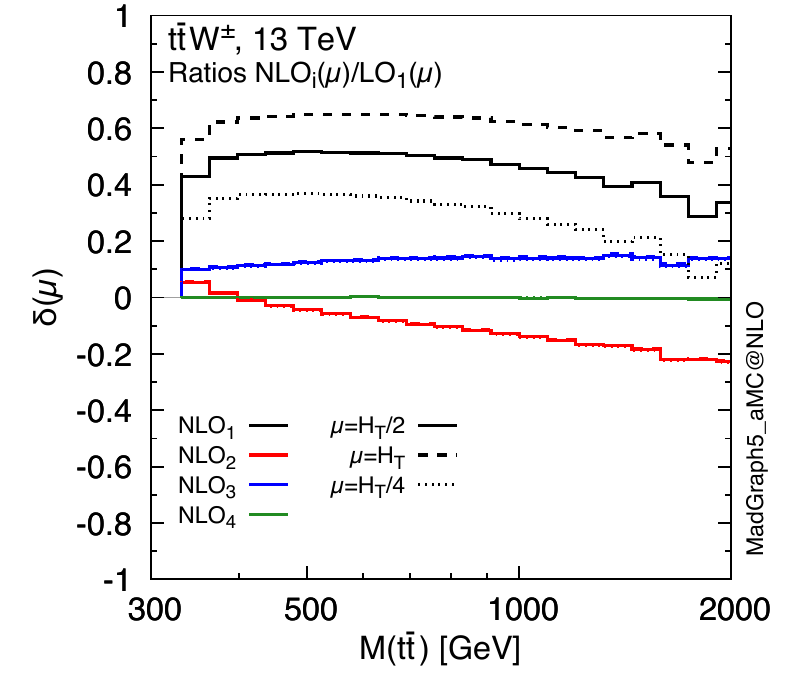}
\vspace{0.4cm}
\includegraphics[width=0.45\textwidth]{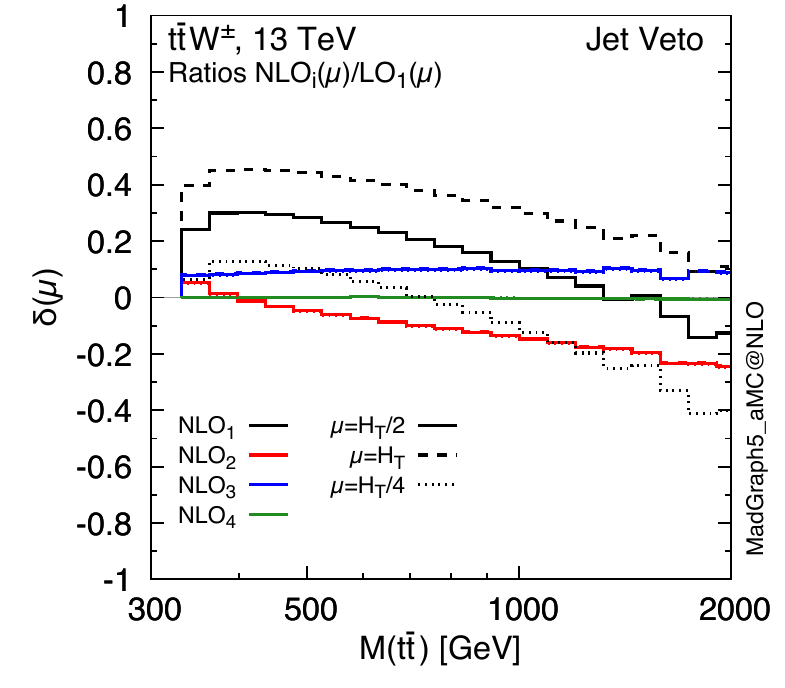}
\vspace{0.4cm}
\includegraphics[width=0.45\textwidth]{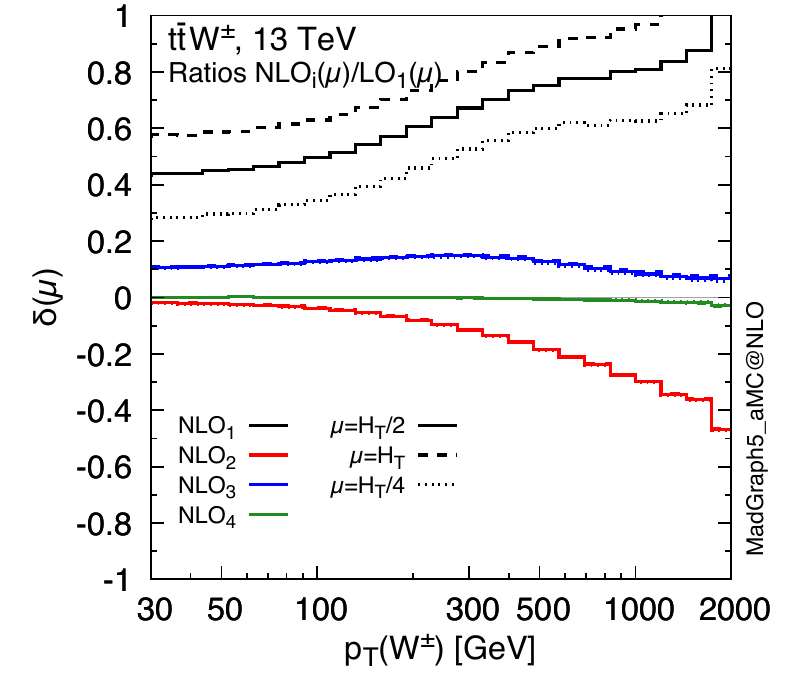}
\includegraphics[width=0.45\textwidth]{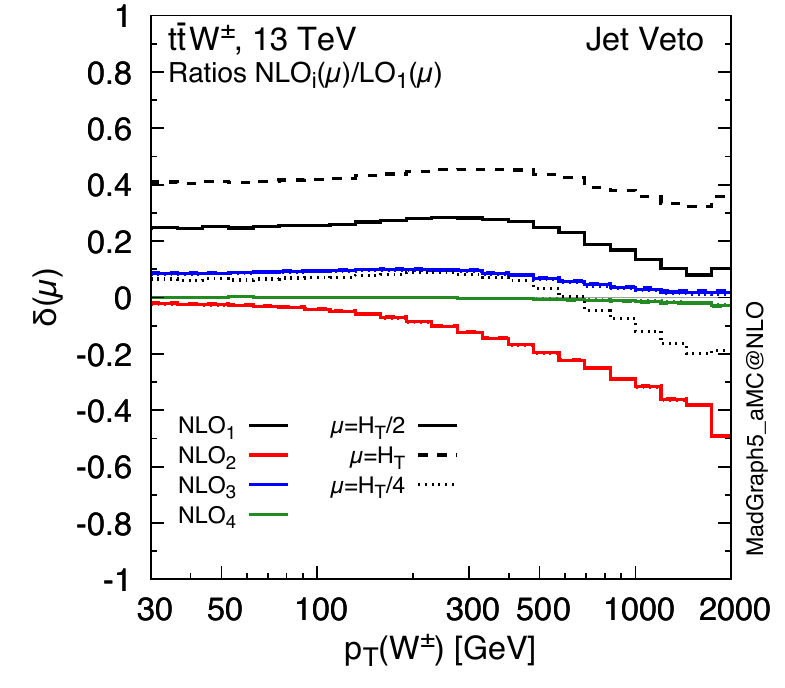}
\includegraphics[width=0.45\textwidth]{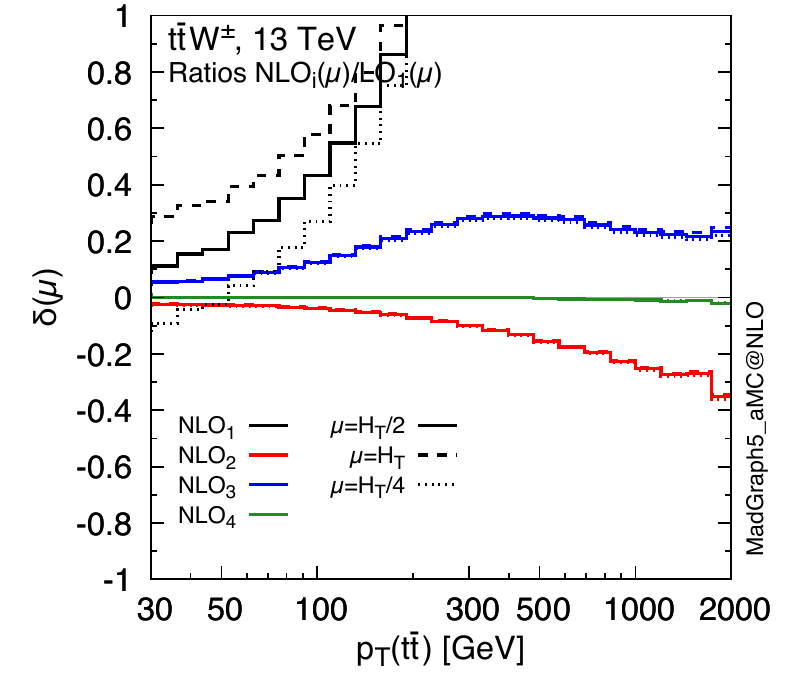}
\includegraphics[width=0.45\textwidth]{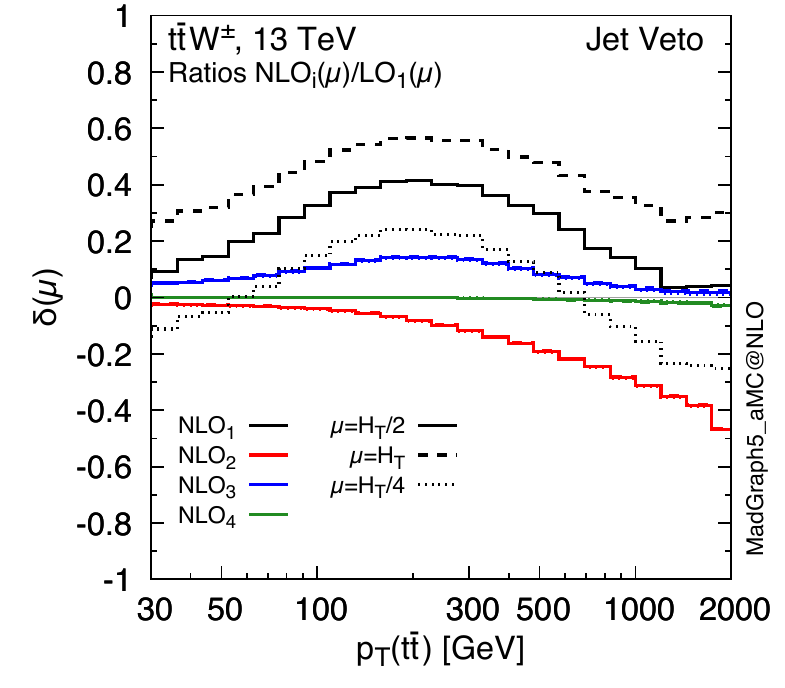}
\caption{Individual $\NLO_i$ contributions to $\ttw$ production at 13 TeV normalised to $\LO_1\equiv\LOQCD$, for different values of the scale $\mu$ for the same distributions as considered in Fig.~\ref{fig:ttw13}. These plots do not directly show scale uncertainties. Note that $\NLO_1\equiv\NLOQCD$ and $\NLO_2\equiv\NLOEW$.}
\label{fig:ttw13orders}
\end{figure}

\begin{figure}[t]
\centering
\includegraphics[width=0.45\textwidth]{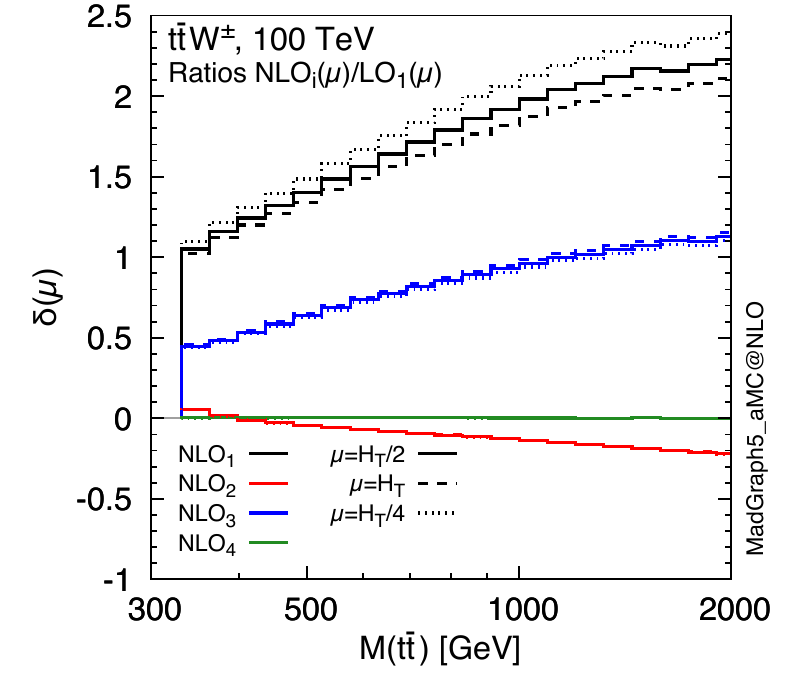}
\vspace{0.4cm}
\includegraphics[width=0.45\textwidth]{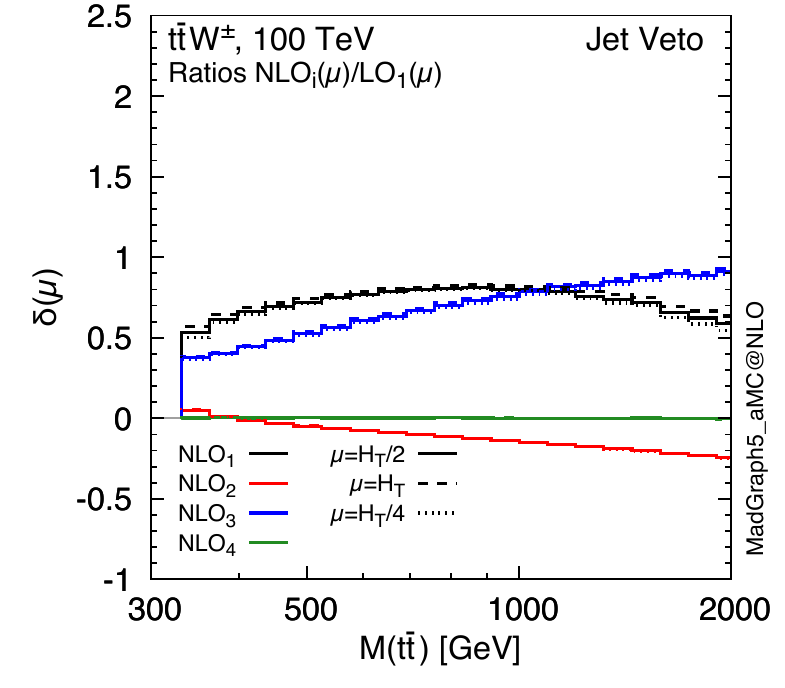}
\vspace{0.4cm}
\includegraphics[width=0.45\textwidth]{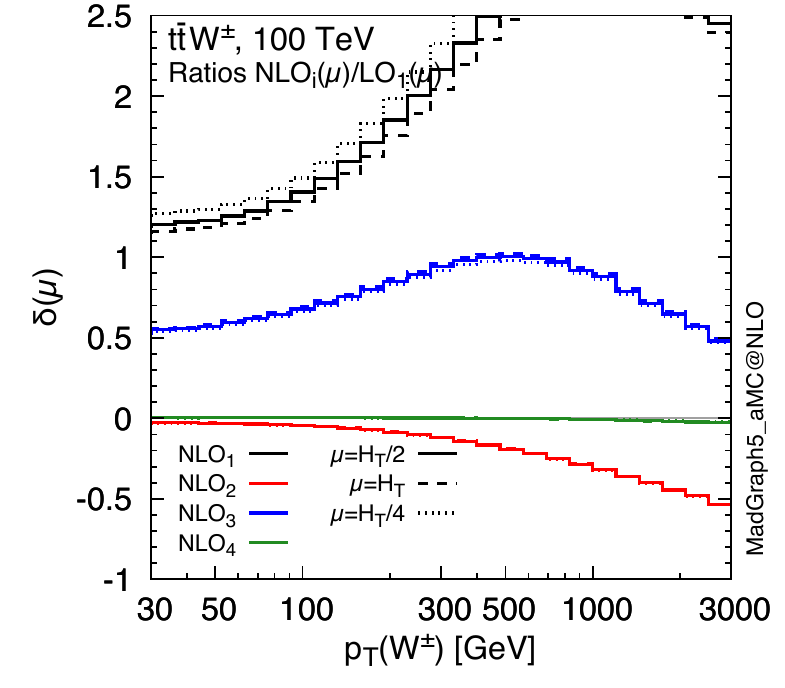}
\includegraphics[width=0.45\textwidth]{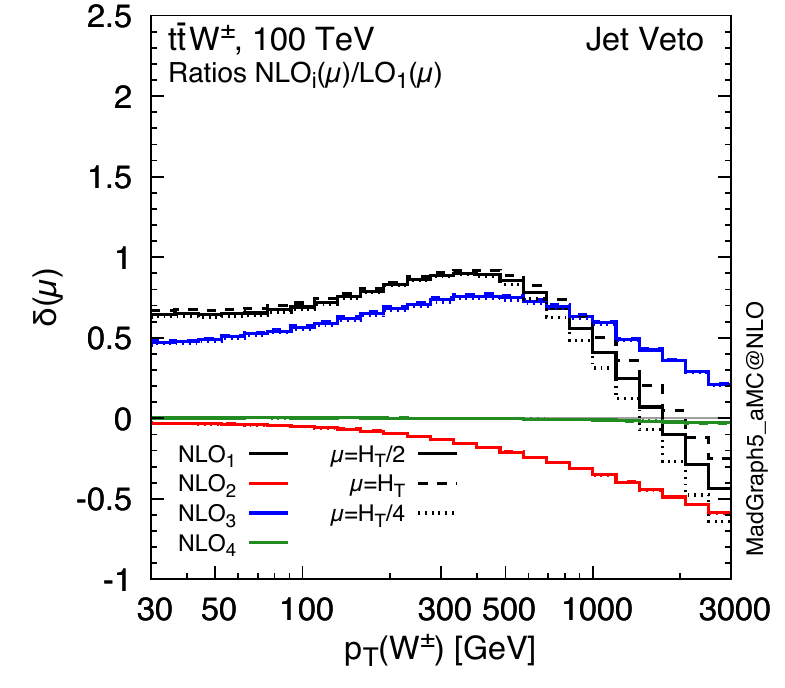}
\includegraphics[width=0.45\textwidth]{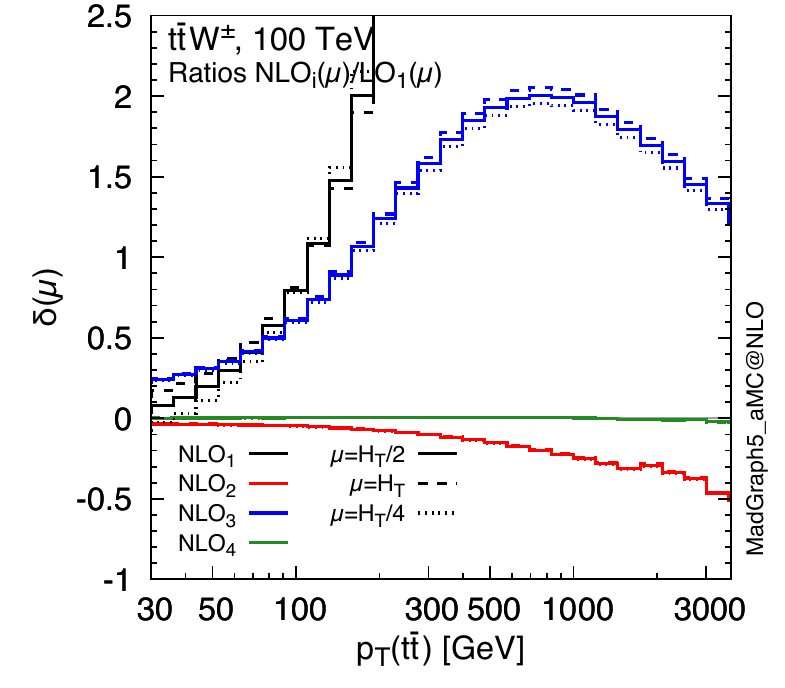}
\includegraphics[width=0.45\textwidth]{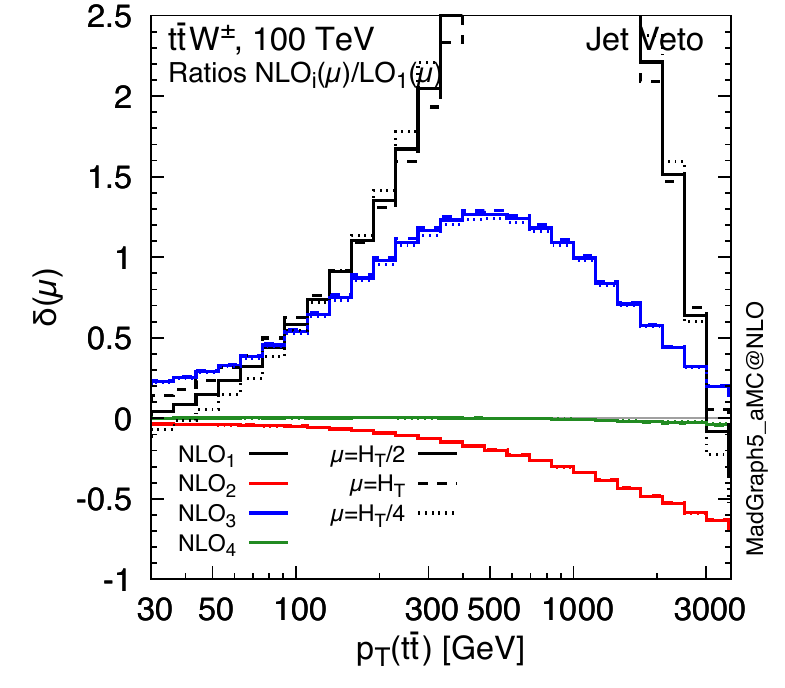}
\caption{Same as Fig.~\ref{fig:ttw13orders} but for 100 TeV collisions.}
\label{fig:ttw100orders}
\end{figure}

Further details about individual $\NLO_i$ contributions at the
differential level are given in Fig.~\ref{fig:ttw13orders} (13 TeV)
and Fig.~\ref{fig:ttw100orders} (100 TeV). In the plots we show all
the $\delta_{\NLO_i}(\mu)$ for $\mu=\mu_c\equiv H_T/2$ (solid line),
$\mu=\mu_c/2$ (dashed line) and $\mu=2\mu_c$ (dotted line). We show
the same distributions (with and without veto) as in
Figs.~\ref{fig:ttw13} and~\ref{fig:ttw100}. We remark again that the
$\delta_{\NLO_i}(\mu)$ do not show directly scale uncertainties since
the value of $\mu$ is varied both in the numerator and the denominator
of $\delta$. On the other hand, we can directly see that also at the
differential level the relative sizes of both $\NLO_2$ and $\NLO_3$
w.r.t.~the $\LOQCD$ are almost insensitive to the value of the scale;
the corresponding solid, dashed and dotted lines are almost
indistinguishable. As expected, also at the differential level the
impact of the $\NLO_4$ is completely negligible for the whole range of
the distributions considered.

As could already have been concluded by comparing the dashed and solid
black lines in Figs.~\ref{fig:ttw13} and~\ref{fig:ttw100}, the NLO QCD
corrections are not at all a constant over phase space. The solid
black lines in Figs.~\ref{fig:ttw13orders} and~\ref{fig:ttw100orders}
make this very clear. In particular for the $\pt(\ttt)$ distributions
without the jet veto (lower left plots), the $\NLO_1\equiv \NLOQCD$ contribution
easily becomes as large as the $\LO_1\equiv \LOQCD$ and increases to more than an
order of magnitude larger than $\LO_1$ at large transverse momenta in
100~TeV collisions. But also for $\pt(W^\pm)$ we see large NLO QCD corrections,
in particular at 100~TeV. On the other hand, for $m(\ttt)$ the NLO QCD
corrections are mostly flat, in particular at 13~TeV. With the jet
veto (plots on the right) the situation changes quite
dramatically. The NLO QCD corrections are, in general, under much
better control, even though one can see that the extreme tails in the
$\pt(W^\pm)$ and $\pt(\ttt)$ at 100 TeV the $\NLOQCD$ contributions
decrease rapidly and are starting to be strongly influenced by
logarithms related to the jet-veto scale. If one would look at even
larger transverse momenta, or, equivalently, reduce the jet-veto
scale, these logarithms will grow and eventually fixed-order
perturbation theory would break down, showing the need for resummation
of these jet-veto logarithms.

Since these plots are normalised w.r.t.~the $\LO_1$ ({\it c.f.}, the
lower insets of Figs.~\ref{fig:ttw13} and~\ref{fig:ttw100} which are
normalised to $\LO_1+\NLO_1$), one can clearly see the effects of the
NLO~EW corrections, {\it i.e.}, the $\NLO_2$, independently from the
NLO~QCD corrections. One sees the typical EW Sudakov logarithms:
negligible effects at the percent level at small and moderate $\ttt$
invariant masses and $W^\pm$ and $\ttt$ transverse momenta, but
growing rapidly with increasing values of the observables, to about
$-$20\% at $m(\ttt)\simeq 2000$~GeV and $-$40\% at
$\pt(W^\pm)\simeq\pt(\ttt)\simeq 2000$~GeV. The fact that the NLO EW
corrections are smaller for $m(\ttt)$ in comparison to $\pt(W^\pm)$
and $\pt(\ttt)$ is no surprise since the impact of the EW Sudakov
logarithms is related to the number of invariants that are large for
the observable considered. Typically, for large invariant masses,
there need to be fewer large invariants than for producing large
transverse momenta. The size of the NLO EW corrections relative to the
$\LO_1$ is quite similar for 13~TeV and 100~TeV collisions.  Moreover,
by comparing the distributions with and without the jet veto we also
see that their sizes are hardly influenced by the jet veto.

At variance with the $\NLO_2$ term, at 13~TeV the $\NLO_3$
contribution is much more constant w.r.t.~the $\LO_1$ over the whole
phase space. Indeed, for the $m(\ttt)$ the $\delta_{\NLO_3}$ is
effectively a constant, increasing the $\LO_1$ cross section by about
12\% (which is reduced by applying the jet veto to about
9\%). Similarly, for the $\pt(W^\pm)$ distribution, the $\NLO_3$
correction is fairly flat. On the other hand, the $\pt(\ttt)$ does
show some kinematic dependence in the $\delta_{\NLO_3}$ ratio. It is
small at small transverse momenta, increases at intermediate values
and, in particular when the jet veto is applied, it decreases again at
large values of $\pt(\ttt)$. This is consistent with what we found in
the comparing the $\LOQCD+\NLOQCD$ and NLO~QCD~+~EW predictions in
Fig.~\ref{fig:ttw13}. At 100~TeV the $\NLO_3$ contributions are large
and the $\delta_{\NLO_3}$ plots are not at all flat in the phase
space. As at 13~TeV, the effects are most dramatic in the $\pt(\ttt)$
distributions, which show a large hump at around 500~GeV (1~TeV) with
(without) the jet veto. However, as discussed before, without a jet
veto, at large $\pt(\ttt)$ the $\NLOQCD$ corrections is giant and is
even the dominant contribution among all the $\LNLO_i$ ones, including
the $\LO_1$. For this reason, although $\delta_{\NLO_2}$ and
$\delta_{\NLO_3}$ are large at high $\pt(\ttt)$, results at $\LOQCD
+\NLOQCD$, $\LOQCD +\NLOQCD + \NLOEW$, and $\LO +\NLO$ accuracies are
very close to each other; the three predictions are all dominated by
$\NLOQCD$, while $\delta_{\NLO_i}$ are normalised to $\LOQCD$.

The application of a jet veto as in eq.~\eqref{jetveto} may be
exploited in BSM analyses such as the one described in
ref.~\cite{Dror:2015nkp}; rather than requiring a forward jet it may
be possible to observe enhancements in the $tW^{\pm}\to tW^{\pm}$
scattering directly in $\ttw$ production by vetoing hard central jets.


\clearpage

\subsection{Results for $pp\to \ft$ production}
\label{sec:4t}
\begin{table}[t]
\small
\renewcommand{\arraystretch}{1.5}
\begin{center}
\begin{tabular}{c c c c c c}
\toprule
$\sigma[\textrm{fb}]$ & LO${}_{\textrm{QCD}}$ & $\LOQCD+\NLOQCD$ & LO & $\LO+\NLO$ & $\frac{\LO(+\NLO)}{\LOQCD(+\NLOQCD)}$\\
\midrule
$\mu=H_T/4$ & $ 6.83 {}^{+70 \%}_{-38 \%}$ & $ 11.12 {}^{+19 \%}_{-23 \%}$ & $ 7.59 {}^{+64 \%}_{-36 \%}$ & $ 11.97 {}^{+18 \%}_{-21 \%}$ & $ 1.11\, ( 1.08 )$ \\
\bottomrule
\end{tabular}
\caption{Cross section for $pp\to\ft$ at 13 TeV in various approximations.}
\label{table:4t}
\end{center}
\end{table}

\begin{table}[t]
\small
\renewcommand{\arraystretch}{1.5}
\begin{center}
\begin{tabular}{c c c c c c}
\toprule
$\sigma[\textrm{pb}]$ & LO${}_{\textrm{QCD}}$ & $\LOQCD+\NLOQCD$ & LO & $\LO+\NLO$ & $\frac{\LO(+\NLO)}{\LOQCD(+\NLOQCD)}$\\
\midrule
$\mu=H_T/4$ & $ 2.37 {}^{+49 \%}_{-31 \%}$ & $ 3.98 {}^{+18 \%}_{-19 \%}$ & $ 2.63 {}^{+44 \%}_{-28 \%}$ & $ 4.18 {}^{+17 \%}_{-17 \%}$ & $ 1.11\, ( 1.05 )$ \\
\bottomrule
\end{tabular}
\caption{Same as in Tab.~\ref{table:4t} but for 100~TeV.}
\label{table:4t100}
\end{center}
\end{table}

Similarly to the previous section, we start by presenting predictions for $\ft$ total cross sections at 13 and 100 TeV proton--proton collisions and then we discuss  results at the differential level.
Using a layout that is similar to Tab.~\ref{table:wpm}, in Tab.~\ref{table:4t} we show 13 TeV predictions at  $\LOQCD$,  $\LOQCD + \NLOQCD$,  $\LO$ and  $\LO + \NLO$ accuracies. We also display the $\LO/\LOQCD$ and, in brackets,  $(\LO + \NLO)/(\LOQCD + \NLOQCD)$ ratios. Results at 100 TeV are in Tab.~\ref{table:4t100}. 
In Tab.~\ref{table:4torders}, similarly to Tab.~\ref{table:wpmorders}, we show 13 TeV predictions for the $\delta_{\LNLO_i}(\mu)$ ratios, and analogous results at 100 TeV are in Tab.~\ref{table:4torders100}.

As can be seen in Tabs.~\ref{table:4t} and \ref{table:4t100}, the
scale dependence is very large at $\LOQCD$ and $\LO$ accuracy and it
is strongly reduced both in the NLO QCD and complete-NLO predictions
to about 20\%. Nevertheless, it is still larger than the impact of  the non-purely-QCD contributions, which is also reduced moving from LO to NLO accuracy,
halved in the 100 TeV case. At the inclusive level, the difference
between $\LO+\NLO$ and $\LOQCD+\NLOQCD$ predictions is well within
their respective scale uncertainties, especially at 100 TeV where this
difference is merely $5 \%$ of the $\LOQCD+\NLOQCD$ result. However, the numbers in
Tabs.~\ref{table:4t} and \ref{table:4t100} hide the most important
feature of the complete-NLO result, {\it i.e.}, very large and
scale-dependent cancellations among the $\LNLO_i$ terms with $i\ge
2$. This will become clear from the discussion in the next paragraph.

 \begin{table}[t]
\small
\begin{center}
\begin{tabular}{c c c c}
\toprule
$\delta[\%]$ & $\mu= H_T/8$ & $\mu=H_T/4$ & $\mu = H_T/2$ \\
\midrule 
LO${}_2$ & $-26.0$ & $-28.3$ & $-30.5$ \\
LO${}_3$ & $32.6$ &   $39.0$ &   $45.9$ \\
LO${}_4$ & $0.2$ &   $0.3$ &   $0.4$ \\
LO${}_5$ & $0.02$ &   $0.03$ &   $0.05$ \\
\midrule 
NLO${}_1$ & $14.0$ & $62.7$ & $103.5$ \\
NLO${}_2$ & $8.6$ & $-3.3$ & $-15.1$ \\
NLO${}_3$ & $-10.3$ & $1.8$ & $16.1$ \\
NLO${}_4$ & $2.3$ & $2.8$ & $3.6$ \\
NLO${}_5$ & $0.12$ & $0.16$ & $0.19$ \\
NLO${}_6$ & $< 0.01$ & $<0.01$ & $<0.01$ \\
\midrule 
$\NLO_2+\NLO_3$ & $-1.7$ & $-1.6$ & $0.9$ \\
\bottomrule
\end{tabular}
\caption{$\ft$: $\sigma_{\LNLO_i}/\sigma_{\LOQCD}$ ratios at 13 TeV, for different values of $\mu=\mu_r=\mu_f$.}
\label{table:4torders}
\end{center}
\end{table}

As anticipated in sec.~\ref{calc:frame}, in $\ft$ production the $\LO_
2$ and $\LO_ 3$ contributions are not so suppressed w.r.t.~the
$\LOQCD$, at variance with $\ttw$ production (see
Tabs.~\ref{table:4torders} and \ref{table:4torders100}, {\it c.f.}
Tabs.~\ref{table:wpmorders} and \ref{table:wpmorders100}). For $\ft$
production,  due to  sizeable
contributions from the EW $tt \to tt$ scattering, $\LO_
2$ and $\LO_ 3$  can induce corrections of the order $-30\%$ and $+40\%$
 on top of the $\LO_1$, respectively.\footnote{Similarly to the case of the
  $\LO_3$ in $\ttw$ production, the scale dependences of the $\LO_2$ and especially of the
  $\LO_3$  are much smaller than that of  $\LO_1$, due to the different powers of $\alpha_s$ associated to them. Hence,
  with larger(smaller) values of the scales and consequently smaller(larger) values of $\LO_1$,  the $\delta_{\LO_2}$ and $\delta_{\LO_3}$ become larger(smaller) in absolute value.}
  Therefore, also the
$\NLO_ 2$ and $\NLO_ 3$ contributions are large, since they contain
``QCD corrections'' to $\LO_ 2$ and $\LO_ 3$ terms, respectively. The
fact that a large fraction of $\NLO_ 2$ and $\NLO_ 3$ contributions is
of QCD origin can be understood by the $\mu$-dependencies of
$\delta_{\NLO_2}$ and $\delta_{\NLO_3}$ ratios, which, as can be seen
in Tabs.~\ref{table:4torders} and \ref{table:4torders100}, are very
large. Indeed, $\NLO_ 2$ and $\NLO_ 3$ terms involve explicit
logarithms of $\mu$ that compensate the PDF and $\alpha_s$ scale
dependence at $\LO_ 2$ and $\LO_ 3$ accuracy, respectively. Thus, in
$\ft$ production, at variance with most of the other production
processes studied in the literature, quoting the relative size of
$\NLOEW\equiv\NLO_2$ or $\NLO_3$ corrections without specifying the
QCD-renormalisation and factorisation scale is simply
meaningless. Moreover, $\delta_{\NLO_ 2}$ and $\delta_{\NLO_ 3}$
corrections can {\it separately} be very large, easily reaching $\pm
15\%$ (depending on the value of $\mu$). Surprisingly, for our central
value of the renormalisation and factorisation scales, the
$\delta_{\NLO_2}$ and $\delta_{\NLO_3}$ are almost zero\footnote{Our
  choice for the central value of the scales has not been tuned in
  order to reduce the effects from the $\NLO_2$ and $\NLO_3$. Rather,
  it is motivated by the study in ref.~\cite{Maltoni:2015ena}, which
  deals only with the $\LO_1$ and $\NLO_1$.}, particularly for
13~TeV. On the other hand, if we had  taken $H_T/2$ or even $m_{\ft}$ as
our central scale choice, the $\NLO_2$ and $\NLO_3$ corrections
relative to the $\LO_1$, $\delta_{\NLO_2}$ and $\delta_{\NLO_3}$, would have been much larger. Still, even for the
central value $\mu=H_T/4$, the corrections are much larger than
foreseen, especially for $\delta_{\NLO_3}$ which naively is expected
to be of order $\alphas^{3}\alpha^{2}/\alphas^{4}=\alpha^{2}/\alphas
\sim 0.1 \%$ level. On the other hand, the relative cancellation
observed between $\NLO_ 2$ and $\NLO_ 3$ contributions is even larger
than in the case of $\LO_ 2$ and $\LO_ 3$. As can be seen in the last
rows of Tabs.~\ref{table:4torders} and \ref{table:4torders100}, at the
inclusive level the sum of the ratios
$\delta_{\NLO_2}+\delta_{\NLO_3}$ is not only small, but also stable
under scale variation,\footnote{We verified this feature also with
  different functional forms for the scale  $\mu$.} resulting in corrections of
at most a few percents w.r.t.~the $\LOQCD$. Furthermore, particularly
at 13 TeV, $\delta_{\NLO_2}+\delta_{\NLO_3}$ receives also additional
cancellations when summed to $\delta_{\NLO_4}$, which itself is much larger than
the expected $\alphas^{2}\alpha^{3}/\alphas^{4}=\alpha^{3}/\alphas^2
\sim 0.01 \%$ level. To the best of our understanding, these
cancellations are accidental.

These large and accidental cancellations among the $\LNLO_i$ terms
with $i>1$ are particularly relevant from a BSM perspective, since the
level of these cancellations may be altered by new physics. As an
example, we can refer to the case of an anomalous $y_t$ coupling,
which, as we have already mentioned, has been considered in
the tree-level analysis of ref.~\cite{Cao:2016wib}. Terms proportional
to $y_t^2$ are present in all the $\LNLO_i$ with $i\ge2$ and terms
proportional to $y_t^4$ are present in all the $\LNLO_i$ with $i\ge3$,
but also terms proportional to $y_t^6$ are present for any
$i\ge3$. Moreover, also contributions proportional to $y_t$, $y_t^3$
and $y_t^5$ are possible.  Similar considerations apply also to other
new physics effects in $\ft$ production (see, {\it e.g.},
ref.~\cite{Zhang:2017mls} and references therein for scenarios already
analysed in the literature).

In order to understand the hierarchy of the different $\LNLO_i$ contributions, it is important to note that at 13 TeV and especially at
100 TeV the total cross section is dominated  by the $gg$ initial state (see, {\it e.g.},
ref.~\cite{Maltoni:2015ena}). For this reason, the $\LO_4$, $\LO_5$,
$\NLO_5$ and $\NLO_6$ contributions, which are vanishing for the $gg$
initial state, are much smaller than the other contributions. The
modest scale dependence of $\delta_{\NLO_ 4}$ is also induced by this
feature; the $\NLO_4$ contribution mainly arises from ``EW
corrections'' to $gg$-induced $\LO_3$ contributions, which do not have
any explicit dependence on $\mu$; and therefore the scale dependence
of the $\NLO_4$ follows the scale dependence of the $\LO_3$ to a large
extent.

\begin{table}[t]
\small
\begin{center}
\begin{tabular}{c c c c}
\toprule
$\delta[\%]$ & $\mu= H_T/8$ & $\mu=H_T/4$ & $\mu = H_T/2$ \\
\midrule 
LO${}_2$ & $-18.7$ & $-20.7$ & $-22.8$ \\
LO${}_3$ & $26.3$ &   $31.8$ &   $37.8$ \\
LO${}_4$ & $0.05$ &   $0.07$ &   $0.09$ \\
LO${}_5$ & $0.03$ &   $0.05$ &   $0.08$ \\
\midrule 
NLO${}_1$ & $33.9$ & $68.2$ & $98.0$ \\
NLO${}_2$ & $-0.3$ & $-5.7$ & $-11.6$ \\
NLO${}_3$ & $-3.9$ & $1.7$ & $8.9$ \\
NLO${}_4$ & $0.7$ & $0.9$ & $1.2$ \\
NLO${}_5$ & $0.12$ & $0.14$ & $0.16$ \\
NLO${}_6$ & $< 0.01$ & $<0.01$ & $<0.01$ \\
\midrule 
$\NLO_2+\NLO_3$ & $-4.2$ & $-4.0$ & $2.7$ \\
\bottomrule
\end{tabular}
\caption{$\ft$: $\sigma_{\LNLO_i}/\sigma_{\LOQCD}$ ratios at 100 TeV, for different values of $\mu=\mu_r=\mu_f$.}
\label{table:4torders100}
\end{center}
\end{table}

\subsubsection*{Differential distributions}

\begin{figure}[t]
\centering
\includegraphics[width=0.45\textwidth]{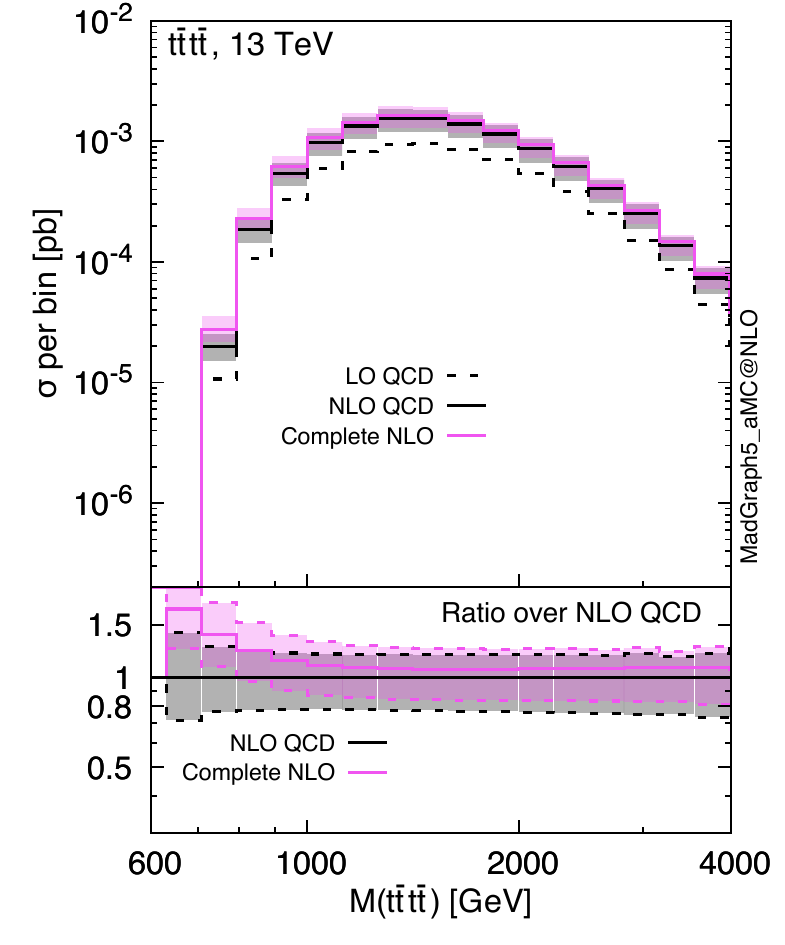}
\vspace{0.4cm}
\includegraphics[width=0.45\textwidth]{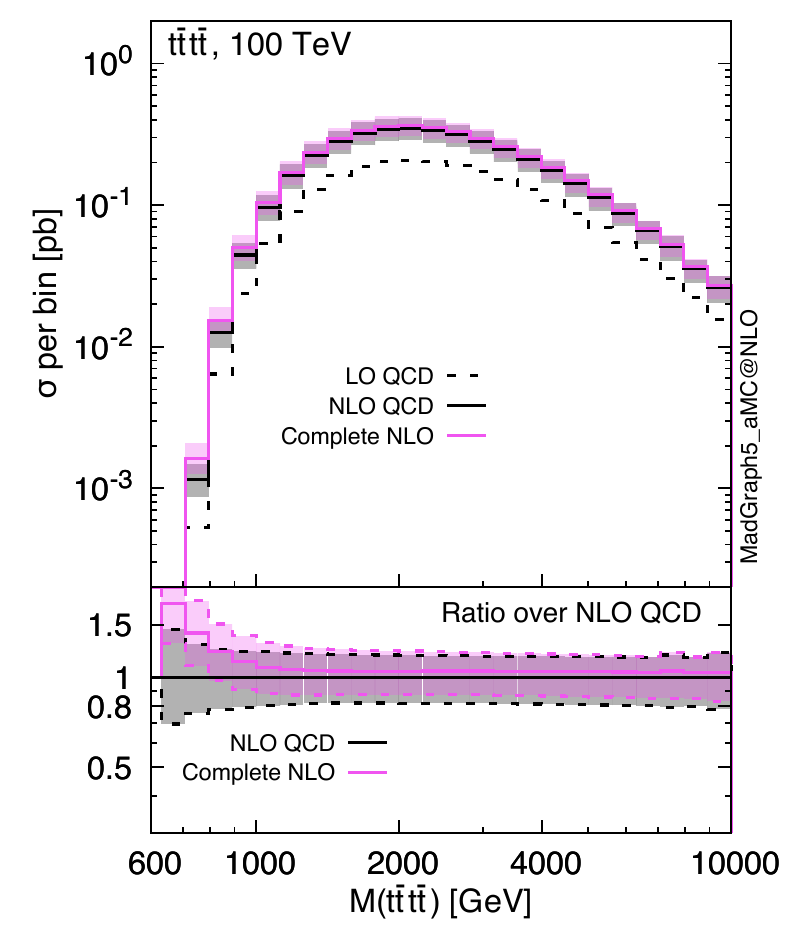}
\vspace{0.4cm}
\includegraphics[width=0.45\textwidth]{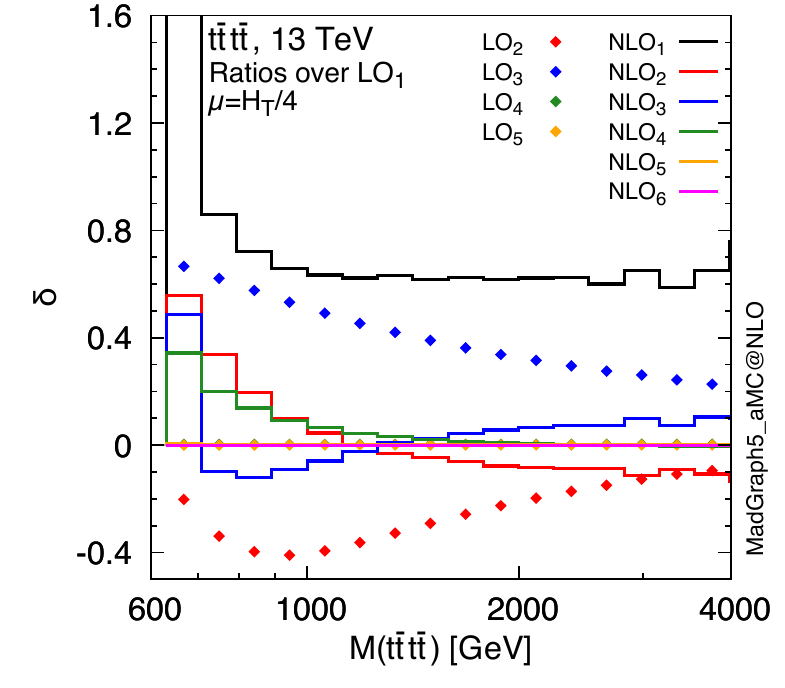}
\includegraphics[width=0.45\textwidth]{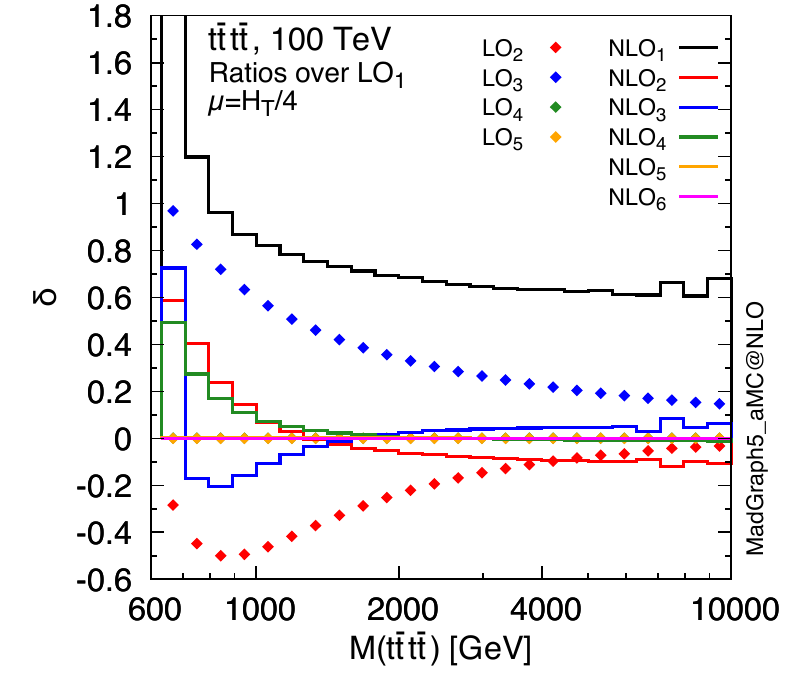}
\includegraphics[width=0.45\textwidth]{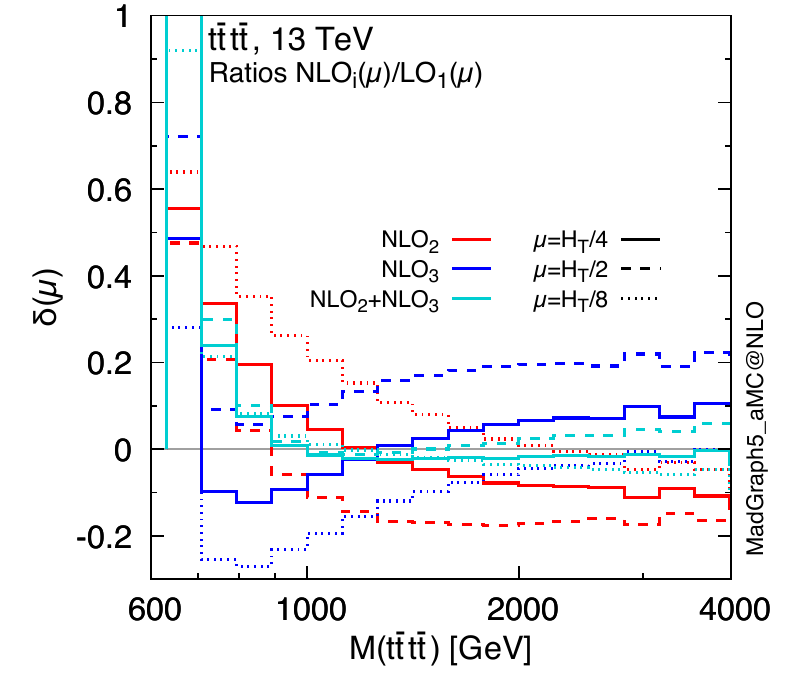}
\includegraphics[width=0.45\textwidth]{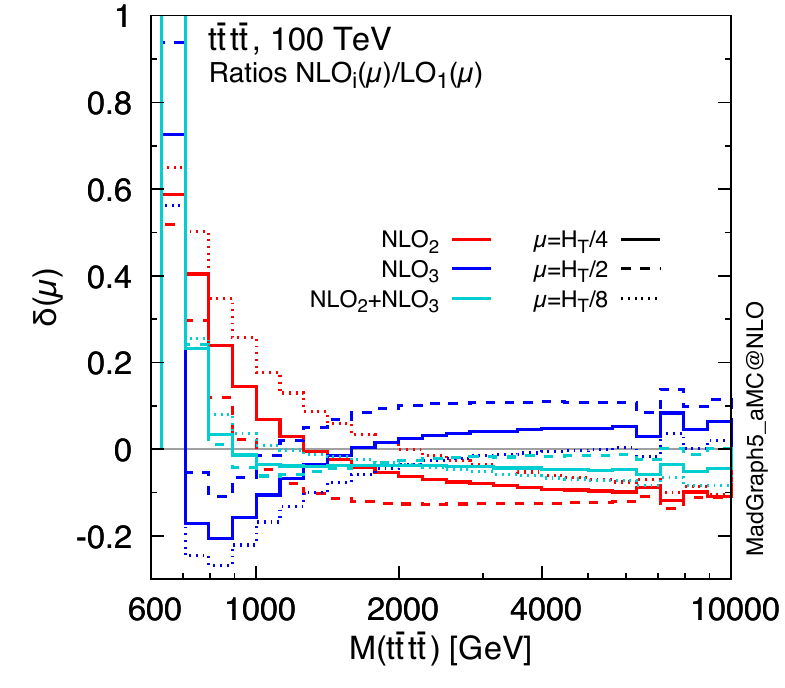}
\caption{The $m(\ft)$ distribution in $\ft$ production. Left: 13 TeV. Right: 100 TeV. Upper plots: scale uncertainty bands (same layout as the plots in Figs.~\ref{fig:ttw13} and \ref{fig:ttw100}). Central plots: individual $\LNLO_i$ contributions  normalised to $\LO_1\equiv\LOQCD$. Lower plots: same as central plots but only with  $\NLO_2$, $\NLO_3$, and their sum, at different values of the scale $\mu$. These lower plots do not show scale uncertainties. Note that $\NLO_1\equiv\NLOQCD$ and $\NLO_2\equiv\NLOEW$.}
\label{fig:4tm4t}
\end{figure}

\begin{figure}[t]
\centering
\includegraphics[width=0.45\textwidth]{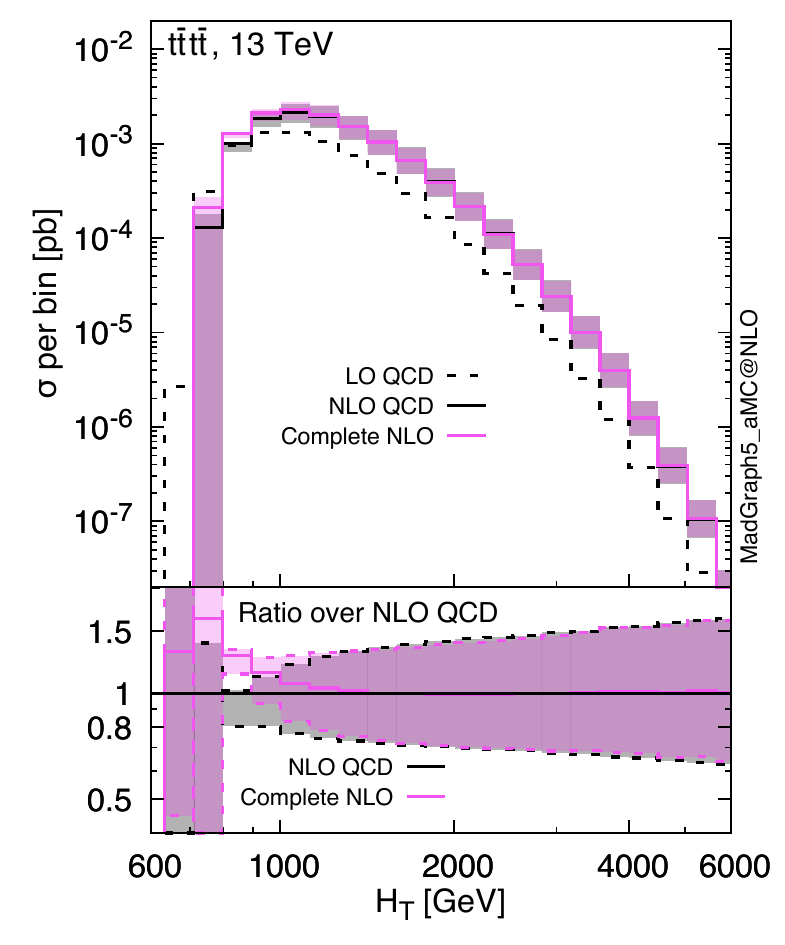}
\vspace{0.4cm}
\includegraphics[width=0.45\textwidth]{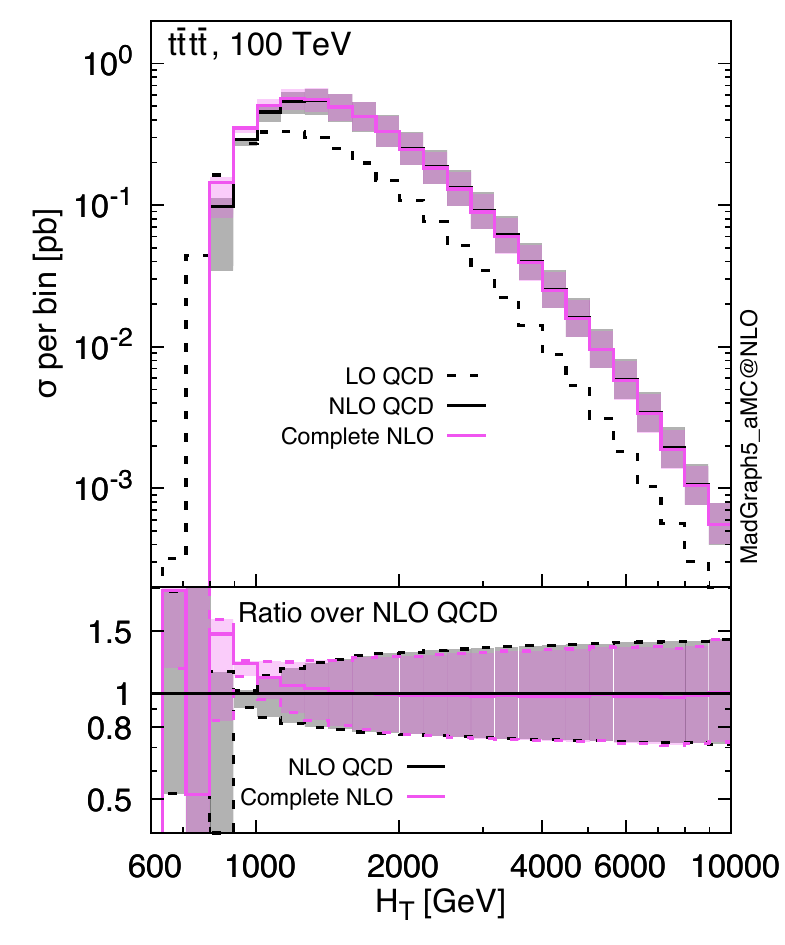}
\vspace{0.4cm}
\includegraphics[width=0.45\textwidth]{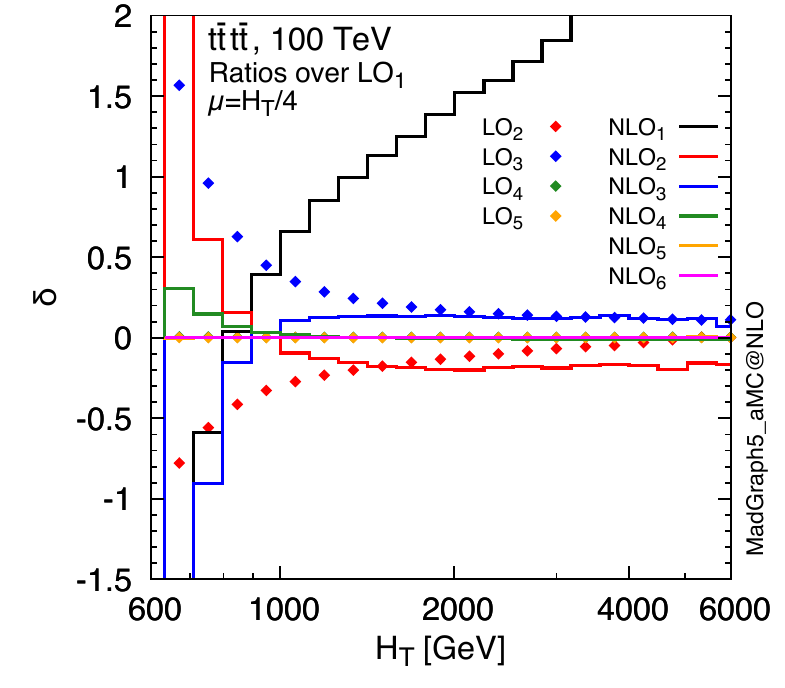}
\includegraphics[width=0.45\textwidth]{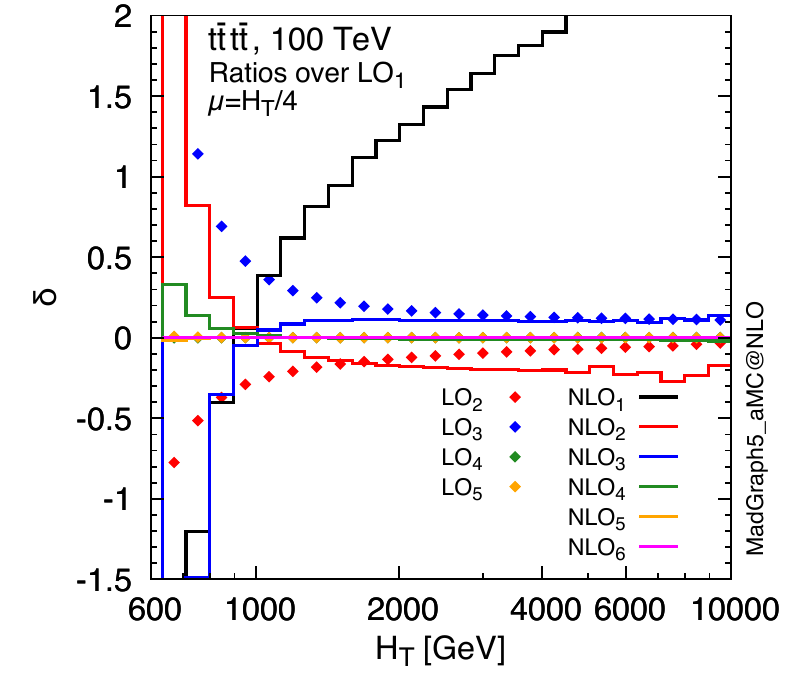}
\includegraphics[width=0.45\textwidth]{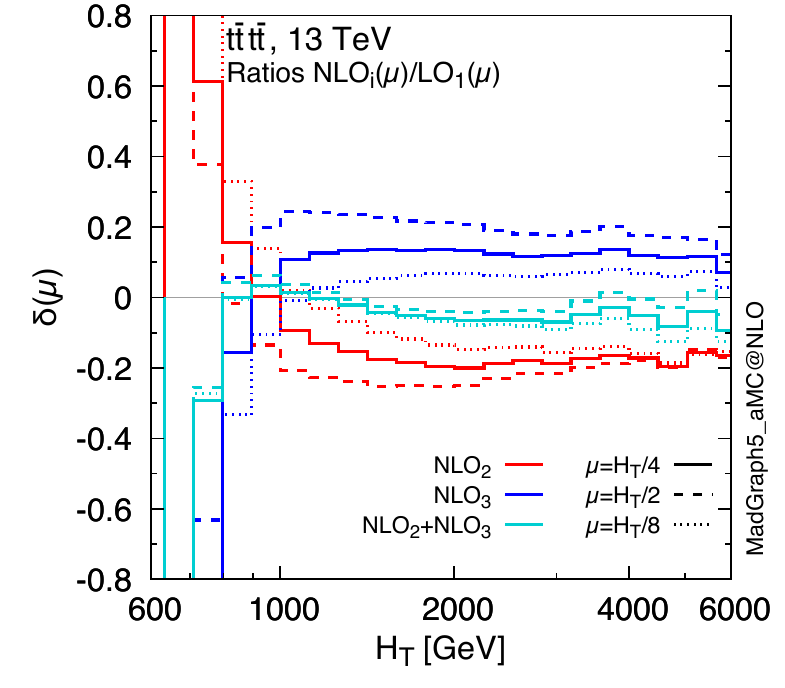}
\includegraphics[width=0.45\textwidth]{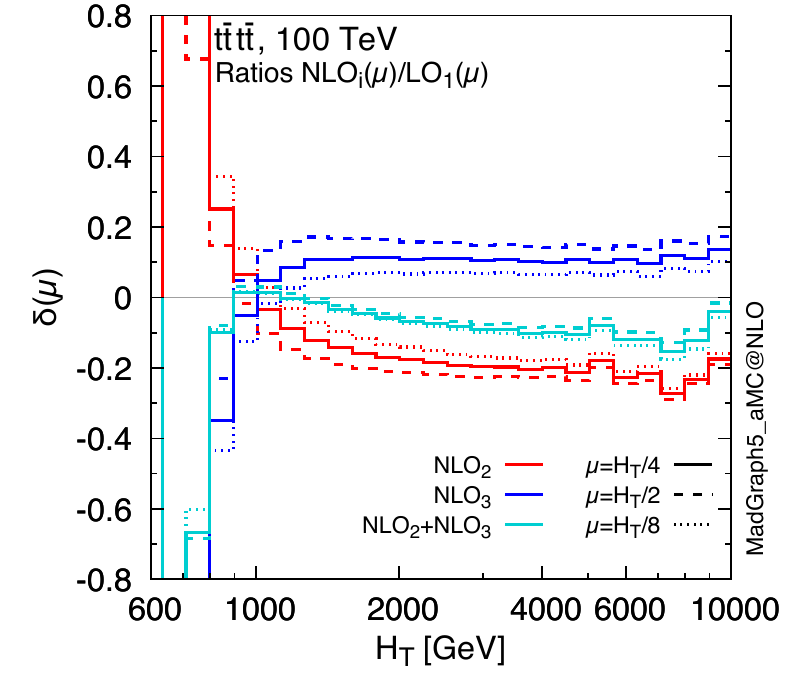}
\caption{The $H_T$ distribution in $\ft$ production. See the caption of Fig.~\ref{fig:4tm4t} for the description of the plots.}
\label{fig:4tht}
\end{figure}

\begin{figure}[t]
\centering
\includegraphics[width=0.45\textwidth]{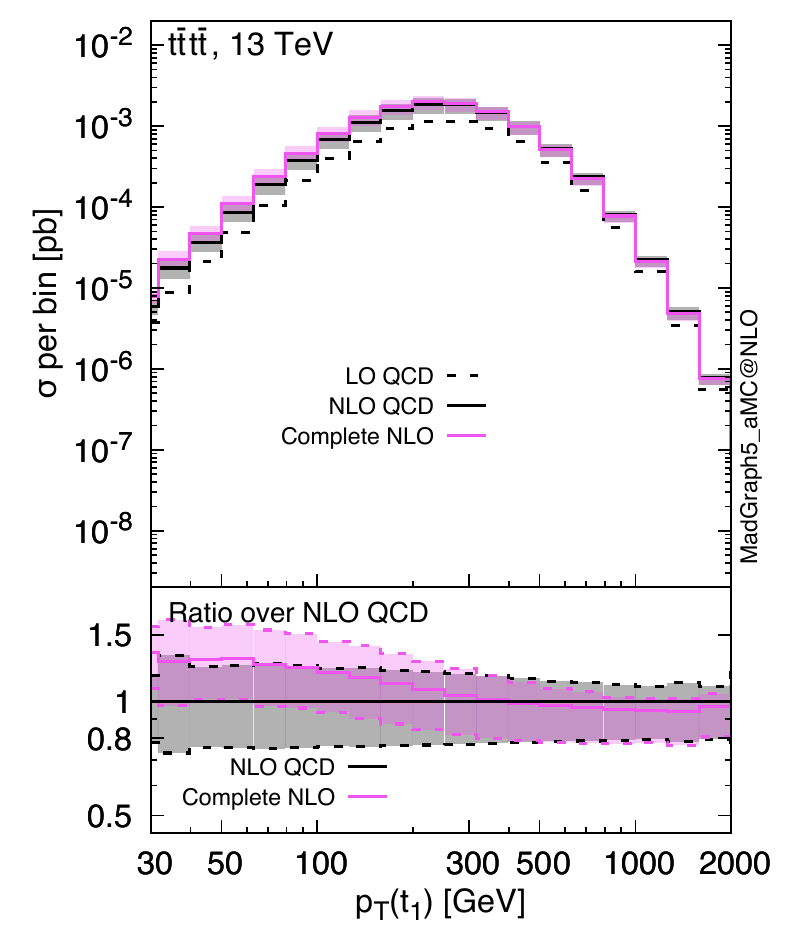}
\vspace{0.4cm}
\includegraphics[width=0.45\textwidth]{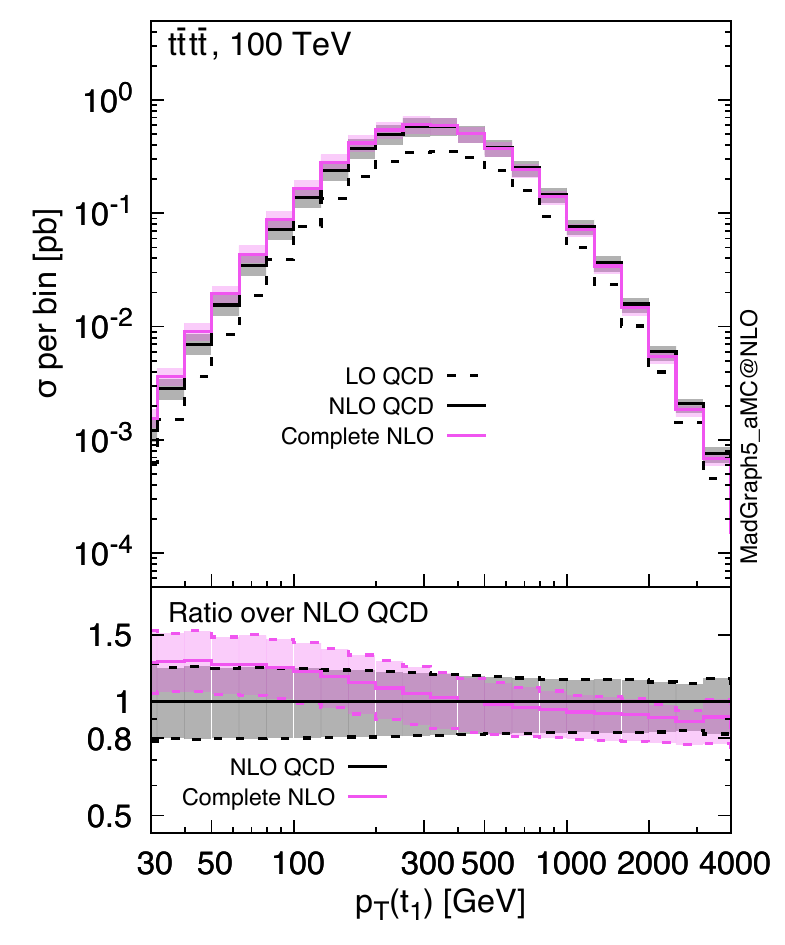}
\vspace{0.4cm}
\includegraphics[width=0.45\textwidth]{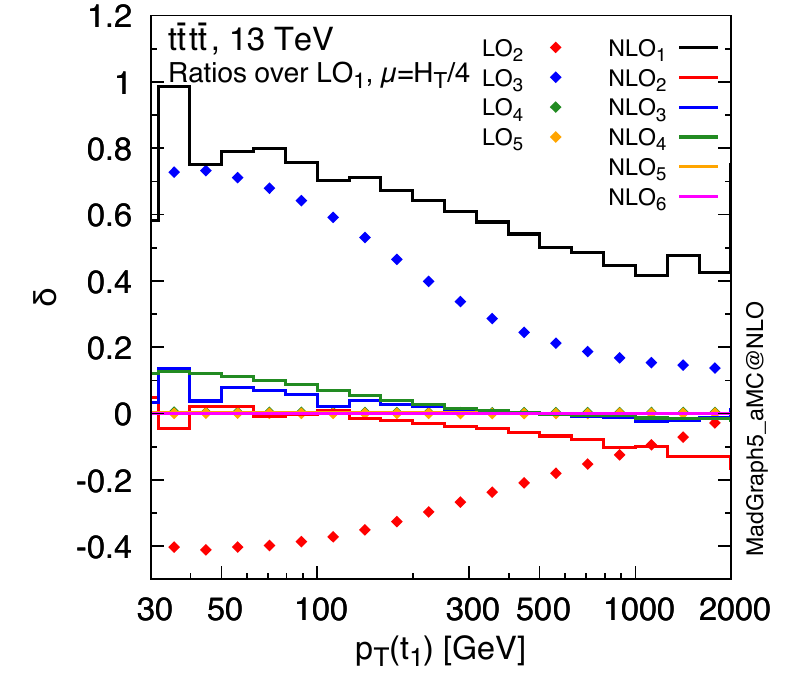}
\includegraphics[width=0.45\textwidth]{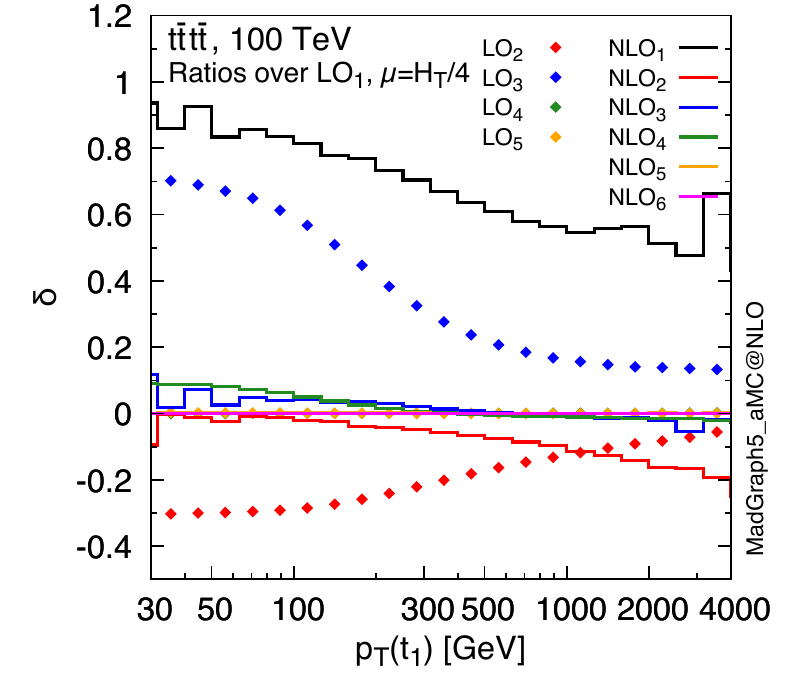}
\includegraphics[width=0.45\textwidth]{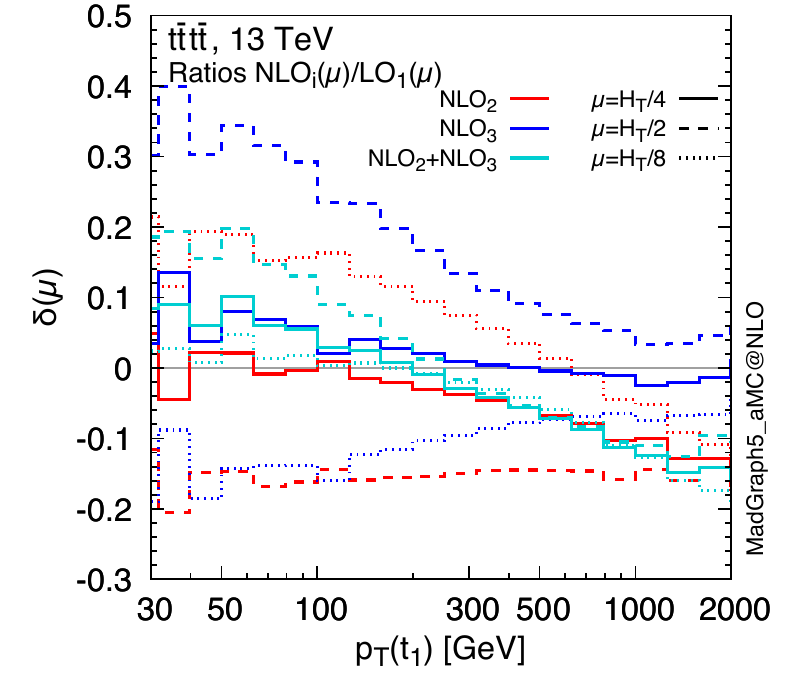}
\includegraphics[width=0.45\textwidth]{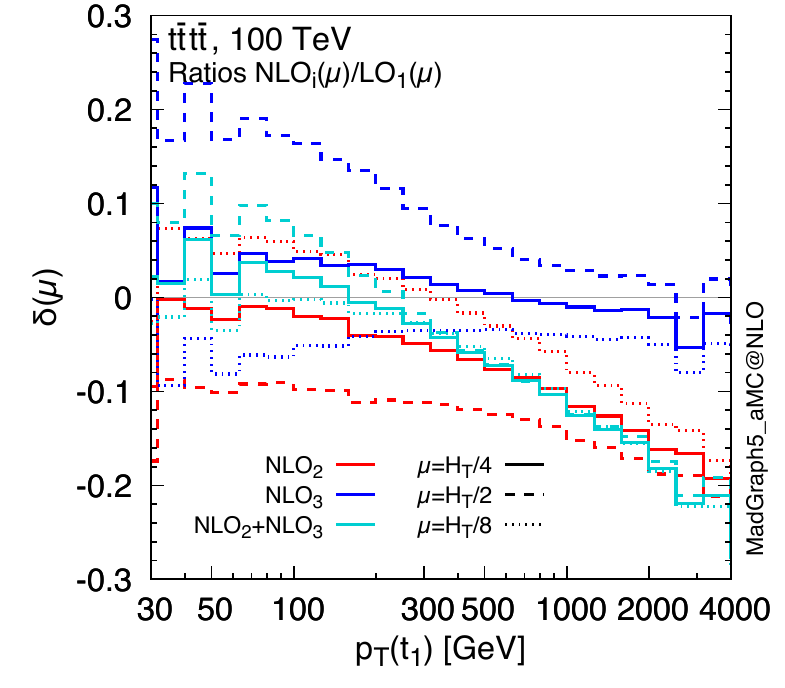}
\caption{The $\pt(t_1)$ distribution in $\ft$ production. See the caption of Fig.~\ref{fig:4tm4t} for the description of the plots.}
\label{fig:4tpt1}
\end{figure}

\begin{figure}[t]
\centering
\includegraphics[width=0.45\textwidth]{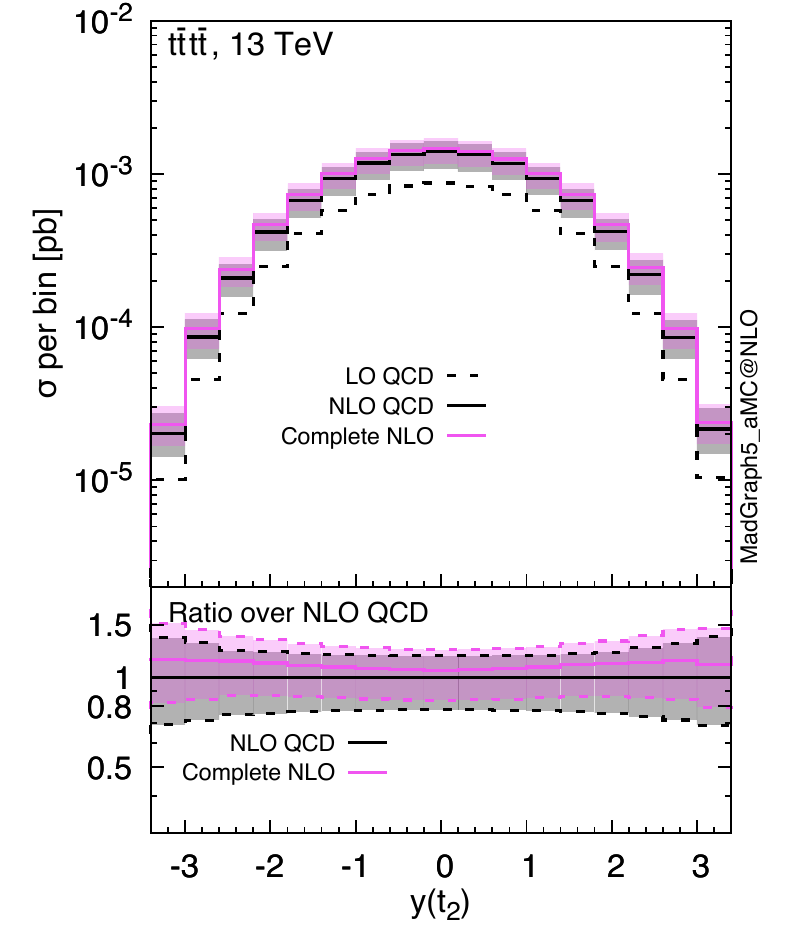}
\vspace{0.4cm}
\includegraphics[width=0.45\textwidth]{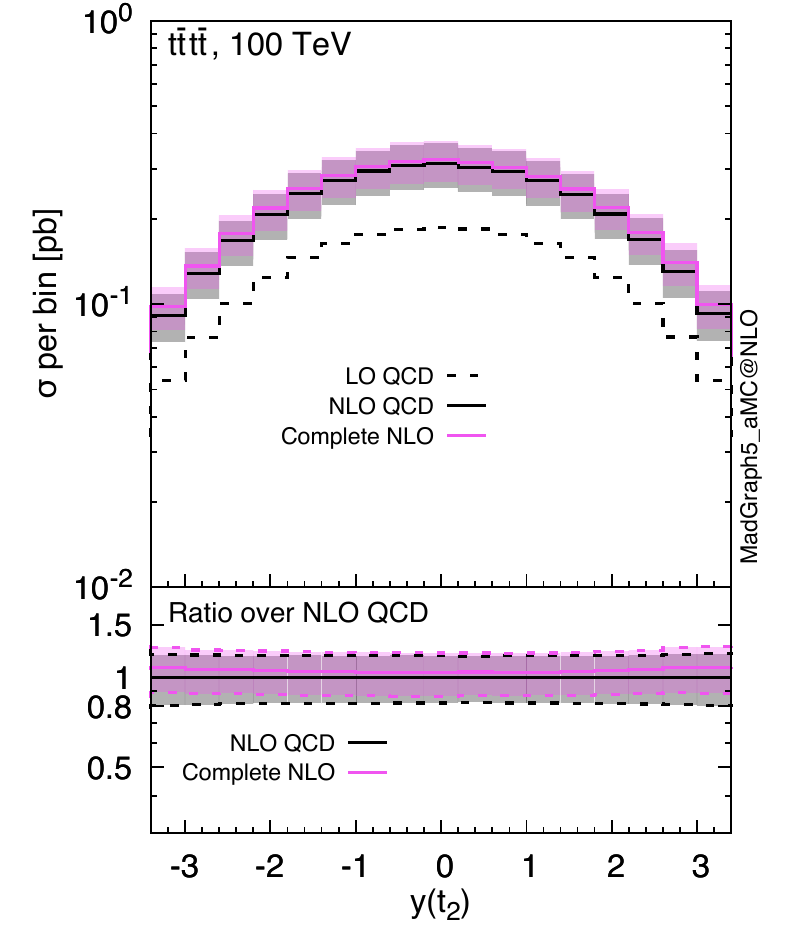}
\vspace{0.4cm}
\includegraphics[width=0.45\textwidth]{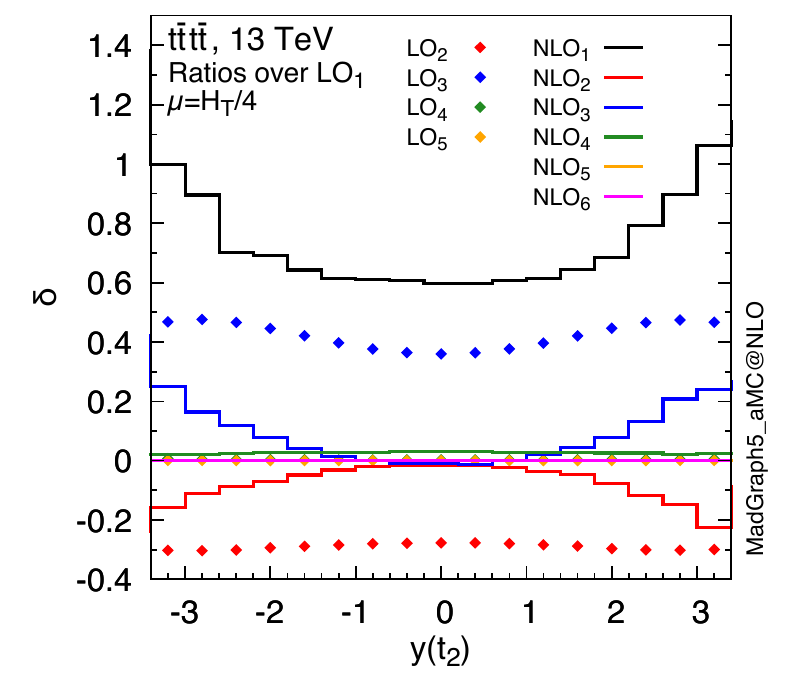}
\includegraphics[width=0.45\textwidth]{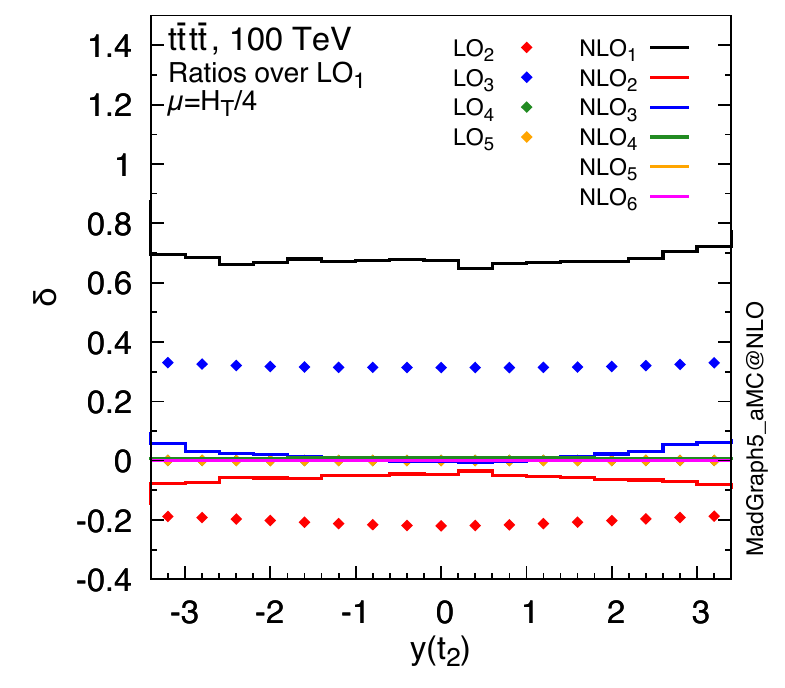}
\includegraphics[width=0.45\textwidth]{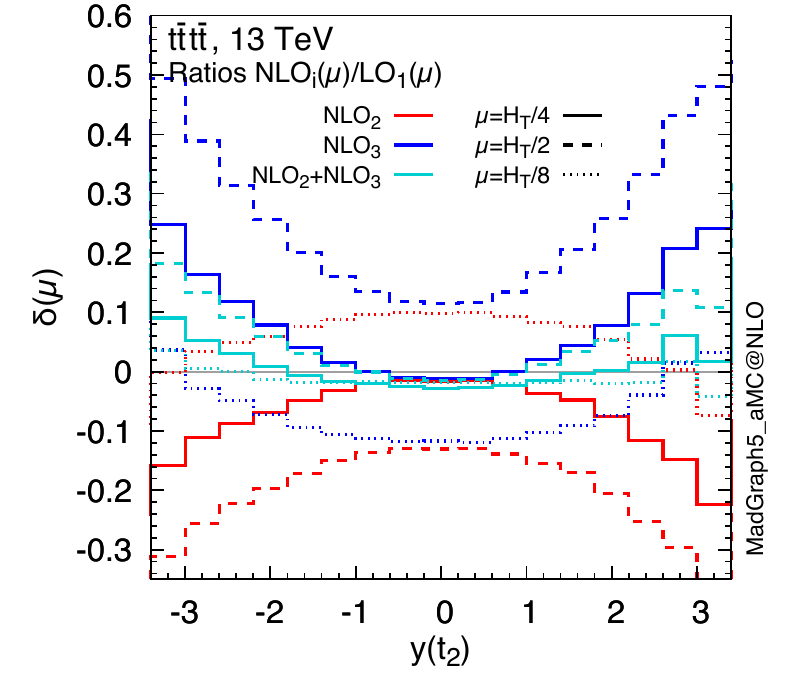}
\includegraphics[width=0.45\textwidth]{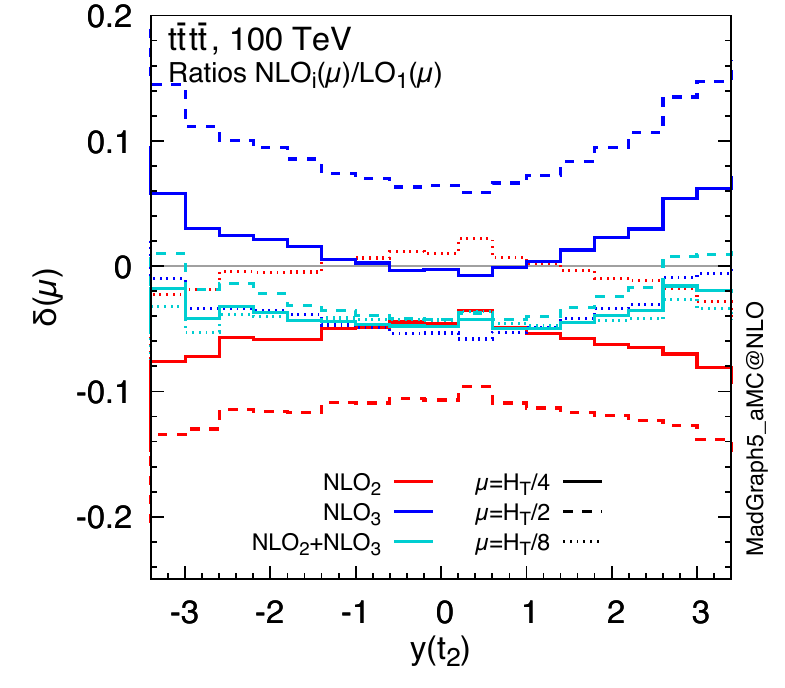}
\caption{The $y(t_2)$ distribution in $\ft$ production. See the caption of Fig.~\ref{fig:4tm4t} for the description of the plots.}
\label{fig:4tyt2}
\end{figure}

We now move to the description of the results at the differential
level, where we consider the following distributions: the invariant
mass of the four (anti)top quarks $m(\ft)$ (Fig.~\ref{fig:4tm4t}), the
sum of the transverse masses of all the particles in the final state
$H_T$ as defined in eq.~\eqref{eq:HT} (Fig.~\ref{fig:4tht}), the
transverse momenta of the hardest of the two top quarks $\pt(t_1)$
(Fig.~\ref{fig:4tpt1}), and the rapidity of the softest one $y(t_2)$
(Fig.~\ref{fig:4tyt2}).  At variance with the case of $\ttw$
production in sec.~\ref{sec:ttw}, we organise plots according to the
observable considered. In the figures we display 13 TeV results on the
left and 100 TeV results on the right. In the upper plots of each of
these figures we provide predictions at different levels of accuracy,
using a similar layout\footnote{At variance with $\ttw$ production, we
  do not show $\LOQCD + \NLOQCD + \NLOEW$ predictions. This level of
  accuracy is rather artificial, since the $\NLOEW\equiv \NLO_2$ terms are
  dominated by ``QCD corrections'' to the $\LO_2$ ones. Hence,
  including $\NLO_2$ without $\LO_2$ would not be very
  consistent. Moreover, there are large cancellations between $\LO_2$
  and $\LO_3$, so, including only the former and not the latter would
  not be giving a correct picture. On top of this, from the inclusive
  results, we already know that there are also large cancellations
  between the $\NLO_2$ and $\NLO_3$ terms. Given the dominance of the
  $gg$-induced contributions, $\sum_{i=1}^3 \LNLO_i$ is already very
  close to the complete-NLO predictions, hence we show only the latter
  and compare them to the pure-QCD NLO predictions.} as in
Figs.~\ref{fig:ttw13} and \ref{fig:ttw100}, which is described in
detail in sec.~\ref{sec:ttw}. Also for $\ft$ production, comparisons
among the scale uncertainties of the $\LOQCD$ and $\LOQCD + \NLOQCD$
result have been documented in detail in ref.~\cite{Maltoni:2015ena}
for 13 TeV, so they are not repeated here. Individual contributions
from the different $\LNLO_i$ terms are instead displayed in the
central and lower plots. In the central plots we show the
$\delta_{\LNLO_i}(\mu)$, see eq.~\eqref{delta}, with $\mu=\mu_c\equiv
H_T/4$, while the lower plots focus on $\NLO_2$ and $\NLO_3$
contributions and their sum featuring large cancellations. In
particular, we show $\delta_{\NLO_2}(\mu)$, $\delta_{\NLO_3}(\mu)$ and
their sum for $\mu=\mu_c$ (solid line), $\mu=\mu_c/2$ (dashed line)
and $\mu=2\mu_c$ (dotted line). In practice, the dark-blue and red
solid lines are the same quantities in the middle and lower plots.
Once again, we remark that the $\delta_{\NLO_i}(\mu)$ ratio does not
show directly the scale uncertainty since the value of $\mu$ is varied
both in the numerator and the denominator of $\delta$.

Away from the threshold region, {\it i.e.}, $m(\ft)>$~900~GeV, the
complete-NLO prediction for the four-top invariant-mass distribution is very
close to the NLO QCD one, with an almost constant increase of
about 10\%, both at 13 and 100~TeV, see upper plots in
Fig.~\ref{fig:4tm4t}. This increase is well within the uncertainty
bands of either of the predictions. On the other hand, in the
threshold region the enhancement of the cross section due to terms
with $\LNLO_i$, with $i>1$, is much larger than for the inclusive
results. In this region the central value of the complete-NLO predictions
lies outside the $\LO_1+\NLO_1$  uncertainty band. From the central
plots of Fig.~\ref{fig:4tm4t}, it can be seen that the $\LNLO_2$ and
$\LNLO_3$ contributions are individually sizeable w.r.t.~$\LOQCD$ and
their relative impact has a large dependence on kinematics, easily
reaching several tens of percents in certain regions of phase
space.

 As anticipated from the inclusive results, there are large
cancellations in the distributions among $\LO_2$ and $\LO_3$
contributions and especially among $\NLO_2$, $\NLO_3$ ones; the latter
are explicitly shown in the lower plots. In particular, although  the corresponding $\delta_{\LNLO_{i}}$ terms individually depend on the value of $m(\ft)$,  they lead for $m(\ft)>$~900~GeV to the aforementioned constant increase of about 10\% of the complete-NLO prediction w.r.t.~the NLO QCD result. As can be seen in the central plots, the $\delta_{\LO_2}$ is negative, it is about $-10\%$ at $m(\ft)\simeq$~4000~GeV
and further decreases for smaller invariant masses, reaching about
$-40\%$ at $m(\ft)\simeq$~900~GeV. On the other hand, the $\delta_{\LO_3}$ is positive, and very close to  the absolute value of $\delta_{\LO_2}$ 
plus a constant 12 (at 13~TeV) or 16 (at 100~TeV) percentage points.  Moreover, even though also the $\delta_{\NLO_2}$ and
$\delta_{\NLO_3}$ are depending quite strongly on the value of $m(\ft)$, they sum to almost a constant $-1\%$ (at 13~TeV) and
$-4\%$ (at 100~TeV). Therefore, indeed, the entire sum
$\LO_2+\LO_3+\NLO_2+\NLO_3$ is almost a constant 10\% correction to
the $\LO_1+\NLO_1$ ---away from the threshold region.

In the threshold region, the situation is quite different. While the
$\delta_{\LO_3}$ keeps increasing closer and closer to threshold, the
derivative of $\delta_{\LO_2}$ reverses sign at
$m(\ft)\simeq$~900~GeV. In other words, the $\delta_{\LO_2}$ also starts
to increase closer and closer to threshold. The same is true for the
corrections induced by $\NLO_2$ and $\NLO_3$ contributions: the $\delta_{\NLO_3}$ sharply increases
close to threshold. Hence, the delicate cancellation among the $\LO_2$
and $\LO_3$ (and $\NLO_2$ and $\NLO_3$) contributions completely breaks down in this
region of phase space. Moreover, also the $\NLO_4$ reaches several
tens of percent close to threshold and should not be neglected when
studying this region of phase space. Conversely, also at the
differential level, $\LO_4$, $\LO_5$, $\NLO_5$ and $\NLO_6$
contributions are negligible.

There are two different physical effects at the origin of the large
NLO corrections in the threshold region. First, also the $\LO_2$ and
$\LO_3$ contributions are larger in this region and thus their ``QCD
corrections'', which respectively enter the $\NLO_2$ and $\NLO_3$
contributions, preserve this increment w.r.t.~the rest of the phase
space. Second, the exchange of $Z$ or Higgs bosons among top quarks,
or in general among heavy particles, can lead to Sommerfeld
enhancements when the top quarks are in a non-relativistic
regime. This effect has already been documented in
refs.~\cite{Kuhn:2013zoa, Beneke:2015lwa} for the case of top-quark
pair production and in refs.~\cite{Degrassi:2016wml, Bizon:2016wgr,
  Maltoni:2017ims} for the exchange of a virtual Higgs boson between
an on-shell Higgs boson and another on-shell heavy particle. The
threshold region forces each $t \bar t$, $tt$ or $\bar t \bar t$ pair
to potentially lead to this kind of effect. These large ``EW
corrections'' on top of $\LO_1$ and $\LO_2$ terms lead to additional
sizeable contributions to $\NLO_2$ and $\NLO_3$,
respectively. Moreover, since also $\LO_3$ is large, via this kind of
``EW corrections'' even $\NLO_4$ is very large and incredibly enhanced
w.r.t.~the result at the inclusive level.

The lower plots in Fig.~\ref{fig:4tm4t} further confirm the QCD origin
of the $\NLO_2$ and $\NLO_3$ contributions. In order to explain this, we
remind the reader that the scale dependence of the $\LO_2$ and $\LO_3$
contributions is the typical one, {\it i.e.}, $\LO_2$ and $\LO_3$ absolute values
become smaller  when the scales are increased.  In
the plots we see that for $\NLO_3$ the (dark blue) dashed lines are
larger than the solid lines, which are in turn larger than dotted
lines, while in the case of $\NLO_2$ the order is the reversed. Since
the $\LO_2$ is negative, the $\NLO_2$ term
reduces the $\mu$ dependence of the $\LO_2$ one and, similarly, the
$\NLO_3$ term reduces the $\mu$ dependence of the $\LO_3$
one. Moreover, these plots confirm that also at the differential level
there are large cancellations among the $\NLO_2$ and $\NLO_3$ terms
and that the $\delta_{\NLO_2}+\delta_{\NLO_3}$ sum has a much smaller
scale dependence than the two separate addends. In other words, the
remarkable cancellations among the $\NLO_2$ and $\NLO_3$ corrections
are not only present for the central value of $\mu$, as already
concluded from the middle plots in the discussion above, but also for their scale dependencies. Notably, these cancellations are present over a very large region of phase space.
Also, if we had chosen, {\it e.g.}, $H_T/2$ as our central scale (dashed
lines in the lower plot), the $\NLO_2$ and $\NLO_3$ curves in the
middle plots would have been much further apart, leading to much
larger cancellations, since their sum would hardly have changed at
all.

Compared to the invariant-mass distribution of the four tops, the case
of the $H_T$ distribution  (Fig.~\ref{fig:4tht}) is similar in many
respects. In particular, from the upper plots, we see that again only
in the threshold region there is a sizeable difference between the NLO
QCD predictions and the complete-NLO ones. It should be noted, though, that
above the peak in the distribution, $H_T\gtrsim$~1500~GeV, the
difference between the two predictions is very small, their central
values as well as the scale uncertainties are lying almost exactly on
top of each other. Just as in the case of the $m(\ft)$, the middle plots
show that this is rather due to large and accidental cancellations among
the various $\LNLO_i$ with $i>1$ contributions, which can individually
reach several tens of percent.

Close to the $H_T\simeq 4m_t$ threshold, the $\NLO_i$ contributions
are in general reverted in sign w.r.t.~the $\LO_i$ ones and receive
particularly large enhancements in absolute value. This feature is due
to large negative QCD Sudakov logarithms that appear in the limit
$H_T\to 4 m_t$. Indeed, since $H_T$ includes in its definition the
momentum of the possible extra jet, it effectively acts as a tight jet
veto in this limit. Thus, ``QCD corrections'' involves large and
negative contributions that have to be resummed. The effect is so
large that in the first bin of the central plots of
Fig.~\ref{fig:4tht}, the $\LOQCD + \NLOQCD$ prediction is negative and
should not be trusted. This is a well-known instability of fixed-order perturbative calculations. Similar but smaller effects originate also from
``EW corrections'', due to the effective veto on the real emission.

It is also interesting to note how the $\mu$-dependence of
$\delta_{\NLO_2}$ reduces for large values of $H_T$ (see bottom plots
of Fig.~\ref{fig:4tht}). We can see in the central plots that
$\delta_{\LO_2}$ is very small in this phase-space region, which means
that the dominant $\NLO_2$ contribution cannot be originated by ``QCD
corrections'' on top of $\LO_2$. Rather, it is mainly induced by ``EW corrections'' on top
of the $\LOQCD$ term. Thus, we recover the typical situation, which we
found also in $\ttw$ production, where
$\delta_{\NLO_2}\equiv\delta_{\NLOEW}$ is almost independent of the
value of $\mu$.

An example of an observable in which the cancellation between the
$\NLO_2$ and $\NLO_3$ is less complete in the whole range considered
is the transverse momentum of the hardest of the two top quarks, shown
in Fig.~\ref{fig:4tpt1}. Similarly to $m(\ft)$ and $H_T$, close to the
threshold region, $\pt(t_1)\lesssim$~300~GeV, the complete-NLO
predictions are above the NLO QCD ones, reaching $\sim\!\!25\%$ at
very small transverse momenta. On the other hand, for
$\pt(t_1)\gtrsim$~300~GeV, the complete-NLO corrections on top of the
NLO QCD are growing negative and become about $-10\%$ in the tails of
the distributions shown. From the middle plots, which refer to the case $\mu=H_T/4$, it becomes clear which
orders are responsible for this behaviour. At small transverse momenta
there are large positive corrections from the $\LO_3$ (up to about
70\% on top of $\LO_1$) and to a lesser extent the $\NLO_4$, which is
itself slightly larger than $\NLO_3$. $\LO_2$ is also large, but
negative, about $-40\%$ on top of $\LO_1$, only partially cancelling
the large positive contribution from $\LO_3$. Accidentally, $\NLO_2$ corrections are instead almost
equal to zero.\footnote{Once again we want to remark that, unless differently specified, all the numbers in the main text refer to $\mu=H_T/4$, but they strongly depend on the scale $\mu$. As can be seen from the lower plots, {\it e.g.}, at 13 TeV for small transverse momenta $\delta_{\NLO_2}(H_T/4) \sim 0\% $, but $\delta_{\NLO_2}(H_T/8) \sim 20\% $ and $\delta_{\NLO_2}(H_T/2) \sim - 20\% $ }  Adding together all these contributions and taking also into account that the
$\NLO_1$ yields a positive 80\% correction, we indeed find close to the threshold  a
 correction of about 25\%  from complete-NLO result on top of the NLO QCD one. On the other hand, with increasing
$\pt$, all the corrections quickly reduce (in absolute value),
although not all in a uniform way. The exception is the $\delta_{\NLO_2}$,
which steadily grows negative. Thus, at transverse momenta in the TeV
range, the $\NLO_2\equiv \NLOEW$ becomes the dominant correction
to the NLO QCD predictions. At first sight, this seems to be the standard situation with NLO EW corrections completely dominated by Sudakov logarithms, which we 
also observed in  the $\NLO_2$ curves for the $pp\to\ttw$ process, see
Figs.~\ref{fig:ttw13orders} and \ref{fig:ttw100orders}. However,
looking at the lower plots, it is clear that this cannot be the complete story. If the
$\NLO_2$ had been completely dominated by ``EW corrections'' on
top of the $\LO_1$, the $\delta_{\NLO_2}$ ratio would have been (almost) scale
independent. Conversely, although the scale dependence of $\delta_{\NLO_2}$ does decrease
with increasing transverse momenta, it remains anyway sizeable even
in the far tail of the distribution. Therefore, a non-negligible part of
$\NLO_2$ is due to ``QCD corrections'' on top of the
$\LO_2$ also in the far tail. For these reasons, although in this phase-space region the individual and summed $\delta_{\NLO_i}$ with 
 $i>1$ are not at all constant,  the scale dependence of 
$\delta_{\NLO_2}+\delta_{\NLO_3}$ remains very small. The non-constant
part seems to be the ``EW corrections'' entering the $\NLO_2$, which are dominated by large and negative Sudakov logarithms and do not
introduce a new scale dependence w.r.t.~the $\LO_1$.

From the $y(t_2)$ distribution (Fig.~\ref{fig:4tyt2}) we can see that, besides the threshold region, a non-negligible difference between NLO QCD and complete-NLO predictions is present also at 13 TeV (not 100 TeV) in the peripheral region of the softest of
the top quark quarks. The $y(t_2)$ distribution
 is also the only one, among those considered, 
where the impact of the different $\LNLO_i$ terms is qualitatively
different at 13 and 100~TeV. While the $\LO_i$ corrections are rather flat at both 13
and 100~TeV,  $\NLO_i$ corrections are flat only at 100~TeV; the
$\NLO_i$ corrections for 13~TeV yield large effects in the peripheral
region.  The origin of this difference is the range of Bjorken-$x$
probed in the PDFs, which is indeed very different at 13 and
100~TeV. While at 13~TeV the peripheral region is typically associated
with tops that have large rapidities also in the $\ft$ rest frame, at
100~TeV it is more likely that they originate from partonic initial
states that are boosted w.r.t.~the proton--proton reference
frame.\footnote{The maximum value for the rapidity of the $\ft$ system
  in a Born-like configuration is $\log\left(\frac{13~\rm TeV}{4
    m_t}\right)\sim 3$ at 13 TeV, while it is $\log\left(\frac{100~\rm
    TeV}{4 m_t}\right)\sim 5$ at 100 TeV.}  For this reason the
$y(t_2)$ distribution is flatter at 100 TeV than at 13 TeV, where large
rapidities are strongly suppressed in a Born-like kinematics and
therefore they are also much more sensitive to effects due to real
emission from $\NLO_i$ contributions. However, as before, the $\NLO_2$
and $\NLO_3$ contributions almost cancel, resulting in at most
$\sim\!\!10\%$ effects w.r.t.~the $\LO_1$ in the far forward and
backward regions.

Given our findings, we suggest that the study of the $\mu$-dependence of $\delta_{\NLO_i}$   can be  a very useful procedure for identifying the nature of $\NLO_i$ corrections in numerical calculations. For higher values of $i$, the $\Sigma_{\NLO_i}(\mu)/\Sigma_{\LO_{i-1}}(\mu)$ may be even more appropriate given the different numerical sizes of the $\LO_i$ terms and of their dependence on the running of $\alpha_s$.\footnote{Note that $\Sigma_{\NLO_i}/\Sigma_{\LO_{i-1}}=\delta_{\NLO_i}/\delta_{\LO_{i-1}}$, so at the inclusive level the necessary information can be obtained from Tabs.~\ref{table:4torders} and \ref{table:4torders100}. } For instance, we verified that in $\ft$ production both $\Sigma_{\NLO_4}/\Sigma_{\LO_{3}}$ and $\Sigma_{\NLO_6}/\Sigma_{\LO_{5}}$ are very mildly scale-dependent at inclusive and differential level. 
Indeed, both can be considered almost purely ``EW corrections''; the latter by construction and the former due to the dominance of the $gg$ initial-state. Conversely, we do not find this feature in the  $\Sigma_{\NLO_5}/\Sigma_{\LO_{4}}$ ratio, since $\LO_4$ and $\LO_5$ contributions are both small but comparable in size and thus $\Sigma_{\NLO_5}$ receives large ``QCD corrections'' on top of $\LO_5$ contributions.

In summary, at the inclusive and the differential levels complete-NLO
results for $\ft$ production are well within the NLO QCD
uncertainties. For the observables presented here, there are no large
qualitative differences between results at 13 and 100~TeV, except in
the peripheral regions of the rapidity of the second hardest top
quark. However, for all observables very large cancellations among the
different perturbative orders are present both at the inclusive and
differential level. Their individual sizes w.r.t.~the $\LOQCD$
prediction are also strongly dependent on the scale definition. All
these arguments point to the fact that in any BSM analysis involving
$\ft$ production contributions from all NLO corrections can be
relevant. Thus, they should be taken into account, at least in the
estimate of the theory uncertainty.

\clearpage 

\section{Conclusions}
\label{sec:conclusions}

In this paper we have presented the complete-NLO predictions for
$\ttw$ and $\ft$ production at 13 and 100 TeV in proton--proton
collisions. All the seven $\ord{\alphas^i \alpha^j}$ contributions
with $i+j=3,4$ and $j \ge 1$ for $\ttw$ production and all the eleven
$\ord{\alphas^i \alpha^j}$ contributions with $i+j=4,5$ have been
calculated exactly without any approximation. We have shown that
complete-NLO corrections involve large contributions beyond the NLO EW
accuracy for both the $\ttw$ and $\ft$ production processes

In $\ttw$ production we find that the $\ord{\alpha_s \alpha^3}$
contributions, denoted as $\NLO_3$ in this article, are larger than
NLO EW corrections and have opposite sign. They are of the order
12(70)\% of the LO at 13(100) TeV, with a strong dependence on
particular kinematic variables such as $\pt(\wpm)$ and $\pt(\ttt)$,
but not $m(\ttt)$. Thus, they are several orders of
magnitude larger than the values naively expected from their coupling orders, {\it i.e.},
$\NLO_3/\LO \gg \alpha^2/\alpha_s \sim 0.1\%$. The main reason is the opening of the
$tW \to tW$ scattering in the $\NLO_3$. Since the NLO QCD corrections
are dominated by hard radiation, applying a jet veto
suppresses the $\NLOQCD$ contributions considerably. Conversely, the $\NLO_3$ (and
the NLO EW corrections) are affected to a much lesser
extent, resulting in large corrections on top of the NLO-QCD result. At 13 TeV, applying a
100~GeV central jet veto, the central value
of the complete-NLO prediction is typically outside the NLO QCD
scale-uncertainty band. At 100~TeV, the uncertainty bands of these two predictions do not
even touch. Besides their relevance for the SM and reliable
comparisons with current and future measurements, these results
further support the proposal of the BSM analysis described in
ref.~\cite{Dror:2015nkp}, showing a possible sensitivity to
higher-dimensional operators in $tW \to tW$ scattering directly in
$\ttw$ production. Rather than requiring a jet and considering $tW \to tW$ scattering as a Born process, our results suggest that 
the sensitivity may be increased by directly considering $\ttw$ production and vetoing additional jets.

In $\ft$ production, LO contributions of $\ord{\alpha_s^3 \alpha}$ are
about $-$25-30\% of the purely-QCD $\ord{\alpha_s^4}$ ones, while
$\ord{\alpha_s^2 \alpha^2}$ contributions are about $+$30-45\%,
depending on the scale choice.  For this reason, we find that the
$\ord{\alpha_s^4 \alpha}$ (the NLO EW corrections, or $\NLO_2$) as
well as the $\ord{\alpha_s^3 \alpha^2}$ (denoted as $\NLO_3$ in this
article) contributions are also large. Moreover, since they receive
large contributions from ``QCD corrections'' (and thus $\alpha_s$ and
PDF renormalisation) on top of respectively $\ord{\alpha_s^3 \alpha}$
and $\ord{\alpha_s^2 \alpha^2}$ terms, they strongly depend on the
scale definition. At 13 TeV, their relative impact w.r.t.~purely-QCD
$\ord{\alpha_s^4}$ contribution varies in both cases between
$\pm15\%$. On the other hand, their sum reduces to a rather small
$\pm$1-2\%, and is almost independent from the QCD scale choice and
kinematics. Qualitatively similar results are found also at 100
TeV. The size of the cancellations is quite remarkable, unexpected,
and, to the best of our knowledge, accidental. Thus, a calculation of
only part of the complete-NLO results would be missing important
contributions. These large cancellations between the corrections and
the reduced scale dependencies of their sum are not present very close
to threshold. In this region of phase space, complete-NLO results are
sizeably different from those at NLO QCD accuracy and even
contributions of $\ord{\alpha_s^2 \alpha^3}$ (denoted as $\NLO_4$ in
this article) are found to be of the order of several tens of percents
of the LO.  Besides their relevance for the SM and reliable
comparisons with current and future measurements, our calculations
show that the possible impact of NLO corrections should be critically
considered for studies such as ref.~\cite{Cao:2016wib}, where $\ft$
production has been proposed as candidate, in conjunction with $\ttt
H$ production, for an independent determination of the Yukawa coupling
of the top quark and the Higgs-boson total decay width. Similar
considerations apply to other BSM studies involving $\ft$ production:
the various contributions from NLO corrections are large and the
cancellations among them could be spoiled by BSM
effects. This should be taken into account at least in the estimate of
the theory uncertainties.

In this work we have also shown that the study of the $\mu$-dependence of the quantity $\delta_{\NLO_i}\equiv \Sigma_{\NLO_i}(\mu)/\Sigma_{\LO_{1}}(\mu) $   can be  a very useful procedure for identifying the nature of $\NLO_i$ corrections in numerical calculations. A large scale dependence is a signal of ``QCD corrections'' on top of the $\LO_i$ contribution, while a scale independence for  $\delta_{\NLO_i}$ points to ``EW corrections'' on top of the $\LO_{i-1}$ contributions. For higher values of $i$, the $\Sigma_{\NLO_i}(\mu)/\Sigma_{\LO_{i-1}}(\mu)$ may be even more appropriate given the possible different numerical sizes of the $\LO_i$ terms and of their dependence on the running of $\alpha_s$.

As a final remark, we want to remind the reader that the three known cases where NLO corrections from supposedly subleading EW contributions are large, $pp\to \ttw$, $pp\to \ft$ and $pp\to W^+W^+ j j$ with leptonic $W^+$ decays \cite{Biedermann:2017bss}, involve very different mechanisms. In $\ttw$ production it is the opening of $tW \to tW$ scattering via the real emission in the $\NLO_3$. In $\ft$ production it is mainly the ``QCD corrections'' on top of EW $tt \to tt$ scattering, which gives large contributions already at the LO. In $W^+W^+jj$ production it is instead the large EW Sudakov logarithmic corrections featured by the formally most subleading NLO contribution \cite{Biedermann:2016yds} together with the relatively large size (especially when standard VBS cuts are applied) of the purely EW $W^+W^+ $ scattering component.

\section*{Acknowledgments }

We are grateful
to Hua-Sheng Shao, Stefano Frixione and Valentin Hirschi for the ongoing collaboration on the automation of the calculation of the complete-NLO corrections in the {\sc\small MadGraph5\_aMC@NLO} framework.
We acknowledge Cen Zhang, Ennio Salvioni, Fabio Maltoni, Ioannis Tsinikos and  Mathieu Pellen for enlightening discussions. The work of R.F.~and D.P.~is supported by the Alexander von Humboldt Foundation, in the framework of the Sofja Kovalevskaja Award Project ``Event Simulation for the Large Hadron Collider at High Precision''. The work of M.Z.~has been supported by the Netherlands
National Organisation for Scientific Research (NWO), by the European Union's Horizon 2020 research and
innovation programme under the Marie Sklodovska-Curie grant
agreement No 660171 and in part by the ILP LABEX (ANR-10-LABX-63),
in turn supported by French state funds managed by the ANR
within the ``Investissements d'Avenir'' programme
under reference ANR-11-IDEX-0004-02.

\bibliographystyle{JHEP}
\bibliography{article_4t_ttwm}

\end{document}